\documentclass[12pt]{article}
\usepackage{jheppub}

\usepackage{subcaption,amsmath,bbm,array,amsfonts,graphicx,wrapfig,lscape,float,slashbox,multirow,longtable,rotating,epstopdf,commath}
\usepackage[normalem]{ulem}

\newcommand{\be}{\begin{equation}}
\newcommand{\ee}{\end{equation}}
\newcommand{\beq}{\begin{equation}}
\newcommand{\beql}[1]{\begin{equation}\label{#1}}
\newcommand{\eeq}{\end{equation}}
\newcommand{\ba}{\begin{array}}
\newcommand{\ea}{\end{array}}
\newcommand{\bea}{\begin{eqnarray}}
\newcommand{\beal}[1]{\begin{eqnarray}\label{#1}}
\newcommand{\eea}{\end{eqnarray}}
\newcommand{\ben}{\begin{enumerate}}
\newcommand{\een}{\end{enumerate}}
\newcommand{\bean}{\begin{eqnarray*}}
\newcommand{\eean}{\end{eqnarray*}}

\newcommand{\btab}[1]{\begin{tabular}{#1}}
\newcommand{\etab}{\end{tabular}}

\newcommand{\comment}[1]{}

\newcommand{\IC}{\mathbb{C}}

\newcommand{\qed}{\nobreak \ifvmode \relax \else
      \ifdim\lastskip<1.5em \hskip-\lastskip
      \hskip1.5em plus0em minus0.5em \fi \nobreak
      \vrule height0.75em width0.5em depth0.25em\fi}

\def\beqa{\begin{eqnarray}}
\def\eeqa{\end{eqnarray}}
\def\NN{{\cal N}}

\newcolumntype{C}[1]{>{\centering\arraybackslash}m{#1}}

\newcommand{\IX}{{\bf X}}
\newcommand{\IS}{{\bf S}}
\newcommand{\IZ}{{\mathbb{Z}}}
\def\II{\relax{\rm I\kern-.18em I}} 

\def\makeatletter{\catcode`\@=11}
\makeatletter
\def\mathbox#1{\hbox{$\m@th#1$}}%
\def\math@ccstyles#1#2#3#4#5#6#7{{\leavevmode
     \setbox0\mathbox{#6#7}%
     \setbox2\mathbox{#4#5}%
     \dimen@ #3%
     \baselineskip\z@\lineskiplimit#1\lineskip\z@
     \vbox{\ialign{##\crcr
            \hfil \kern #2\box2 \hfil\crcr
            \noalign{\kern\dimen@}%
            \hfil\box0\hfil\crcr}}}}
\def\mathaccstyles{\math@ccstyles\maxdimen}
\def\maththroughstyles{\math@ccstyles{-\maxdimen}}
\def\unity%
{\maththroughstyles{.45\ht0}\z@\displaystyle {\mathchar"006C}\displaystyle 1}

\newcommand{\drawsquare}[2]{\hbox{%
\rule{#2pt}{#1pt}\hskip-#2pt
\rule{#1pt}{#2pt}\hskip-#1pt
\rule[#1pt]{#1pt}{#2pt}}\rule[#1pt]{#2pt}{#2pt}\hskip-#2pt
\rule{#2pt}{#1pt}}

\newcommand{\fund}{~\raisebox{-.5pt}{\drawsquare{6.5}{0.4}}~}
\newcommand{\antifund}{~\overline{\raisebox{-.5pt}{\drawsquare{6.5}{0.4}}}~}

\def\IC{{\bf C}}
\def\IS{{\bf S}}
\def\IR{{\bf R}}
\def\IZ{{\bf Z}}
\def\IX{{\bf X}}

\def\IT{{\bf T}}

\title{Backreacting D-brane instantons \\on branes at singularities}

\author[a,b]{Eduardo Garc\'{i}a-Valdecasas Tenreiro}
\author[a]{, Angel Uranga}

\affiliation[a]{Instituto de F\'isica Te\'orica UAM-CSIC \\
C/ Nicol\'as Cabrera 13-15, Campus de Cantoblanco,  28049 Madrid, Spain}
\affiliation[b]{ Departamento de F\'isica Te\'orica, Universidad Aut\'onoma de Madrid,\\ Campus de Cantoblanco, 28049 Madrid, Spain }
\emailAdd{eduardo.garcia.valdecasas@gmail.com, angel.uranga@uam.es}

\abstract{Non-perturbative D-brane instanton effects in 4d $\NN=1$ string compactifications can be geometrized in terms of a backreacted generalized geometry. We extend earlier results to setups in which the D-brane instanton is charged under the 4d gauge symmetries, and show that the backreacted topology yields the correct charged field theory operators in the 4d effective action. In type IIA models with D6-branes, the backreaction of D2-brane instantons forces the recombination of D6-branes, such that the 4d charged field theory operators arise from basic worldsheet instantons in the backreacted geometry. We provide large classes of examples of D2-brane instanton effects on intersecting D6-brane systems in local models mirror to D3-branes at singularities. The backreacted geometry and the field theory operators are easily encoded in terms of simple operation in the graphs arising from the underlying dimer diagrams. This description agrees, in the appropriate cases, with the complex deformations triggered by certain fractional branes at the bottom of duality cascades.
}

\preprint{
\begin{flushright}IFT-UAM/CSIC-17-036
\end{flushright} \vspace{-0.9cm}
}

\begin{document}

\maketitle


\section{Introduction} \label{sec:intro}

\bigskip

Non-perturbative effects in string theory are at the core of a fascinating interplay between formal developments  (see e.g. \cite{Becker:1995kb,Witten:1996bn,Harvey:1999as,Witten:1999eg}) and phenomenological applications, e.g. in moduli stabilization \cite{Kachru:2003aw,Balasubramanian:2005zx}. 

These non-perturbative effects include field theoretical gauge instantons, as well as genuinely stringy ones such as D-brane instantons. Recently \cite {Garcia-Valdecasas:2016voz} succeeded in reformulating the non-perturbative potentials from the latter in the language of 3-forms of \cite{Dvali:2005an, Dvali:2005ws} (see also \cite{Kaloper:2008fb, Kaloper:2011jz,Marchesano:2014mla,McAllister:2014mpa,Retolaza:2015sta,Bielleman:2015ina,Ibanez:2015fcv} for recent applications). This follows from considering the  modification of the geometry due to the instanton backreaction in the spirit of \cite{Koerber:2007xk,Koerber:2008sx}. Specifically, the backreaction modifies the topology of the compactification space, producing additional 3-forms, whose coupling to the corresponding axion produces the non-perturbative potential.

In string compactifications with 4d gauge sectors arising from D-branes, some of these axions are gauged by the $U(1)$ symmetries. In those cases, the D-brane instantons produce 4d effective couplings violating the (perturbatively exact global) $U(1)$ symmetries. In the language of type IIA D2-brane instantons in compactifications with 4d gauge sectors from D6-branes, this can be understood microscopically as follows. The insertions corresponding to the 4d charged fields in the effective operator induced by the instanton amplitude, arise form the saturation of charged instanton fermion zero modes arising from open strings at intersections between the D2-brane instanton and the 4d gauge D6-branes \cite{Blumenhagen:2006xt,Ibanez:2006da,Florea:2006si}, see \cite{Blumenhagen:2009qh,Ibanez:2012zz} for reviews. The required coupling of fermion zero modes to 4d charged fields is mediated by a world sheet instanton bounded by the D2- and D6-branes.

There is a large set of scenarios exploiting this mechanism (or its dual versions). For instance, they can generate neutrino masses \cite{Blumenhagen:2006xt,Ibanez:2006da} (see also \cite{Ibanez:2007rs}), Yukawa couplings \cite{Blumenhagen:2007zk}, the $\mu$-term in SUSY extensions of the Standard Model \cite{Cvetic:2010dz,Cvetic:2010mm}, or be crucial in SUSY breaking \cite{ Argurio:2006ny,Argurio:2007qk,Cvetic:2008mh} or its mediation \cite{Buican:2008qe}, as well as in rare processes \cite{Blumenhagen:2010dt,Addazi:2014ila,Addazi:2015rwa,Addazi:2015hka} (see \cite{Blumenhagen:2009qh,Ibanez:2012zz} for reviews of these and other applications). Achieving these effects typically requires introducing orientifold projections to remove additional fermion zero modes associated to the otherwise underlying $\NN=2$ supersymmetry, see \cite{Argurio:2007vqa,Bianchi:2007wy,Ibanez:2007rs} and \cite{Blumenhagen:2009qh,Ibanez:2012zz} for reviews.

A natural question is thus the description of the backreaction of D-brane instantons in these systems  with 4d gauge D-branes, and the mechanism by which the backreacted geometry manages to produce the corresponding charged field theory operators in the 4d effective action. This is the the subject of the present paper.

In short, in type IIA language, the instanton backreaction produces a change in the topology of the compactification space, in which the instanton cycle becomes trivial, and any cycles intersecting it  acquire a boundary. In particular, the gauge D6-branes formerly intersecting the D2-brane instanton acquire a boundary, and must necessarily recombine among themselves to form actual wrapped D6-branes in the backreacted geometry. In this process, the intersections between the gauge D-branes and the D-brane instanton disappear, and so do the charged instanton fermion zero modes. Finally, the original worldsheet instantons supported by the gauge and instanton D-branes, and mediating the coupling of charged  fermion zero modes and 4d charged fields, become worldsheet instantons supported, in the backreacted geometry, by just gauge D6-branes, yet producing the same 4d charged field operators in the effective action.

We consider this proposal in the large class of models corresponding to (type IIA mirrors of) D3-branes at toric singularities, and provide a simple combinatoric recipe to obtain the backreacted geometry and recombined gauge D-branes. This provides a new kind of geometric transition for systems of branes at singularities, which reduces to the familiar Klebanov-Strassler type gauge/gravity dualities for brane systems admitting complex deformations of the underlying geometry. Other geometric transitions relate to non-CY geometries, which still seem to admit a simple combinatoric description. 

The paper is organized as follows. In Section \ref{sec:instantons} we review D-brane instantons, both from the open string perspective in section \ref{sec:open-story} and from the backreaction viewpoint in section \ref{sec:backreact}. In Section \ref{sec:flavour} we describe the backreaction description of instantons charged under 4d gauge symmetries: in section \ref{sec:recombination} we show that the non-perturbative backreaction forces the recombination of the gauge branes formerly intersecting the instanton. In section \ref{sec:FW} we provide a topological  derivation, based on the Freed-Witten conditions, of the appearance of worldsheet instantons dressing the D2-brane instanton intersected by D6-branes, and show it also explains the appearance of charged field theory operators in the backreaction picture. In section \ref{sec:multiple-inters} we study several generalizations, including non-abelian gauge symmetries, and argue that the backreaction picture implies a sum over brane recombination geometries.

In Section \ref{sec:dimers} we turn to presenting explicit realizations in terms of systems of D3-branes at singularities, or rather their type IIA mirrors, described using dimer diagrams. These are reviewed in section \ref{sec:dimer-review}. In section \ref{sec:dimer-inst} we introduce the D-brane instantons we consider and their properties and in section \ref{sec:recipe} we provide the graphical recipe to describe the backreacted geometry, the D6-brane recombination and the 4d charged field theory operator produced. In section \ref{sec:examples} we present some illustrative examples of this technique, for single instantons, and generalize to non-compact instantons in section \ref{sec:non-compact}, and to multi-instanton effects in section \ref{sec:multiple}. In Section \ref{sec:complex-def} we consider instantons whose backreaction corresponds to complex deformations of the toric singularity and relate them to the deformation fractional branes producing  duality cascades generalizing the Klebanov-Strassler conifold. Some generalities are discussed in section \ref{sec:complex-gen}, and concrete examples are provided in section \ref{sec:complex-examples}. We provide some final remarks in section \ref{sec:conclu}.

\section{Review of instantons}
\label{sec:instantons}

\subsection{The open string story}
\label{sec:open-story}

In string theory, euclidean $p$-branes wrapped on $(p+1)$-cycles lead to non-perturbative instanton effects. The best understood set of such effects corresponds to wrapped D-brane instantons, for which a microscopic description can be obtained in terms of open string sectors. In the following we sketch their structure, and their role in the induced terms in the 4d effective action, in 4d $\NN=1$ type IIA CY compactifications with D6-branes on intersecting 3-cycles \cite{Aldazabal:2000cn,Aldazabal:2000dg}, see \cite{Ibanez:2012zz} for a review.

Recall that before the introduction of the instantons, each stack of $N_a$ D6-branes wrapped on a 3-cycle $\Pi_a$ leads to a gauge factor $U(N_a)$, and the D6$_a$-D6$_b$ open sectors produce $I_{ab}$ chiral fields in the bifundamentals $ (\fund_a,\antifund_b)$, for all $a,b$, with $I_{ab}=[\Pi_a]\cdot[\Pi_b]$ being the topological intersection number. Although O6-planes are often present (and in fact, are necessary for globally preserved supersymmetry), we omit the corresponding refinement of the discussion, as it will not be ultimately needed for our later purposes.

Upon introduction of a D2-brane instanton on a 3-cycle $\Pi_{3}$, there are D2-D2 and D2-D6$_a$ open string sectors, which in particular produce the instanton zero modes over which we should integrate. In the D2-D2 sector, the most relevant is a universal set of four bosonic zero modes $x^\mu$, defining the instanton position in 4d, and two goldstinos $\theta^\alpha$ of the 4d supersymmetries broken by the instanton. Integration over these produces the superspace measure for the 4d term in the effective action.
\beqa
\int\, d^4x \,d^2\theta\, A\, e^{-Z}\, (...)_{\rm neutral}\, {\cal O}_{\rm charged}
\eeqa
where $A$ is a (Kahler moduli dependent) prefactor, and $Z=\frac{V_3}{g_s} + i \int_{\Pi_{3}} C_3$ is the instanton action. Finally, $(...)_{\rm neutral}$ and  $ {\cal O}_{\rm charged}$ denote additional extra structures from additional zero modes, to be discussed next.

In the absence of additional fermion zero modes,  the instanton  produces superpotential terms, as suggested above. However, in general, the D2-D2 open sector can lead to additional\footnote{Additional bosonic zero modes may also be present, but they are simply integrated over yielding some numerical prefactor.} fermion zero modes, whose saturation introduces the extra structure $(...)_{\rm neutral}$ and implies that the 4d term is actually a higher F-term, see \cite{Beasley:2004ys,GarciaEtxebarria:2008pi}, also \cite{Blumenhagen:2007bn}. These extra fermion zero modes can be lifted by a variety of effects, including orientifold projection \cite{Ibanez:2007rs}, fluxes \cite{Tripathy:2005hv,Bergshoeff:2005yp,Blumenhagen:2007bn}, overlapping gauge branes  \cite{Petersson:2007sc}, etc. 

\begin{figure}
    \centering
    	\def\svgwidth{0.4\linewidth}
		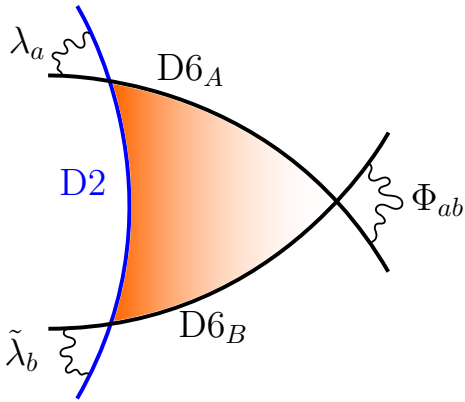
    \caption{A D2-brane instanton intersects two D6-brane stacks. The fermion zero modes $\lambda_a$, $\tilde{\lambda}_b$ and bifundamental matter $\Phi_{ab}$ live at the intersections. There is a worldsheet instanton in the disk bounded by the cycles, shaded in orange.}
    \label{Fig:D2D6D6}
\end{figure}

Our work is however more focused on the structure of the operator $ {\cal O}_{\rm charged}$, due to charged fermion zero modes from the D2-D6 open string sectors,  and which is present both for superpotentials or for higher F-terms.	
In the D2-D6$_a$ sector, there are $I_{a,\Pi_3}$ net fermion zero modes in the fundamental or antifundamental representation of the 4d gauge factor $U(N_a)$, according to the sign of the topological intersection number $I_{a,\Pi_3}=[\Pi_a]\cdot[\Pi_3]$.  Integration over these fermion zero modes typically leads to insertion of 4d charged chiral multiplets in the instanton amplitude \cite{Blumenhagen:2006xt,Ibanez:2006da,Florea:2006si}, see \cite{Blumenhagen:2009qh,Ibanez:2012zz} for reviews. Let us illustrate this in a simple example. Consider a type IIA compactification with two stacks of $N$ D6-branes wrappings 3-cycles $\Pi_a$, $\Pi_b$ and a D2-brane instanton wrapping a 3-cycle $\Pi_3$ such that the intersection numbers are $I_{a,\Pi_3} = 1$, $I_{b,\Pi_3} = -1$, $I_{ab} = 1$. The intersections between the D2-instanton and the D6-branes produce fermion zero modes  $\lambda_a$ in the $\antifund_a$, and $\tilde{\lambda}_b$ in the $\fund_b$. There is also a bi-fundamental chiral multiplet $\Phi_{ab}$ from strings stretching between the D6 stacks. These ingredients are coupled through a worldsheet  instanton supported (and actually required for consistency, see section \ref{sec:FW}) in a disk bounded by these cycles, see Figure \ref{Fig:D2D6D6}, contributing to the instanton worldvolume action as
\beqa
S_{\rm z.m.} \sim \lambda_a \Phi_{ab} \tilde{\lambda}_b
\eeqa
 
In this setup the amplitude of the D2-instantons contains a piece,
\begin{equation}
{\cal O}_{\rm charged}\, \sim\, \int \dif \lambda_a \dif \tilde{\lambda}_b \,e^{ S_{z.m.}} \,=\,  \det (\Phi_{ab}), \label{Eq:ZeroModeInt}
\end{equation}
This operator breaks the anti-diagonal $U(1)$ subgroup of the group $U(N)_a \times U(N)_b$. In fact, this is the mechanism in which string theory manages to break perturbatively exact $U(1)$ global symmetries from gauge $U(1)$ made massive by Stuckelberg couplings \cite{Blumenhagen:2006xt,Ibanez:2006da,Florea:2006si}.

As anticipated, the focus of this paper is the derivation of ${\cal O}_{\rm charged}$ in the description of D-brane instantons by backreacted geometries \cite{Koerber:2007xk,Koerber:2008sx,Garcia-Valdecasas:2016voz}.

\subsection{Backreacting instantons} \label{sec:backreact}

In this section we review the description of D-brane instanton effects in terms of a backreacted geometry. The description in terms of generalized geometry was provided in \cite{Koerber:2007xk,Koerber:2008sx} for D3-brane instantons in type IIB CY compactifications. We review this discussion adapting the basic points to D2-brane instantons in type IIA CY compactifications, sketched in \cite{Garcia-Valdecasas:2016voz}. Inclusion of D6-brane sectors and the corresponding fermion zero modes will provide new results in later sections.

The effect of D-brane instantons can be encoded in the underlying CY geometry by means of a change of the $SU(3)$ holonomy to (in general) an $SU(3)\times SU(3)$ structure, associated to the existence of  two (in genreal, not covariantly constant) spinors corresponding to a 4d $\NN=1$ supersymmetry (possibly in AdS). Focusing already in the type IIA case, the two spinors are written
\beqa
\epsilon_1=\zeta_+\otimes \eta_+^{(1)}+\zeta_-\otimes \eta_-^{(1)}\quad , \quad
\epsilon_2=\zeta_+\otimes \eta_-^{(2)}+\zeta_-\otimes \eta_+^{(2)}
\eeqa
here $\zeta_+$ and $\eta_+$ are complex conjugate of $\zeta_-$, $\eta_-$, and $\zeta_+$ is the 4d spinor specifying the $\NN=1$ supersymmetry, and satisfying $\nabla_\mu \zeta_-=\frac 12 W_0 \gamma_\mu \zeta_+$, where $W_0$ is the superpotential at the AdS minimum, and $W_0=0$ for the Minkowski case.

The spinors $\eta^{(1,2)}$ can be used to define two polyforms, 
\beqa
\Psi_{\pm}=-\frac{i}{||\eta^{(1)}||^2}\, \sum_l \frac{1}{l!} \,\eta_{\pm}^{(2)}{}^\dagger\gamma_{m_1\ldots m_l} \eta_+^{(1)}\, dy^{m_l}\wedge \ldots \wedge dy^{m_1}
\eeqa
From the chirality of the spinors in the sandwich, the polyform $\Psi_+$ contains even degree forms and $\Psi_-$ contains odd degree forms. The following notation is also often used (for type IIA) $\Psi_1=\Psi_-$ and $\Psi_2=\Psi_+$.

The familiar case of $SU(3)$ structure corresponds to $\eta^{(2)}\sim \eta^{(1)}$ and leads to  $\Psi_+\sim e^{iJ}$ and $\Psi_-\sim \Omega$. For $SU(3)$ holonomy the spinors are covariantly constant and the polyforms are closed.

The compactification ansatz is
\beqa
ds^2\, =\, e^{2A(y)}g_{\mu\nu}(x) \,dx^\mu dx^\nu \, +\, h_{mn}(y)\, dy^mdy^n
\eeqa
The 10d fields can be organized in complex quantities, in agreement with the 4d susy structure. The two holomorphic quantities are given by 
\beqa
{\cal Z}\equiv e^{3A-\Phi} \Psi_2 \quad , \quad {\cal T}\equiv e^{-\Phi } {\rm Re}\, \Psi_1+i\Delta C \label{eq:TPolyform}
\eeqa
where $\Phi$ is the 10d dilaton, and $\Delta C$ describes the RR backgrounds not encoded in the background fluxes ${\bar F}$. The definitions (\ref{eq:TPolyform}) are motivated because they provide the calibration for certain BPS objects. More concretely, in type IIA ${\cal Z}$ calibrates even-dimensional cycles, which define BPS D-brane domain walls; while ${\cal T}$ calibrates odd-dimensional cycles, which define BPS D-brane instantons. 

\medskip

As discussed in \cite{Koerber:2007xk,Koerber:2008sx,Garcia-Valdecasas:2016voz} the backreaction of the D-brane instanton is encoded in a modification of the exterior derivative of the quantities ${\cal Z}$ or ${\cal T}$, in type IIB or type IIA respectively. Let us focus on the simplest instance of D2-brane instantons in type IIA compactifications. 

Consider type IIA compactified on a CY $\IX_6$, and an instanton given by a D2-brane wrapped on a 3-cycle $\Pi_3$. For convenience we consider its dual 3-cycle ${\tilde \Pi_3}$, and denote their Poincare dual classes by $\beta_3=\delta_3(\Pi_3)$ and its dual $\tilde\beta_3=\delta_3(\tilde\Pi_3)$.

The equation determining the backreaction effect of the D2-brane instanton is \cite{Koerber:2007xk,Koerber:2008sx} 
\beqa
d{\cal T}_2=W_{\rm np} \delta_3(\Pi_3)
\label{backreacted-geometry}
\eeqa
where $W_{\rm np}$ is the non-perturbative superpotential. For clarity we reabsorb it into a suitable 2-form $\alpha_2\sim {\cal T}_2$,  and write 
\beqa
d\alpha_2=\beta_3
\eeqa

\begin{figure}[htb]
\begin{center}
\includegraphics[scale=.45]{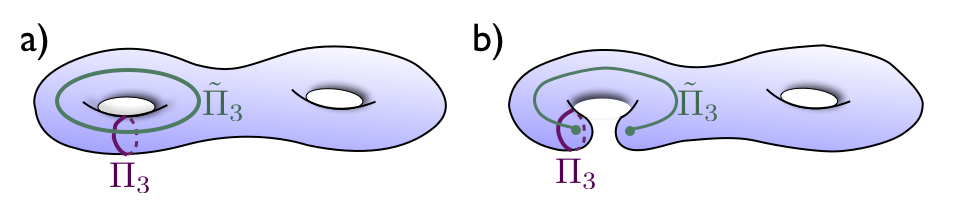}
\caption{\small In a compactification (a) with a D2-brane instanton on a cycle $\Pi_3$, the instanton backreaction modifies the topology (b) such that $\Pi_3$ becomes homologically trivial, and its dual ${\tilde \Pi}_3$ turns into a 3-chain. }
\label{fig:transition}
\end{center}
\end{figure}

This equation encodes a change in the topology of the compactification manifold, see Figure \ref{fig:transition}. In particular, the fact that $\beta_3$ is exact implies that the 3-cycle $\Pi_3$ has become trivial, i.e. the boundary of a 4-chain which is associated to the non-closed form $\alpha_2$.
Conversely, the dual 3-form $\tilde\beta_3$ is now non-closed, i.e. it satisfies
\beqa
d\tilde\beta_3={\tilde \alpha}_4
\eeqa 
for some $\tilde\alpha_4$, which is therefore exact. This implies that the dual 3-cycle $\tilde \Pi_3$, which had intersection number one with $\Pi_3$ has now become a 3-chain, whose boundary is some 2-cycle $\Sigma_2$, which morally corresponds to the exact form $\tilde\alpha_4=\delta_4(\Sigma_2)$.

In the following section we initiate the discussion of the backreaction in the presence of additional ingredients, specifically gauge D-branes.

\section{Backreacting flavoured instantons}
 \label{sec:flavour}

The above description accounts for the generation of non-perturbative terms for isolated instantons. This is interesting for the question of moduli stabilization, as explained in \cite {Garcia-Valdecasas:2016voz}. However, many applications of D-brane instanton involve instantons which intersect non-trivially the sectors of D-branes giving rise to the 4d gauge group. Such instantons generate non-perturbative field theory operators involving 4d charged matter multiplets \cite{Blumenhagen:2006xt,Ibanez:2006da,Florea:2006si}, (see \cite{Blumenhagen:2009qh,Ibanez:2012zz} for reviews), and have been proposed for the non-perturbative generation of several phenomenologically interesting field theory operators, as mentioned in the introduction.

In this paper we address the geometric description of such instantons in terms of their backreacted geometry. In this section we introduce the basic ingredients, and postpone concrete examples to the coming sections.

Consider a type IIA CY compactification, with stacks of $N_a$ D6-branes wrapped on non-trivial 3-cycles $\Pi_a$. For simplicity we consider the theory without O6-planes, so that the RR tadpole condition is
\beqa
\sum_a N_a [\Pi_a]=0
\label{rr-tadpole}
\eeqa
Models with O6-planes can be subsequently considered, but they will not imply any further complication (at least, as long as the O6-planes do not intersect the D2-brane instantons to be introduced shortly). We leave the discussion of O6-plane models for future work\footnote{Orientifold planes are also relevant to remove additional fermion zero modes and allow the instanton to generate non-perturbative superpotential terms. As explained before, we are interested just in the field theory operator structures, which are common to superpotential terms and to higher F-terms.}.

We recall from \ref{sec:open-story} that the gauge group of the theory is $\otimes_a U(N_a)$ and there are $I_{ab}$ chiral fields in the bifundamentals $ (\fund_a,\antifund_b)$, for all $a,b$. Here $I_{ab}=[\Pi_a]\cdot[\Pi_b]$ is the topological intersection number (i.e. weighted with $\pm 1$ according to orientation) of the 3-cycles $\Pi_a$, $\Pi_b$. In supersymmetric intersections, these correspond to chiral multiplets; although compact models without O6-planes are necessarily non-supersymmetric, we keep the supersymmetric terminology in mind, since the ideas apply similarly when supersymmetric models are considered, either by the introduction of O6-planes or in the non-compact setups in later sections.

\subsection{Brane recombination}
\label{sec:recombination}

Consider a D2-brane instanton wrapped on a non-trivial 3-cycle $\Pi_3$. The intersection of the D6-brane 3-cycles with $\Pi_3$ leads to fermion zero modes $\lambda$ in the fundamental representation of the corresponding D6-brane gauge group for positively oriented intersections, and ${\tilde \lambda}$ in the antifundamental for intersections with negatively oriented intersections. Because of the condition (\ref{rr-tadpole}), we obtain 
\beqa
\sum_a N_a\, I_{a,\Pi_3}\, =\, 0
\label{positive-negative-zero}
\eeqa
Namely, there are equal total numbers of fermion zero modes of type $\lambda$ or $\tilde\lambda$ (counted with multiplicity given by the rank $N$ of the corresponding D6-brane groups).

Consider for simplicity only two D6-branes on 3-cycles $[\Pi_+]$, $[\Pi_-]$, respectively, intersecting the D2-brane with $[\Pi_3]\cdot [\Pi_\pm]=\pm 1$. This means that $[\Pi_\pm]$ are essentially given by
\beqa
[\Pi_\pm]\,=\, \pm\, [{\tilde \Pi}_3] \, +\, \ldots
\eeqa
where $[{\tilde \Pi}_3]$ is the 3-cycle dual to $[\Pi_3]$, and the dots indicate additional 3-cycle components not intersecting the D2-brane, and which behave as spectators in what follows.

\begin{figure}[htb]
\begin{center}
\includegraphics[scale=.45]{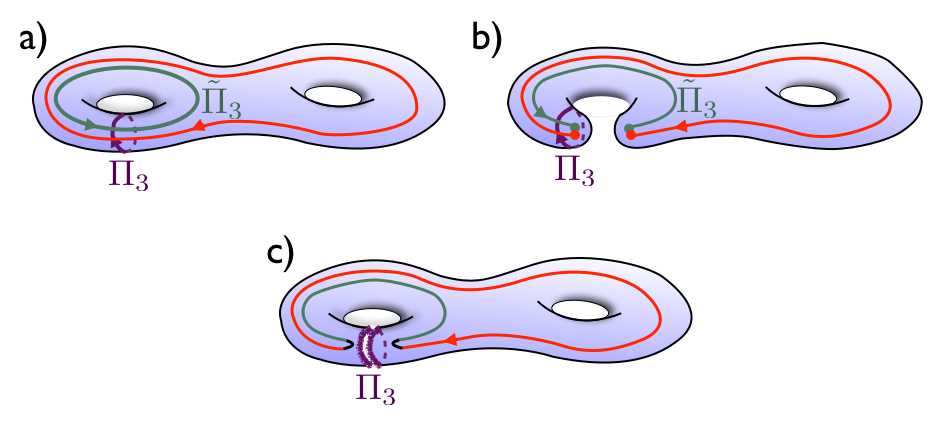}
\caption{\small Consider  (a) a D2-brane instanton on a cycle $\Pi_3$, in the presence of gauge D-branes wrapped on cycles (shown as green and red) intersecting $\Pi_3$ with opposite intersection numbers. After the instanton backreaction (b) making $\Pi_3$ trivial, the red and green cycles become chains, so the corresponding branes must recombine to wrap a consistent combined cycle. In (c) we display a different representation, obtained by excising a local neighbourhood around $\Pi_3$ in the original one; the backreacted geometry is obtained by contracting the left-over boundaries to a point.}
\label{fig:recombination}
\end{center}
\end{figure}

In the backreacted geometry, the D2-brane 3-cycle becomes trivial $[\Pi_3]=0$, and the dual 3-cycle becomes a chain, whose boundary is a new 2-cycle $\Sigma_2$ in the backreacted geometry, $\partial \tilde\Pi_3=\Sigma_2$. Therefore $\partial\Pi_\pm=\pm \Sigma_2$ in homology. This implies that the D6-brane formerly wrapped on $\Pi_+$ is no longer consistent, and similarly for the D6-brane wrapped on $\Pi_-$; however, their recombination  formerly corresponds to the class $[\Pi_+]+[\Pi_-]$, which does not have intersection with $[\Pi_3]$, and therefore defined a consistent 3-cycle without boundaries in the backreacted geometry. The effect of the backreaction in the flavour D6-branes is therefore to trigger their recombination. This is depicted in Figure \ref{fig:recombination}.

The lesson generalizes easily. Given a general set of $N_a$ D6-branes on 3-cycles $\Pi_a$, any 3-cycle with non-zero intersection with $\Pi_3$ becomes a 3-chain with boundary $\partial\Pi_a=([\Pi_a]\cdot[\Pi_3])\,\Sigma_2$ in homology in the backreacted geometry. The condition (\ref{positive-negative-zero}), then implies that the different 3-chains can recombine into a 3-cycle (possibly with several components) without boundary. The effect of the instanton backreaction on the flavour D6-branes is to recombine them into sets defining consistent 3-cycles in the backreacted geometry.

In general, there may be several inequivalent topologically consistent ways to recombine the different D6-brane stacks. This will be discussed in section \ref{sec:multiple-inters}. 

\subsection{Saturation of charged fermion zero modes revisited}
\label{sec:FW}

In the following we move on to the next question, namely the appearance of the charged chiral multiplet insertions in the instanton amplitude. In the open string perspective of section \ref{sec:open-story}, this arises from the saturation of fermion zero modes via couplings arising from worldsheet instantons. As these worldsheet instantons extend away from the instanton locus, we expect them to continue playing a role in the appearance of the field theory operator in the backreacted picture. Indeed we now show that the backreacted configuration necessarily contains euclidean fundamental strings producing the relevant charged field insertions.

For that purpose it is useful to develop a spacetime perspective on the appearance of worldsheet instantons in the presence of the D2-brane instanton. In string compactifications, we are used to think about worldsheet instantons as possible contributions over which one has to sum to obtain a complete quantum amplitude. Namely, we sum contributions from processes with one instanton, two instantons... or none. In other words, the worldsheet instantons may be present or not, and their number is arbitrary, or rather is a quantity over which we should sum. In the presence of a D2-brane instanton, on the other hand, the argument about saturating charged fermion zero modes implies that the presence of the D2-brane requires specific patterns of euclidean worldsheets to be simultaneously present.  	

There is a simple explanation for this from the spacetime perspective, which to our knowledge has not appeared in the literature. In short, the flux sourced by the presence of the D2-brane produces a Freed-Witten (FW) \footnote{Despite the familiar name coined from the analysis in \cite{Freed:1999vc}, one actually means the conditions for D-branes in presence of non-torsion NSNS or RR fluxes derived in \cite{Maldacena:2001xj,Witten:1998xy}.} anomaly on the D6-branes, which forces them to expel a semi-infinite euclidean worldsheet; similarly, the flux sourced by the D6-branes produces a FW anomaly on the D2-brane, forcing them to expel a semi-infinite euclidean worldsheet. Finally, when the D6-branes meet at the intersection point, the euclidean worldsheet can connect with an open string worldsheet escaping to infinity in the 4d spacetime.

In more detail, consider type IIA in flat 10d spacetime and a configuration of an infinite D6-brane along the directions 0123456 and an infinite D2-brane along the directions 789. This choice of directions leads to a non-supersymmetric intersection, but captures the essential topology  which holds also in the supersymmetric case, as the interested reader can easily check.

\begin{figure}[htb]
\begin{center}
\includegraphics[scale=.45]{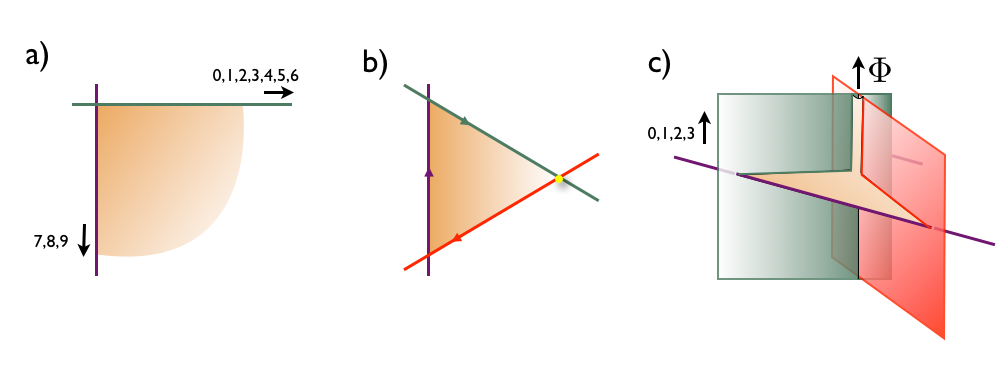}
\caption{\small (a) Intersecting  one D2-brane (violet) and one D6-brane (green) leads to FW anomalies enforcing the appearance of a fundamental string worldsheet bounded by the D2- and the D6-branes. (b) In a configuration of two intersecting D6-branes (red and green), each intersecting the D2-brane (violet), the emitted fundamental string defines a worldsheet instanton spanning the triangle defined by the branes in the internal space, exactly as in the open string argument of saturation of fermion zero modes. (c) Depicting the 4d spacetime dimensions, the boundaries of the fundamental string worldsheet must extend to infinity, implying the emission of the chiral multiplet $\Phi$ at the intersection of the D6-branes.}
\label{fig:FW}
\end{center}
\end{figure}

The D6-brane is a source of RR 2-form field strength, with 
\beqa
dF_2= \delta(x^{7,8,9}) \, dx^7 dx^8 dx^9
\eeqa
where we use the notation $\delta(x^{7,8,9})=\delta(x^7)\delta(x^8)\delta(x^9)$.
Let us consider the integrated flux $\int_{89}F_2$ and its change along the direction $x^7$. The above equation implies that it jumps from e.g. zero at $x^7<0$ to one flux unit at $x^7>0$. Since the D2-brane is wrapped on 789, at $x^7>0$ there is a non-trivial flux of $F_2$ over the D2-brane worldvolume. This leads to a Freed-Witten inconsistency, which must be solved by the emission of a semi-infinite fundamental string with worldsheet along a curve on the 89 plane and the semi-infinite direction $x^7>0$. This string should not be present at $x^7<0$, where no $F_2$ flux and so no FW anomaly is induced on the D2-brane. This is indeed the case because the worldsheet is forced to end on the D6-brane, by the converse FW effect. Namely, the D2-brane is a source of RR 6-form field strength, with 
\beqa
dF_6= \delta(x^{0,1,2,3,4,5,6} )\, dx^0 dx^1 dx^2 dx^3 dx^4 dx^5 dx^6
\eeqa
We can argue as above using the flux integral of $F_6$ along 012356, which jumps as we move from $x^4<0$ to $x^4>0$, to obtain a FW anomaly forcing the D6-brane to expel a fundamental string with worldsheet along a curve in 012356 and $x^4=0$. The worldsheets emitted by the D2- and the D6-brane can nicely combine in a corner near the D2-D6 intersection point, see Figure \ref{fig:FW}. Finally, worldsheets ending on different D6-branes and approaching a D6-D6 intersection escape off to infinity, thus describing the insection of charged 4d matter fields.

\medskip

This spacetime picture allows to provide a derivation of the structure of worldsheet instantons in the backreacted geometry. The key point is that the instanton backreaction produces not only a change in the geometry, but also produces a (sourceless) RR 6-form $F_6$ field string flux, which forces the (recombined) D6-branes to emit fundamental string worldsheets, which imply the emission of chiral multiplets supported at D6-brane intersections.

More concretely, in the original geometry the source equation for $F_6$ is
\beqa
dF_6= \delta_4(x)\, \delta_3(\Pi_3)
\eeqa
where $\delta_4(x)$ is a bump 4-form around the location of the instanton in 4d spacetime. Recalling the discussion following (\ref{backreacted-geometry}), in the backreacted geometry we have
\beqa
F_6 = \delta_4(x) \, \delta_2(\Sigma_4)
\eeqa
where $\Sigma_4$ is the 4-cycle dual to the 2-cycle $\Sigma_2$ defined by $\partial\tilde\Pi_3=\Sigma_2$. For any D6-brane wrapped on a 3-cycle $\Pi$ intersecting $\Pi_3$ in the original geometry, $\Pi=\tilde\Pi_3+\ldots$, and $\Pi$  acquires a boundary $\Sigma_2$ in the backreacted geometry. Focusing on a local neighbourhood of this boundary, we have $\Pi=\Sigma_2\times \IR^+$. Integrating $F_6$ over the 4d spacetime times $\Sigma_2$ gives
\beqa
\int_{M_4\times \Sigma_2} F_6=\int_{\Sigma_2} \delta_2(\Pi_4)=1
\eeqa
This shows that a D6-brane wrapped on $\Pi$ has a FW anomaly, and must expel a fundamental string. The argument about the structure of worldsheet instantons and emission of charged chiral multiplets at D6-brane intersections follows as in the earlier open string description.

The fact that the D2-brane instanton intersecting the D6-branes is forced by FW to expel charged matter fields in open string sectors fits nicely with the M-theory description of such instantons. The D2-brane instanton lifts to an euclidean M2-brane wrapped on a 3-chain with boundaries; the existence of boundaries is the M-theory version of the FW inconsistency of the D2-brane, and closing the boundaries with M2-branes wrapped on certain 2-cycles is the M-theory version of the emission of massless open strings in type IIA.

\subsection{Multiple intersections and non-abelian case}
\label{sec:multiple-inters}

Configurations with multiple intersections can be discussed similarly. We sketch their description emphasizing the novel features that may arise.

In this section we consider certain novel features arising when the D2-brane is intersected by multiple D6-branes with intersection numbers $+1$ (and the same number with intersection number $-1$).

For simplicity we consider the abelian case, in which all the D6-branes intersecting the D2-brane instanton have $N=1$, i.e. $U(1)$ gauge factor (for which the fundamental or antifundamental translate into $\pm 1$ charges); we comment on the non-abelian case later on. 

A preliminary remark is that by the FW arguments, each D6-brane crossing with the D2-brane leads to the emission/absorption of a fundamental string worldsheet, according to the orientation of the intersection. The fact that the number of intersections with positive and negative intersection are equal ensures that all such fundamental string worldsheets close onto each other providing consistent worldsheet instantons (leading to emission of an appropriate set of open string fields at D6-brane intersections). Thus in the presence of the D2-brane instanton, all such worldsheet instantons are necessarily present.
In general, the geometry of the worldsheet instanton can lead to insertions of different numbers of charged multiplets. Given that each fundamental string involves two corners at D2-D6 intersections with opposite orientations, it is easy to convince oneself that, after the instanton backreaction and D6-brane recombination, the structure of worldsheet instantons required by the FW consistency conditions reproduces the saturation of fermion zero modes in the open string description. 

\begin{figure}[htb]
\begin{center}
\includegraphics[scale=.6]{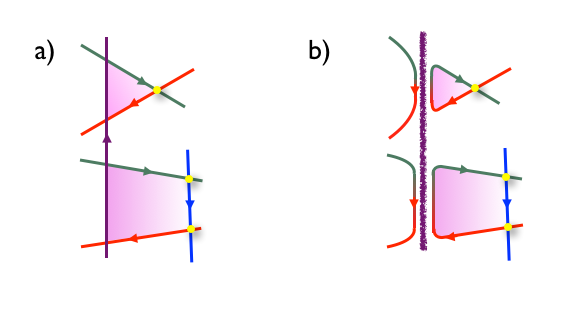}
\caption{\small  Graphical depiction of the worldsheets attached to an instanton process (a) in the open string picture and (b) in the backreacted geometry picture. In the latter, the two pieces of D-branes on both sides of the cut are actually connected on the whole 3-cycle.}
\label{fig:multiple}
\end{center}
\end{figure}

For instance, consider the example in Figure \ref{fig:multiple}(a). The configuration produces fermion zero modes $\lambda_{i}$, ${\tilde\lambda}_i$, for $i=1,2$, and the following couplings in the instanton worldvolume action
\beqa
c_1\, \lambda_1 \Phi_{1} \tilde \lambda_1\, +\, c_2\, \lambda_2 \Phi_2\Phi_3 \tilde\lambda_2
\eeqa
where the coefficients $c_1,c_2$ encode the worldsheet instanton amplitude, and the remaining notation is hopefully clear. Saturating fermion zero modes leads to the field theory operator in the D2-brane instanton amplitude
\beqa
(\Phi_1)\, (\Phi_2\Phi_3)
\label{multiple-fields}
\eeqa
Here and in what follows, we introduce the parentheses notation. It is irrelevant in the present abelian case, but is used as a reminder of the index structure relevant in the non-abelian case (e.g if each D6-brane is promoted to a stack, the field theory operator has the structure $\det\Phi_1\det(\Phi_2\Phi_3)$, with color index contraction in the operator $\Phi_2\Phi_3$).

The same result is recovered in the backreacted geometry. The operator (\ref{multiple-fields}) arises from standard worldsheet instanton amplitude, taking into account that the FW consistency due to the flux in the backreacted configuration demands the simultaneous presence of worldsheet instantons on all (recombined) D6-branes. 

\begin{figure}[htb]
\begin{center}
\includegraphics[scale=.5]{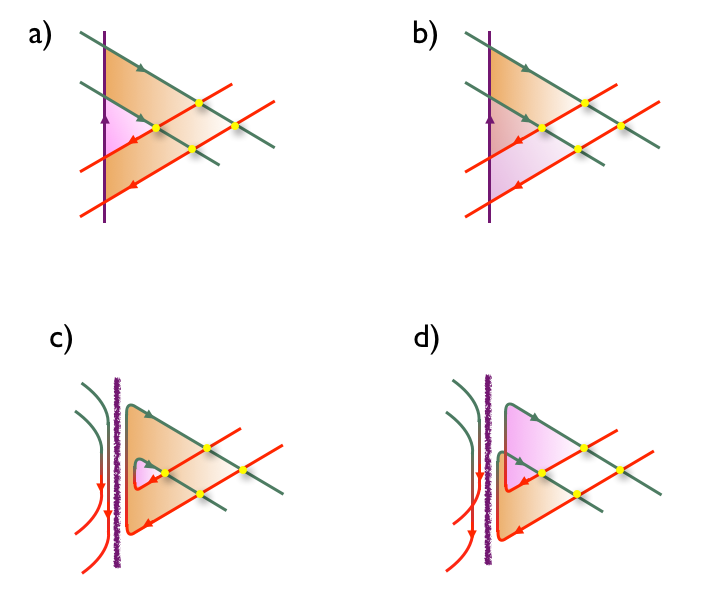}
\caption{\small D2-brane instanton in the presence of D6-branes, in which the charged fermion zero modes can be saturated in two inequivalent ways, shown as shaded triangles in figures (a) and (b).
In the instanton backreaction picture they are described by two possible ways to recombine the D6-branes, sketched in figures (c) and (d).  In the latter, the two pieces of D-branes on both sides of the cut are actually connected on the whole 3-cycle.}
\label{fig:two-ways}
\end{center}
\end{figure}

\medskip

One novelty is the possibility of several inequivalent ways of recombining branes. Consider for instance the example in Figure \ref{fig:two-ways}. In the open string description, the fermion zero mode couplings on the D2-brane worldvolume action are
\beqa
(\lambda_1,\lambda_2)\begin{pmatrix} \Phi_1 & \Phi_2 \cr \Phi_3 & \Phi_4 \end{pmatrix} \begin{pmatrix} \tilde\lambda_1 \cr \tilde\lambda_2 \end{pmatrix}
\label{matrix-case}
\eeqa
 where we omit the coefficients, and where the notation is hopefully clear. There are two different ways to saturate the fermion zero modes, which are depicted in Figures \ref{fig:two-ways} (a) and (b). The resulting field theory operator in the D2-brane instanton amplitude is
 \beqa
 (\Phi_1)( \Phi_4)\, -\,(\Phi_2)(\Phi_3)
 \label{pseudo-det}
 \eeqa
 
 In the backreaction picture, the appearance of two possible terms arises because, in the backreacted geometry there are two inequivalent ways to recombine the flavour D6-branes, so we should sum over both possibilities. Therefore the complete amplitude is a sum over different geometric configurations of D6-branes. On top of the relative coefficients for each amplitude (which we recall have been omitted above), a proper explanation of the relative sign between the two contributions would require further microscopic understanding about the sum over such geometries. We leave this as an open question, and simply learn the thumb rule that the D6-brane boundaries behave as Grassman objects, such that their exchange in recombination processes leads to extra signs.
 
 The generalization to larger numbers of intersections is straightforward, and is left as an exercise to the reader. 
 
 \medskip
 
 The above local picture of the recombination of two pairs of half D6-branes is the basic ingredient in other global configurations. For instance, the two incoming half D6-branes may actually belong to the same D6-brane stack, which thus has intersection number 2 with the D2-brane. The above analysis therefore provides the description of setups with multiple intersection numbers.
 
Also, it encodes the description of non-abelian case. For instance, consider the configuration when the two incoming D6-branes are coincident and so are the two outgoing D6-branes. We have two  $U(2)$ stacks of D6-branes intersecting the D2-brane with intersection number $\pm 1$ respectively. Each pair of fermion zero modes forms a doublet under the corresponding $U(2)$, and the four different charged chiral multiplets $\Phi_1,\ldots,\Phi_4$ form a bi-fundamental $\Phi_{ab}$. The matrix structure in (\ref{matrix-case}) encapsulates the $U(2)^2$ indices, and (taking into account the coefficients of the worldsheet instantons are equal for coincident branes) the expression (\ref{pseudo-det}) becomes $\det(\Phi_{ab})$. The above argument instructs us that, in the geometric description by the instanton backreaction, the non-abelian case requires a sum over possible recombination patterns, weighted by the relative sign upon exchange of boundaries.

We put these ideas to work in a large class of explicit models in the next section.

\section{Instanton backreaction and geometric transitions in dimers}
\label{sec:dimers}

In this section we present a large number of explicit examples of the above ideas, in the setup of local models related to systems of D3-branes at toric singularities, or rather, their type IIA mirror duals. 
The set of supersymmetric D-branes, the structure of of the chiral multiplets in open string sectors, and the worldsheet superpotential couplings are nicely encoded in a graph, the dimer diagram, to be reviewed shortly. The same diagram can be used to encode the set of fermion zero modes and the instanton worldvolume couplings on D-brane instantons. Moreover, there is a systematic way to construct the type IIA mirror in which we recover a picture of D6-branes and D2-brane instantons, in which we can flesh out the proposal of instanton backreaction in the previous sections.

\subsection{Overview of dimers}
\label{sec:dimer-review}

In this Section we review some background material on dimer diagrams as a tool to describe systems of D3-branes at toric singularities.

\subsubsection{Quiver gauge theories and dimer diagrams}
\label{quiver}

The gauge theory of type IIB D3-branes probing toric CY  threefold singularities is given by a bunch of unitary gauge factors,  bi-fundamental chiral multiplets, and a superpotential. The structure of these gauge theories, and many properties of the underlying D-brane system, are nicely encoded in a dimer diagram, see e.g. \cite{Hanany:2005ve,Franco:2005rj}, and \cite{Kennaway:2007tq} for a review. A dimer diagram is a tiling of $\IT^2$ defined by a bipartite graph (i.e. with black and white nodes, such that no edges connect nodes of the same color). Faces in the dimer diagram correspond to gauge factors in the field theory, edges describe bi-fundamental fields, and nodes provide superpotential terms. The bipartite character of the diagram is important in that it defines an orientation for edges (e.g. from black to white nodes), which determines the chirality of the bi-fundamental fields. Also, the color of a node determines the sign of the corresponding superpotential term.
Several well-known examples are described in the examples later on.

We will be interested in the type IIA mirror configuration to the D3-brane systems, which can be constructed as follows \cite{Feng:2005gw}. The mirror geometry is specified by a double fibration over the complex plane W given by
 \beqa
 W \ &=&\ P(z,w) \ 
\\
 W \ &=& \ uv
 \eeqa
with $w,z \ \in \IC^*$ and $u,v \ \in \IC$.\
Here $P(z,w)$ is the Newton polynomial of the toric diagram of  
$\mathcal{M}$, but this description is not necessary for our purposes. 

\begin{figure}
    \centering
    \begin{subfigure}[t]{0.3\textwidth }
        \begin{center} 
		\includegraphics[width=\textwidth]{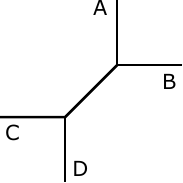}
		\caption{}
		\label{Fig:WebDiagramConifold2}
		\end{center}
    \end{subfigure} \hspace{10mm}
    \begin{subfigure}[t]{0.3\textwidth }
        \begin{center} 
		\includegraphics[width=\textwidth]{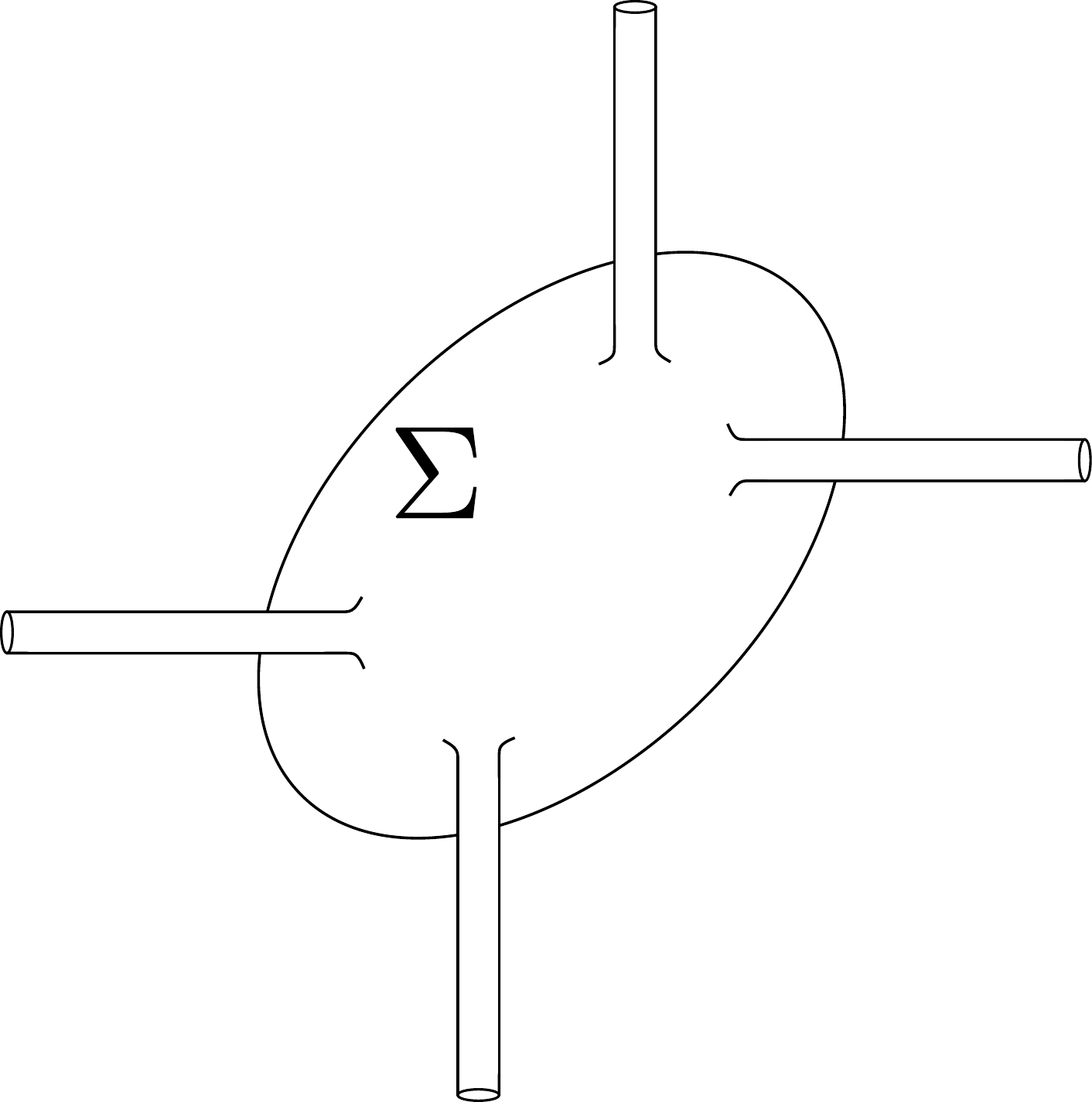}
		\caption{}
		\label{Fig:RiemannConifold3d2}
		\end{center}
    \end{subfigure}  
    \caption{The mirror Riemann surface $\Sigma$ as the thickening of the web diagram. Figure (a) shown the web diagram for the conifold, and Figure  b) depicts the mirror Riemman surface $\Sigma$.}\label{Fig:SigmaThickening}
\end{figure}

The surface $W = P(z,w)$ describes a genus $g$ Riemann surface $\Sigma_W$ with punctures, fibered over $W$ (the genus $g$ equals the number of internal points of the toric diagram). The fiber over $W=0$, denoted simply $\Sigma$, corresponds to a smooth Riemann surface which can be thought of as a 
thickening of the web diagram \cite{Aharony:1997ju,Aharony:1997bh,Leung:1997tw} dual to the toric 
diagram, see Figure \ref{Fig:SigmaThickening}. As discussed later on, it is easily constructed from the dimer.

\begin{figure}
	\centering
	\def\svgwidth{0.6\linewidth}
	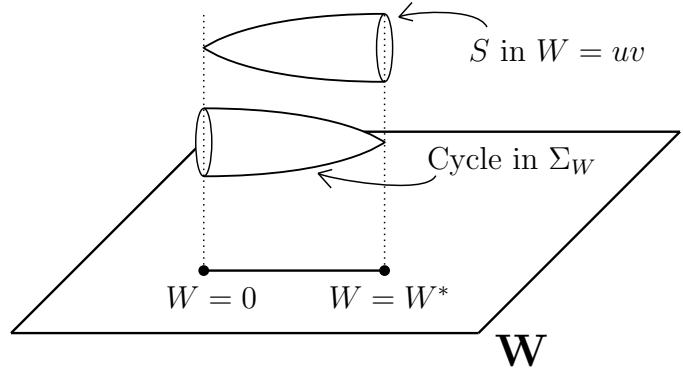
    \caption{Construction of 3-cycles in the mirror geometry. One of the two fibers degenerates at each point ($W=0$, $W=W^{*}$). These together with the segment between the two points result in a 3-cycle with $S^3$ topology. }
    \label{Fig:FibraW}
\end{figure}

At critical points $W=W^*$, a cycle in $\Sigma_W$ degenerates and pinches 
off. Also, at $W=0$ the $S^1$ in $W=uv$ degenerates. One can use these 
degenerations to construct non-trivial 3-cycles in the mirror geometry as 
follows. Consider the segment in the $W$-plane which joins $W=0$ with one 
of the critical points $W=W^*$, and fiber over it the $\IS^1$ in $W=uv$ 
times the 1-cycle in $\Sigma_W$ degenerating at $W=W^*$,
see Figure \ref{Fig:FibraW}. The result is a 3-cycle with an $\IS^3$ 
topology.
Mirror symmetry specifies that the different gauge factors on the 
D3-branes in the original singularity arise from D6-branes 
wrapping these different 3-cycles. The basic geometry of the 3-cycles is encoded in the structure of the 1-cycles in the punctured Riemann surface $\Sigma$.

Such D6-branes reproduce also the chiral matter and superpotential couplings, as follows. The 3-cycles on which the D6-branes wrap 
intersect over $W=0$, precisely at the intersection points of the 
1-cycles in $\Sigma_{W=0}$. Open strings 
at such intersections lead to the chiral bi-fundamental fields. Moreover, 
disks in $\Sigma$ bounded by pieces of different 1-cycles lead to 
superpotential terms generated by world-sheet instantons.

\begin{figure}
    \centering
         \begin{subfigure}[t]{0.28\textwidth }
        \begin{center} 
		\includegraphics[width=\textwidth]{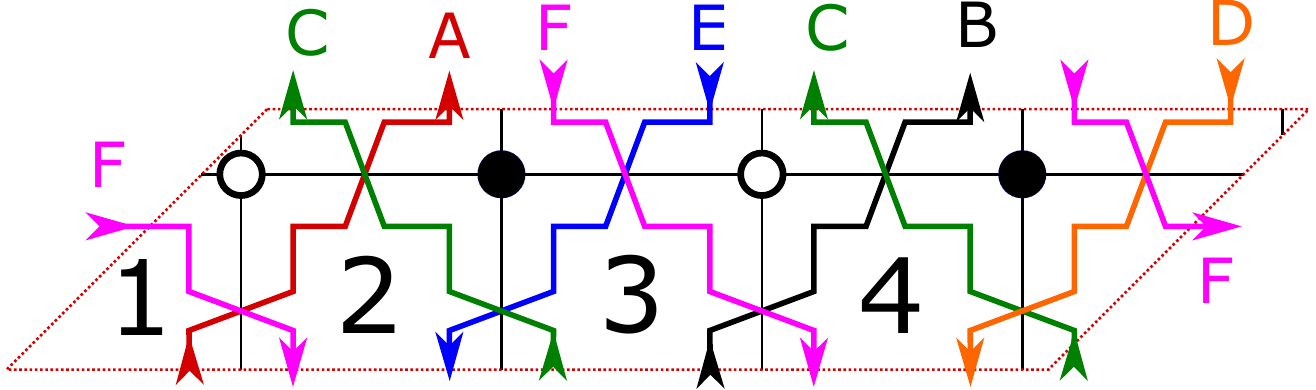}
		\caption{}
		\label{Fig:DimmerDoubleConifoldUnit2}
		\end{center}
    \end{subfigure}  \hspace{10mm}
    \begin{subfigure}[t]{0.13\textwidth }
        \begin{center} 
		\includegraphics[width=\textwidth]{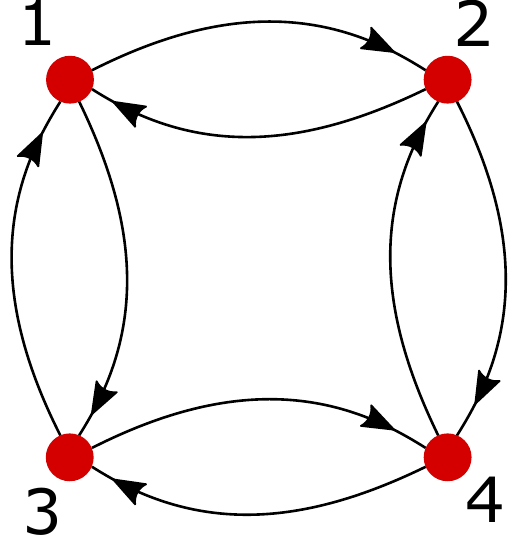}
		\caption{}
		\label{Fig:QuiverDiagramDoubleConifold}
		\end{center}
    \end{subfigure}  \hspace{10mm}
    \begin{subfigure}[t]{0.20\textwidth }
        \begin{center} 
		\includegraphics[width=\textwidth]{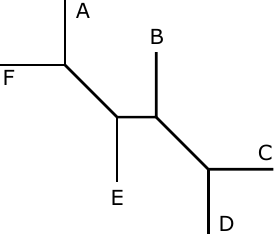}
		\caption{}
		\label{Fig:WebDiagramDoubleConifold}
		\end{center}
    \end{subfigure} \hspace{10mm}
    \begin{subfigure}[t]{0.15\textwidth }
        \begin{center} 
		\includegraphics[width=\textwidth]{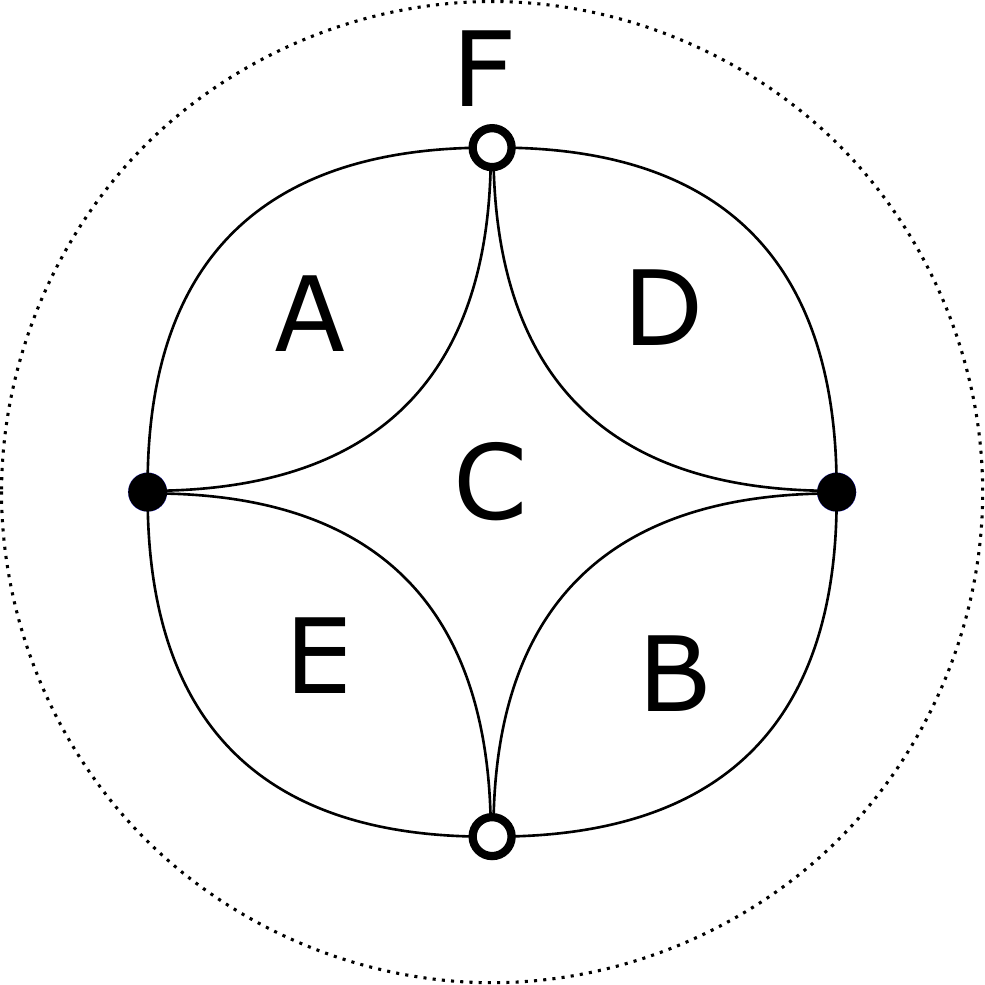}
		\caption{}
		\label{Fig:RiemannDoubleConifold}
		\end{center}
    \end{subfigure}
    \caption{\small Diagrams for the double conifold, a) shows the dimer diagram, b) the quiver diagram, c) the web diagram and d) the mirror Riemann surface $\Sigma$ as a complex plane with the point at infinity added.}\label{Fig:DiagramsDoubleConifold}
\end{figure}

The Riemann surface $\Sigma$ and these 1-cycles can be systematically constructed from the dimer diagram of the  gauge theory, as follows. Given a dimer diagram, one can define zig-zag paths \cite{Hanany:2005ss}, as paths composed of edges, and which turn maximally to the right at e.g. black nodes and maximally to the left at white nodes. They can be conveniently shown as oriented lines that cross 
once at each edge and turn at each vertex, as shown in Figure \ref{Fig:DimmerDoubleConifoldUnit2} for an illustrative example. Notice that the two zig-zag paths  at each edge must have opposite orientations. The winding numbers of the paths along the two basis cycles of the $\IT^2$ define the $(p,q)$ labels of the external legs in the web diagram for the singularity. The Riemann surface $\Sigma$ can be regarded as a thickening of this web diagram into a genus $g$ surface.

As shown in \cite{Feng:2005gw}, the zig-zag paths of the dimer diagram 
associated to D3-branes at a singularity lead to a tiling of the Riemann 
surface $\Sigma$ in the mirror geometry. Specifically, each zig-zag path 
encloses a face of the tiling of $\Sigma$ which includes a puncture, and 
the $(p,q)$ charge of the associated leg in the web diagram is the $(p,q)$ 
homology charge of the zig-zag path in the $\IT^2$. See for example Figures \ref{Fig:DimmerDoubleConifoldUnit2} and \ref{Fig:WebDiagramDoubleConifold}.

The dimer diagram moreover encodes the 1-cycles in the mirror Riemann 
surface, associated to the different gauge factors in the gauge theory. They are essentially given by zig-zag paths of the tiling of $\Sigma$ itself! This description allows to easily classify supersymmetric wrapped branes (either gauge D-branes or D-brane instantons) in  toric singularities and their mirrors, as extensively exploited in the next sections.

\subsection{Instantons at singularities and dimer diagrams}
\label{sec:dimer-inst}

Dimer diagrams are extraordinary tools to characterize the holomorphic properties of supersymmetric D-branes in local toric CY threefolds, and even closely related non-toric geometries. A particular example of the latter is the description of geometric transitions by  complex deformations in \cite{Franco:2005fd,GarciaEtxebarria:2006aq}, which we revisit in section \ref{sec:complex-def}.

Dimer diagrams can be exploited to describe systems of D-brane instantons in configurations with gauge D-branes. As already noticed in e.g. \cite{Franco:2007ii} (see also \cite{Argurio:2007vqa} for related analysis in orbifold singularities), the dimer diagram diagram encodes the content of instanton charged fermion zero modes and their couplings to the 4d chiral multiplets in the gauge D-branes\footnote{See also e.g. \cite{Aharony:2007pr,Amariti:2008xu,Argurio:2012iw,Franco:2015kfa} for physics of D-brane instantons in systems of branes at singularities in a variety of contexts.}. In short, instantons are associated to faces in the dimer, whose edges correspond to fermion zero modes charged under neighbouring objects, and nodes involving those edges describe the fermion zero mode couplings. This picture again becomes more physical in the mirror configuration: as the gauge D-branes in the dimer turn into D6-branes wrapped on 3-cycles (characterized in terms of 1-cycles in the mirror Riemann surface $\Sigma$), the D-brane instantons turn into D2-branes wrapped on the 3-cycles associated to the corresponding faces (and defined in terms of 1-cycles in $\Sigma$). Disks defining worldsheet instantons 
bounded by D6- and D2-branes now define fermion zero mode couplings, while those bounded by D6-branes only still define superpotential terms. The open string description of the system is thus encoded in a set of gauge and instanton 1-cycles on $\Sigma$.

In this section, we take one further step and describe the backreaction of these D2-brane instantons, in terms of simple graph operations in the mirror picture. This provides a fairly explicit description of the system after the backreaction, as described in previous sections. In particular, including the explicit D6-brane recombination, and the appearance of the 4d superpotential in terms of purely worldsheet instanton effects in the backreacted geometry. The results nicely agree with the open string analysis, as we show in several explicit examples. 

Before describing the detailed recipe and the examples, we would like to note certain points:

$\bullet$ We are considering non-compact intersecting D6-brane systems \cite{Uranga:2002pg}, in which cancellation of charges whose fluxlines can escape to infinity should not be required. However, one should still impose cancellation of RR tadpoles in local homolog; this amounts to requiring that the total homology class of the gauge D-branes should have zero intersection with any compact 3-cycle in the geometry. As is familiar, the condition is equivalent to the cancellation of gauge anomalies for all nodes in the quiver, including the empty ones. For nodes actually occupied by instantons, this condition corresponds to matching the  number of fermion zero modes of type $\lambda$ with the number of those of type $\tilde\lambda$, as in the compact case.

$\bullet$ Since we have no orientifold planes, there are additional neutral fermion zero modes and the instantons in general do not contribute to the superpotential. We are however focusing our attention to the generation of charged field theory operators, which occurs even for instantons that generate higher F-terms. In any event, the lessons learnt in this case will apply in any setup in which the extra fermion zero modes are absent, either by removal by orientifold projection, or by other lifting mechanisms.

$\bullet$ For simplicity we consider the case where there are no gauge D-branes on top of the cycle wrapped by the D-brane instantons. Similar conclusions hold in the alternative case, in which the D-brane instantons are related to gauge instantons. One particular instance is related to the discussion of complex deformations in section \ref{sec:complex-def}, where the D2-brane instantons whose backreaction produces a complex deformation of the geometry are of the same kind as the gauge instantons on deformation fractional branes in the sense of \cite{Franco:2005fd,Franco:2005zu}.

We start our analysis with compact D2-brane instantons, in the presence of compact D6-branes. The analysis with non-compact D6-branes (mirror to D7-branes) is similar and we do not include it explicitly in the present work. Finally we discuss non-compact D2-brane instantons (mirror to non-compact D3-brane instantons).

\subsection{The recipe for backreaction}
\label{sec:recipe}

In this section we present our prescription to find the geometry after the backreaction of an instanton placed in a dimer model. As described above, the configuration is encoded in a tiling of the mirror Riemann surface $\Sigma$, whose zig-zag paths (i.e. faces in the original dimer diagram) correspond to either D2-brane instantons or gauge D6-branes. 

For simplicity, we consider starting with the case of a single instanton, namely one 1-cycle, which defines the instanton 3-cycle $\Pi_3$. We also consider the abelian case, and introduce precisely one D6-brane in the remaining cycles. The backreacted configuration can be generated with very simple steps on this graph:

\begin{itemize}

\item {\bf Step 1. Cut:} Since the 3-cycle $\Pi_3$ wrapped by the instanton should disappear from the geometry, we cut $\Sigma$ by removing a small string around the instanton 1-cycle. After cutting off the strip, each of the two boundaries of $\Sigma$ should be identified to a point. Any  1-cycle intersecting the cut is split,  and turns into a chain with boundary points, weighted by signs according to the orientation of the original intersection; this reproduces the fact that any 3-cycle intersecting $\Pi_3$ turns into a 3-chain in the backreacted geometry.  These 1-chains will be glued in the next step.

\item {\bf Step 2. Recombine:} 
Since $\Pi_3$ is compact in the original geometry, there is an equal number of positive and negative orientation intersections with the whole set of D6-branes. At the level of $\Sigma$, this implies that on each side of the cut, there is an equal number of incoming and outgoing 1-chains. The recipe amounts to matching opposite orientation pieces to define recombined 1-cycles. Most importantly, the recombination should be carried out without crossing edges of the underlying tiling of $\Sigma$, see examples later on. Remarkably, such matchings are unambiguous as a consequence of the bipartite nature of the dimer diagram, as manifest in the examples below.

\item {\bf Step 3. Field theory operators:} The above two steps already define the backreacted geometry. This last step merely establishes that the 4d non-perturbative field theory operator induced by the original D2-brane instanton is easily readable as a worldsheet instanton in the backreacted geometry, bounded by the recombined D6-branes. These are easily found by picking out disks bounded by recombined 1-cycles (and the cut, which recall is regarded as shrunk to a point); the coupling involves the chiral multiplets associated to possible D6-brane intersections still present after recombination.

\end{itemize}

Similar recipes apply to the case of multiple instantons, see section \ref{sec:multiple}.

\subsection{Explicit examples}
\label{sec:examples}

We now turn to consider several examples of D-brane instanton backreactions. We point out that they can be classified in two broad classes. 

$\bullet$ The first corresponds to cases in which the resulting geometric transition is (the mirror of) the complex deformation of the initial toric singularity. This occurs precisely when the D-brane instanton wraps the same nodes as the deformation fractional branes in the sense of \cite{Franco:2005fd, Franco:2005zu}, namely the fractional branes triggering duality cascades ending in smooth complex deformed geometries, generalizing the Klebanov-Strassler throat \cite{Klebanov:2000hb}. The complex deformations are associated to splittings of the web diagram into subwebs in equilibrium (i.e. whose external legs have $(p,q)$ labels adding up to zero); therefore, the 1-cycle defined by the instantons (or deformation fractional branes) are homologically trivial in $\Sigma$, which ends up split in several components. 

$\bullet$ The second class correspond to the more generic situation in which the backreacted geometry does not correspond to a complex deformation of the original configuration. Nevertheless, the proposed recipe to obtain the backreacted geometry applies to more general cases, as we show in explicit examples as well. In these cases, the 1-cycle corresponding to the instanton surrounds a set of punctures corresponding to web diagram external legs whose $(p,q)$ labels do not add up to zero; in these circumstances the topology does not admit a consistent split, which signals that the 1-cycle is homologically non-trivial in $\Sigma$. This agrees with the fact that there is no complex deformation associated to the backreaction. An implication is that the resulting geometries are non-CY (since in the IIB language we are growing 3-cycles despite the absence of CY complex deformation). We recall that this poses no problem with supersymmetry, since the framework is that of generalized geometries, as described in section \ref{sec:backreact}.

\subsubsection{The conifold}
\label{sec:conifold}

\begin{figure}
    \centering
         \begin{subfigure}[t]{0.25\textwidth }
        \begin{center} 
		\includegraphics[width=\textwidth]{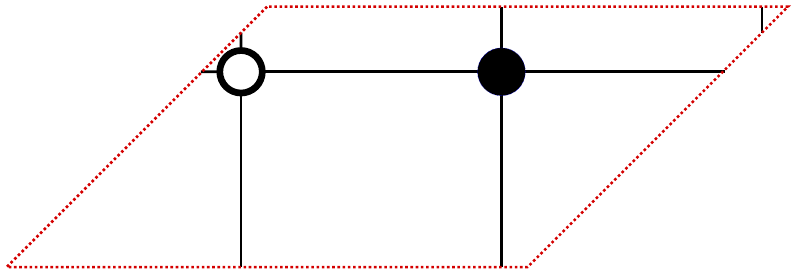}
		\caption{}
		\label{Fig:DimmerConifoldUnit}
		\end{center}
    \end{subfigure}  \hspace{10mm}
    \begin{subfigure}[t]{0.15\textwidth }
        \begin{center} 
		\includegraphics[width=\textwidth]{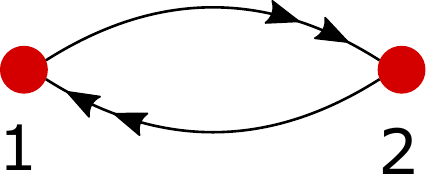}
		\caption{}
		\label{Fig:QuiverDiagramConifold}
		\end{center}
    \end{subfigure}  \hspace{10mm}
    \begin{subfigure}[t]{0.15\textwidth }
        \begin{center} 
		\includegraphics[width=\textwidth]{WebDiagramConifold.pdf}
		\caption{}
		\label{Fig:WebDiagramConifold}
		\end{center}
    \end{subfigure} \hspace{10mm}
    \begin{subfigure}[t]{0.20\textwidth }
        \begin{center} 
		\includegraphics[width=\textwidth]{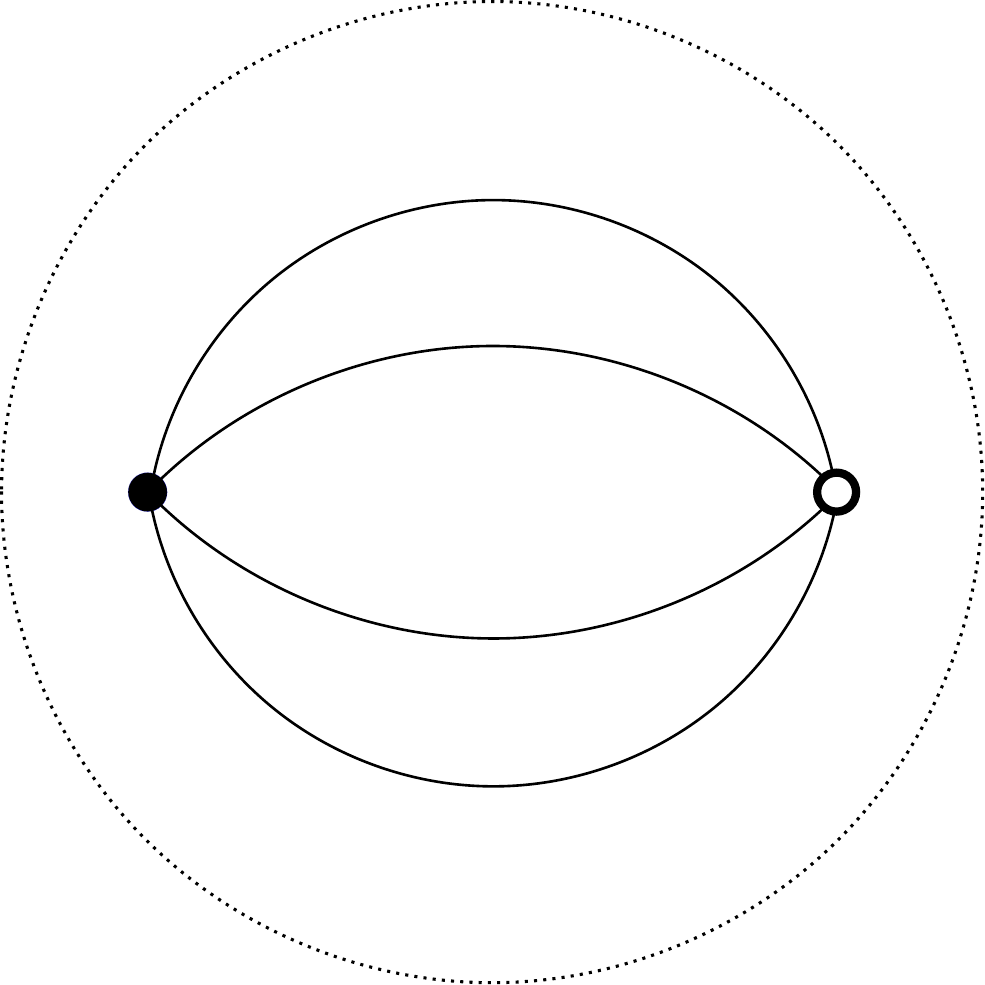}
		\caption{}
		\label{Fig:RiemannConifold}
		\end{center}
    \end{subfigure}
    \caption{\small Diagrams for the conifold, a) shows the dimer diagram, b) the quiver diagram, c) the web diagram and d) the mirror Riemann surface $\Sigma$ is a sphere with 4 punctures, which we depict as a complex plane (with the point at infinity added).
    }\label{Fig:DiagramsConifold}
\end{figure}

We now describe a simple example, which illustrates mainly steps 1 and 2. Consider a system of D-branes at a conifold singularity. In Figures \ref{Fig:DimmerConifoldUnit}, \ref{Fig:QuiverDiagramConifold}, \ref{Fig:WebDiagramConifold} and \ref{Fig:RiemannConifold} we display the dimer diagram, the quiver diagram, the web diagram, and the mirror Riemann surface, a sphere with four punctures, respectively.

\begin{figure}
    \centering
    \begin{subfigure}[t]{0.3\textwidth }
        \begin{center} 
		\includegraphics[width=\textwidth]{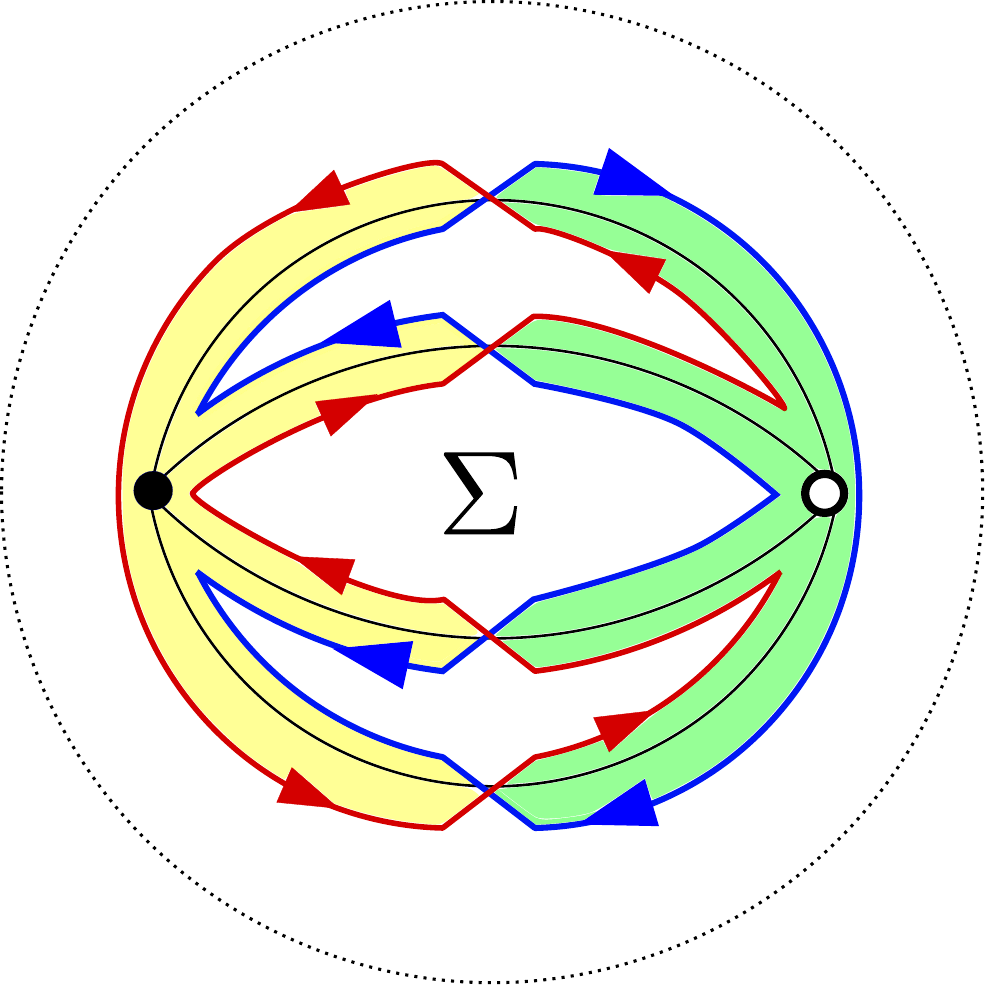}
		\caption{}
		\label{Fig:RiemannConifoldBackreaction}
		\end{center}
    \end{subfigure} \hspace{10mm}
    \begin{subfigure}[t]{0.3\textwidth }
        \begin{center} 
		\includegraphics[width=\textwidth]{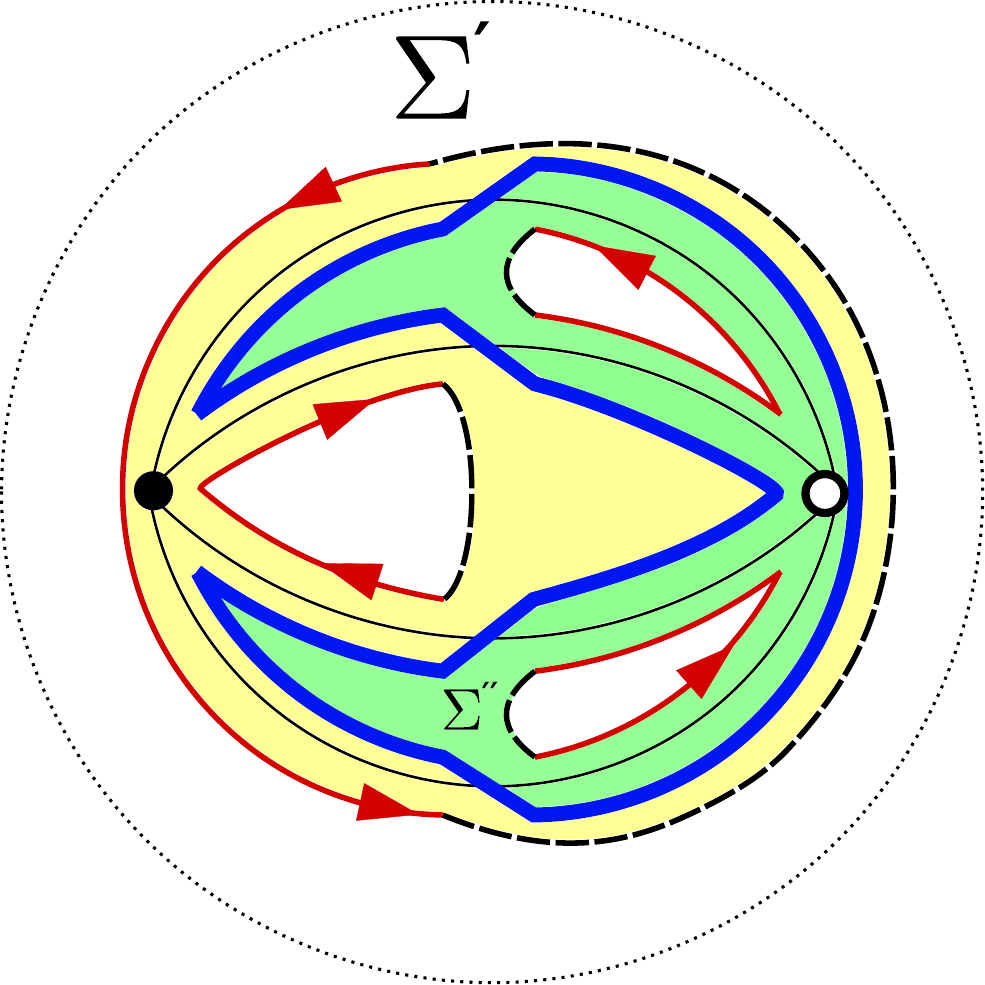}
		\caption{}
		\label{Fig:RiemannConifoldZigZagDeformation}
		\end{center}
    \end{subfigure}  \vspace{10mm}
    
    \begin{subfigure}[t]{0.3\textwidth }
        \begin{center} 
		\includegraphics[width=\textwidth]{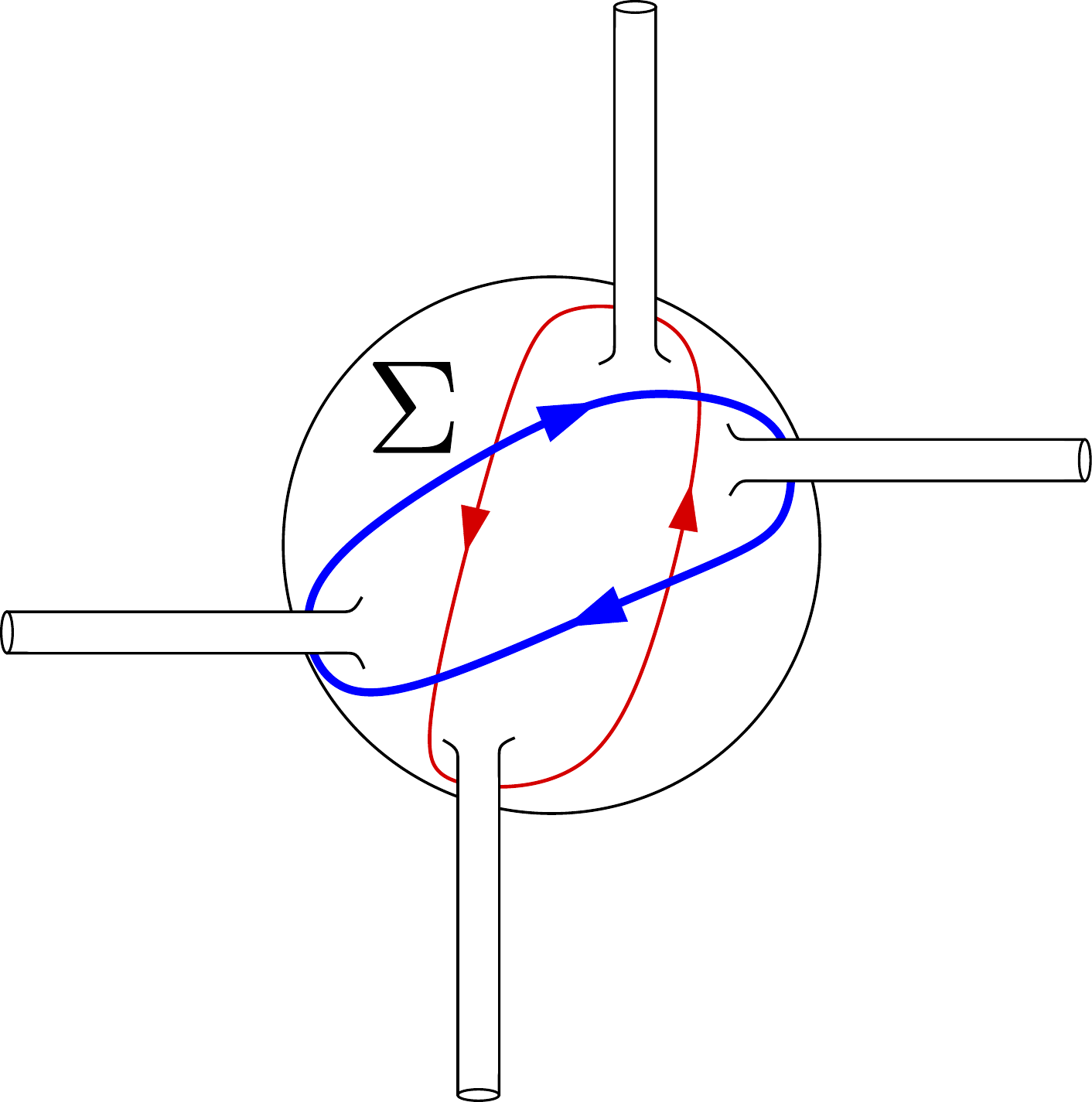}
		\caption{}
		\label{Fig:RiemannConifold3d}
		\end{center}
    \end{subfigure} \hspace{10mm}
    \begin{subfigure}[t]{0.3\textwidth }
        \begin{center} 
		\includegraphics[width=\textwidth]{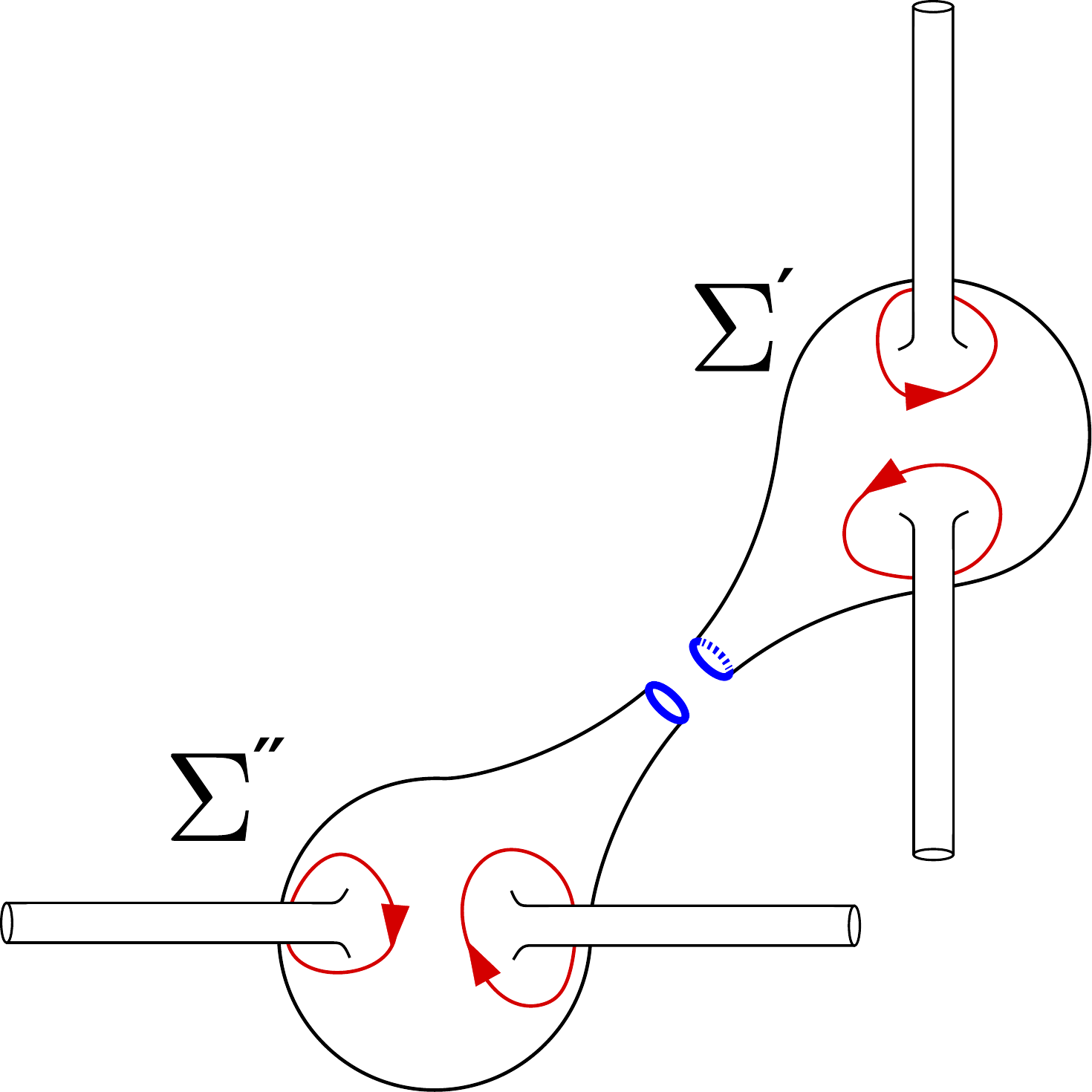}
		\caption{}
		\label{Fig:RiemannConifold3dBackreaction2}
		\end{center}
    \end{subfigure}  
    \caption{\small Instanton backreaction on a conifold. In a) $\Sigma$ and its zig-zag paths are shown. The red one is wrapped by a D6-brane and the blue one by a D2-brane instanton. In b) the Riemann surface is cut along the blue strip wrapped by the D2-brane instanton, splitting the red 1-cycle at four points; the corresponding open endpoints are duly recombined to form four recombined 1-cycles, each surrounding a puncture. c) and d) provide alternative graphical views of the process.  }\label{Fig:SigmaBackreactionConifold}
\end{figure}

As explained, the cycles corresponding to the two nodes of the quiver (faces of the dimer diagram) are given by zig-zag paths of the tiling of $\Sigma$, as shown in blue and red in Figure \ref{Fig:RiemannConifoldBackreaction}.

Let us put one D6-brane in the red 1-cycle (abusing language, as it actually denotes a 3-cycle) and a D2-brane instanton on the blue 1-cycle. In the open string description, the arrows in the quiver (edges in the dimer) correspond to two fermion zero modes of type $\lambda$ and two $\tilde\lambda$, with quartic couplings (essentially, as in the conifold superpotential), which allow to saturate them. The instanton amplitude therefore reduces to the instanton exponential, with no field theory operator (as expected, since there are no charged chiral multiplets in the theory).

Let us turn to the description of the system in terms of the instanton backreaction, and let us describe it using our recipe for the Riemann surface picture.
According to step 1, we cut the sphere along the blue line, splitting the red 1-cycle at four points (two with each orientation). We recombine such pieces according to step 2, resulting in the picture in Figure \ref{Fig:RiemannConifoldZigZagDeformation}.

Concerning step 3, there are two worldsheet instantons, which are obtained by certain recombination of the original disks. They however do not involve any chiral multiplets (as there is none left). This is actually the required result, since the non-perturbative contribution by the original D-brane instanton should include no field-theoretical operator, just the instanton exponential.

In Fig. \ref{Fig:RiemannConifold3dBackreaction2} we see that, in fact, the deformed geometry corresponds to the complex deformation of the conifold \cite{Klebanov:2000hb}, agreeing  with the fact that the two punctures in each daughter Riemann surface correspond to sub-webs in equilibrium \cite{Franco:2005fd}. In the next section we will see that this is only the case whenever the D-brane instantons introduced match the corresponding fractional brane triggering the complex deformation. In other cases, the backreacted geometry does not correspond to a CY deformation.

Finally let us mention one feature of our recipe. Since it is based on the graphical properties of the tiling of the mirror Riemann surface, with all 1-cycles present, the recipe provides an accurate description of instantons in the presence of D-branes in all nodes (at least, all nodes other than those occupied by the instanton). This will become more clear in richer examples in the next sections.

\subsubsection{Double Conifold}
\label{sec:double-con}

Let us now turn to a slightly more involved example, to illustrate in more detail step 3, since we will obtain a non-trivial superpotential. Consider a system of branes in a double conifold singularity. The dimer diagram, quiver diagram, web diagram and mirror Riemann surface are shown in Figs. \ref{Fig:DimmerDoubleConifoldUnit2}, \ref{Fig:QuiverDiagramDoubleConifold}, \ref{Fig:WebDiagramDoubleConifold}, \ref{Fig:RiemannDoubleConifold}, respectively.

\begin{figure}
    \centering
         \begin{subfigure}[t]{0.15\textwidth }
        \begin{center} 
		\includegraphics[width=\textwidth]{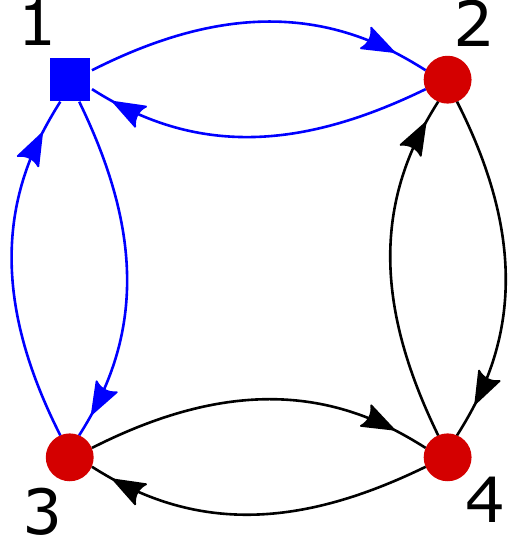}
		\caption{}
		\label{Fig:QuiverDiagramDoubleConifoldDeformation}
		\end{center}
    \end{subfigure}  \hspace{10mm}
    \begin{subfigure}[t]{0.30\textwidth }
        \begin{center} 
		\includegraphics[width=\textwidth]{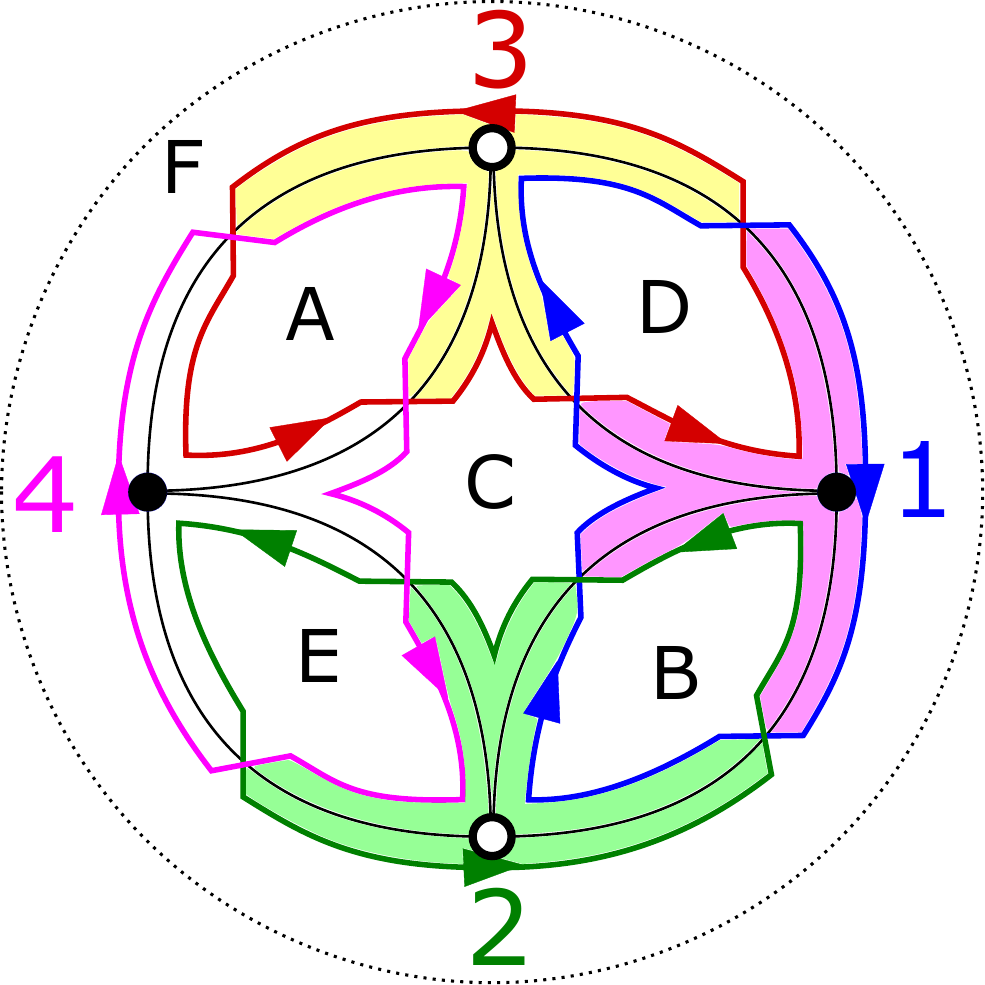}
		\caption{}
		\label{Fig:RiemannDoubleConifoldZigZag}
		\end{center}
    \end{subfigure}  \hspace{10mm}
    \begin{subfigure}[t]{0.30\textwidth }
        \begin{center} 
		\includegraphics[width=\textwidth]{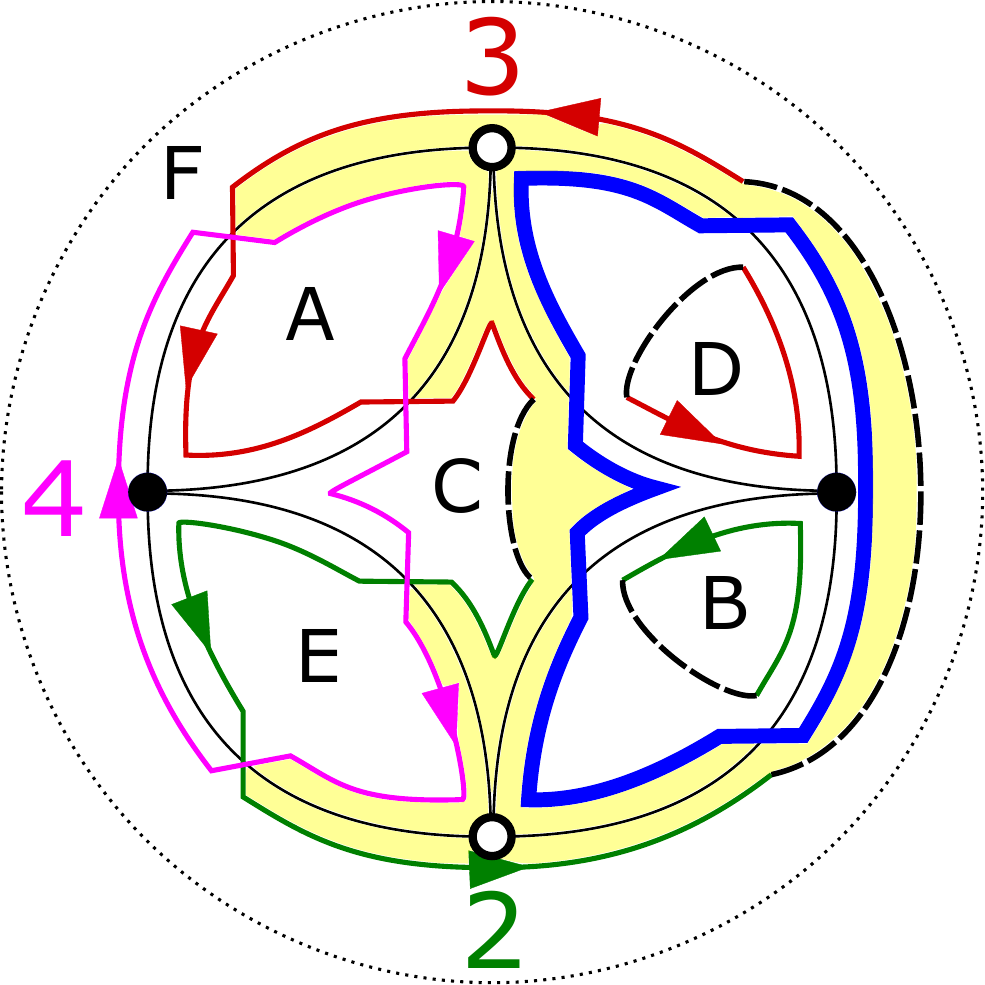}
		\caption{}
		\label{Fig:RiemannDoubleConifoldZigZagBackreactionRe}
		\end{center}
    \end{subfigure}
    \caption{\small Double conifold backreaction. a) Quiver diagram with a D-brane instanton in node 1 (blue square) and gauge D-branes on the remaining nodes (red circles); blue arrows denote instanton charged fermion zero modes, while black arrows denote 4d charged chiral multiplets. Figure b) shows $\Sigma$ before the backreaction, where the D-brane instanton (resp. gauge D-branes) is the blue (resp. red) zig-zag path and there are three worldsheet instantons shaded in light colors. In c) $\Sigma$ after the backreaction is shown. The D-brane instanton has torn $\Sigma$ apart and the gauge D-branes have recombined. }
    \label{Fig:DoubleConifoldBackreaction}
\end{figure}

The four cycles in $\Sigma$ corresponding to the quiver nodes are shown in  Fig. \ref{Fig:RiemannDoubleConifoldZigZag}. Let us put 3 D6-branes, one in each cycle but the blue one, and a D2-brane instanton in the blue cycle. The bifundamental matter fields charged under the``would-be-D6-brane" of node 1 become now fermion zero modes $\lambda_{12}, \lambda_{13}, \tilde{\lambda}_{21}, \tilde{\lambda}_{31}$, see Fig. \ref{Fig:QuiverDiagramDoubleConifoldDeformation}. These have couplings given by the worldsheet instantons stretching from the D2-instanton to the D6's. As shown in Fig. \ref{Fig:RiemannDoubleConifoldZigZag}, these couplings are :

\begin{equation}
\sim \lambda_{13} X_{34} X_{43} \tilde{\lambda}_{31} + \lambda_{12} X_{24} X_{42} \tilde{\lambda}_{21} ,
\end{equation}
where the $\sim$ indicates that we are omitting coefficients for these couplings.

These couplings can be used to saturate the zero modes, and there is only one way of doing so, yielding the following 4d non-perturbative field theory operator,
\begin{equation}
\mathcal{O}_{\rm charged} \sim ( X_{34} X_{43}) (X_{24} X_{42}) \label{Eq:DoubleConifoldWnp}
\end{equation}
where recall that parenthesis indicate determinants when promoting to the non-abelian case.

Let us now consider the description of the instanton in the backreaction picture.
According to \textbf{step 1} of our recipe, the cycle wrapped by the instanton shrinks to a point, and this splits $\Sigma$ into two pieces and splits the D6-brane 1-cycles. In \textbf{step 2}, the latter are recombined, as shown in Fig. \ref{Fig:RiemannDoubleConifoldZigZagBackreactionRe}. Finally, \textbf{step 3} 
instructs us to use worldsheet instantons on disks in the backreacted geometry, shaded in the Figure, to obtain the 4d field theory operator, which indeed reproduces (\ref{Eq:DoubleConifoldWnp}).

\medskip

Incidentally, we note that the remaining geometry is just the conifold. Indeed this coincides with the removal of punctures D and B, which corresponds to a subweb in equilibrium and, thus, to a complex deformation. This agrees with the fact that node 1, in which the instanton sits, actually corresponds to a deformation fractional brane in the sense of \cite{Franco:2005fd, Franco:2005zu}, as we develop in section \ref{sec:complex-def}.

\subsubsection{A dP$_0$ example} 

\begin{figure}
    \centering
         \begin{subfigure}[t]{0.25\textwidth }
        \begin{center} 
		\includegraphics[width=\textwidth]{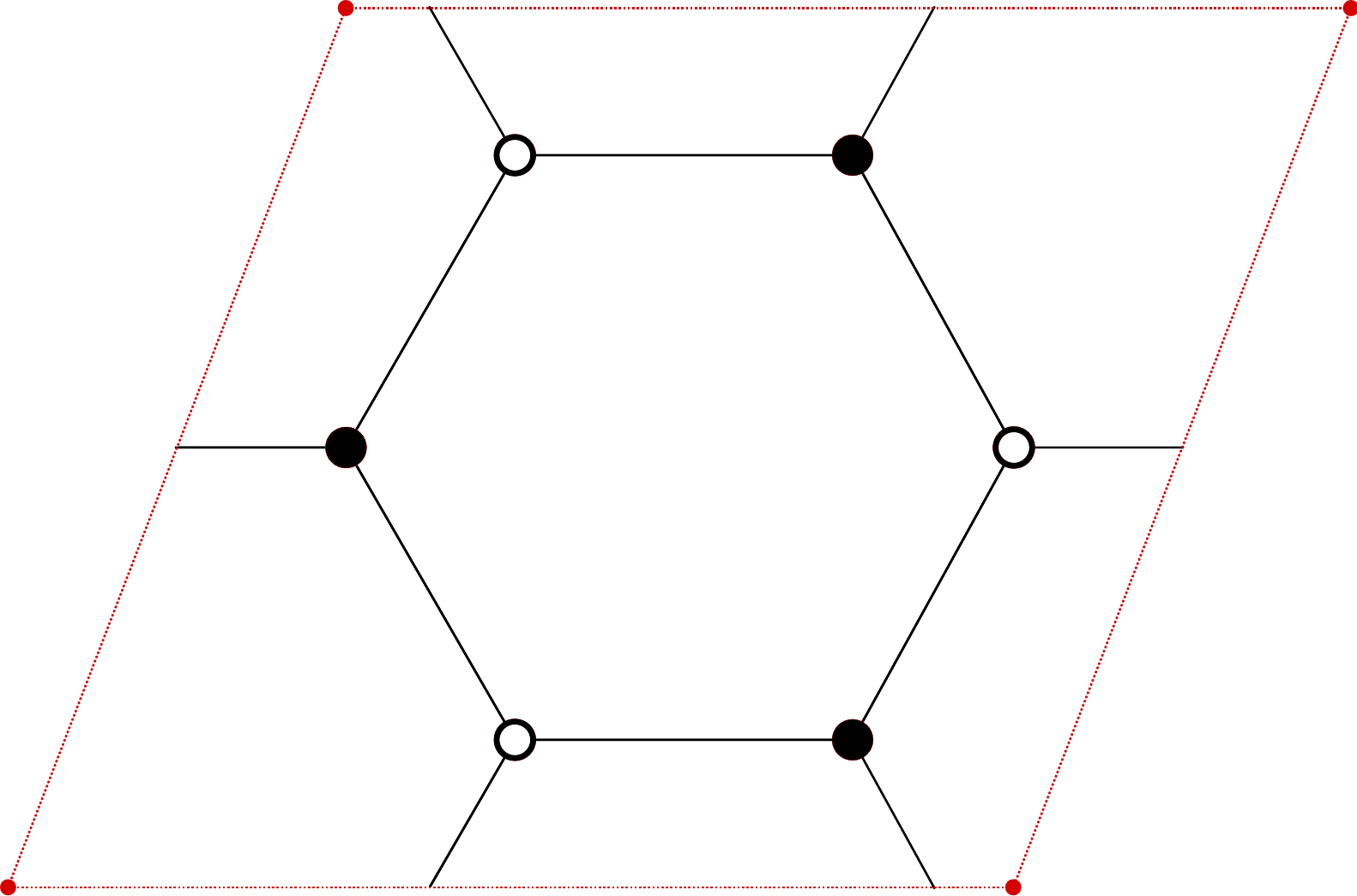}
		\caption{}
		\label{Fig:DimmerdP0Unit}
		\end{center}
    \end{subfigure}  \hspace{10mm}
    \begin{subfigure}[t]{0.20\textwidth }
        \begin{center} 
		\includegraphics[width=\textwidth]{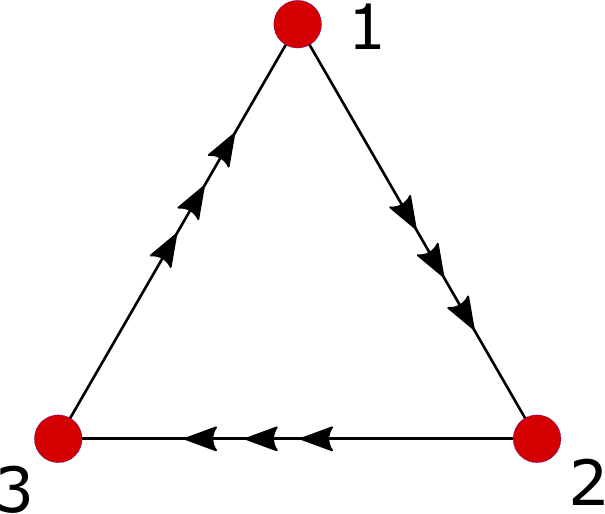}
		\caption{}
		\label{Fig:QuiverDiagramdP0}
		\end{center}
    \end{subfigure}  \hspace{10mm}
    \begin{subfigure}[t]{0.20\textwidth }
        \begin{center} 
		\includegraphics[width=\textwidth]{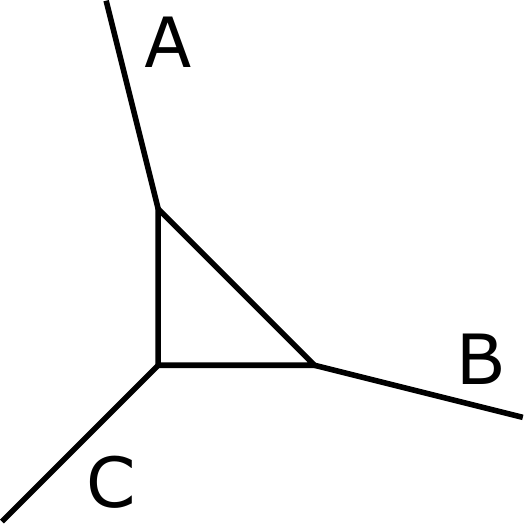}
		\caption{}
		\label{Fig:WebDiagramdP0}
		\end{center}
    \end{subfigure} \hspace{10mm}
    \caption{\small Diagrams for dP0 singularity, a) shows the dimer diagram, b) the quiver diagram and c) the web diagram. The mirror riemann surface tiling $\Sigma$ coincides with the dimer diagram.}\label{Fig:DiagramsdP0}
\end{figure}

So far all examples have secretly been  related to complex deformations, in that the instanton occupies the node of deformation fractional branes, in the sense of \cite{Franco:2005fd, Franco:2005zu} (namely, the total 1-cycle class surrounds the punctures associated to a subweb in equilibrium in the web diagram), as we discuss in more detail in section \ref{sec:complex-def} . However, our recipe applies to arbitrary instantons, and to illustrate this more generic situation, we now study the case of the $\IC^3/\IZ_3$ orbifold, also known as the dP$_0$ quiver theory. The dimer diagram of this theory, its quiver diagram and its web diagram are shown in Fig. \ref{Fig:DiagramsdP0}. Incidentally (as in all models whose web diagram has exactly one internal face), the mirror Riemann surface tiling coincides with the dimer diagram.

\begin{figure}
    \centering
         \begin{subfigure}[t]{0.10\textwidth }
        \begin{center} 
		\includegraphics[width=\textwidth]{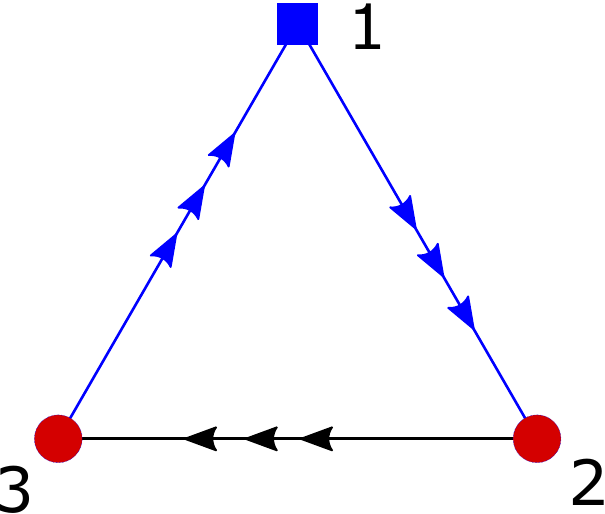}
		\caption{}
		\label{Fig:QuiverDiagramdP0Backreaction}
		\end{center}
    \end{subfigure}  \hspace{10mm}
    \begin{subfigure}[t]{0.35\textwidth }
        \begin{center} 
		\includegraphics[width=\textwidth]{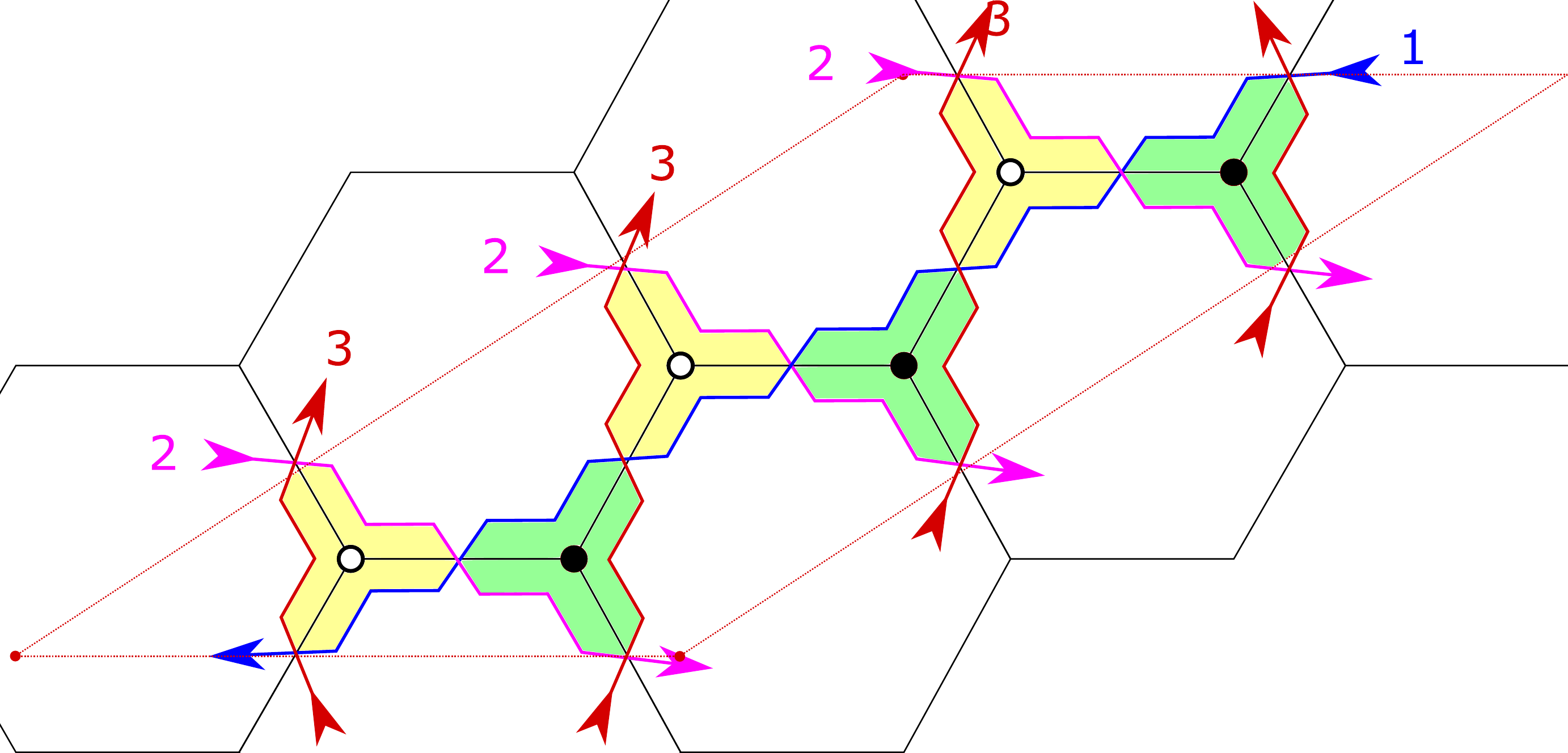}
		\caption{}
		\label{Fig:RiemanndP0ZigZag3Inst}
		\end{center}
    \end{subfigure}  \hspace{10mm}
    \begin{subfigure}[t]{0.35\textwidth }
        \begin{center} 
		\includegraphics[width=\textwidth]{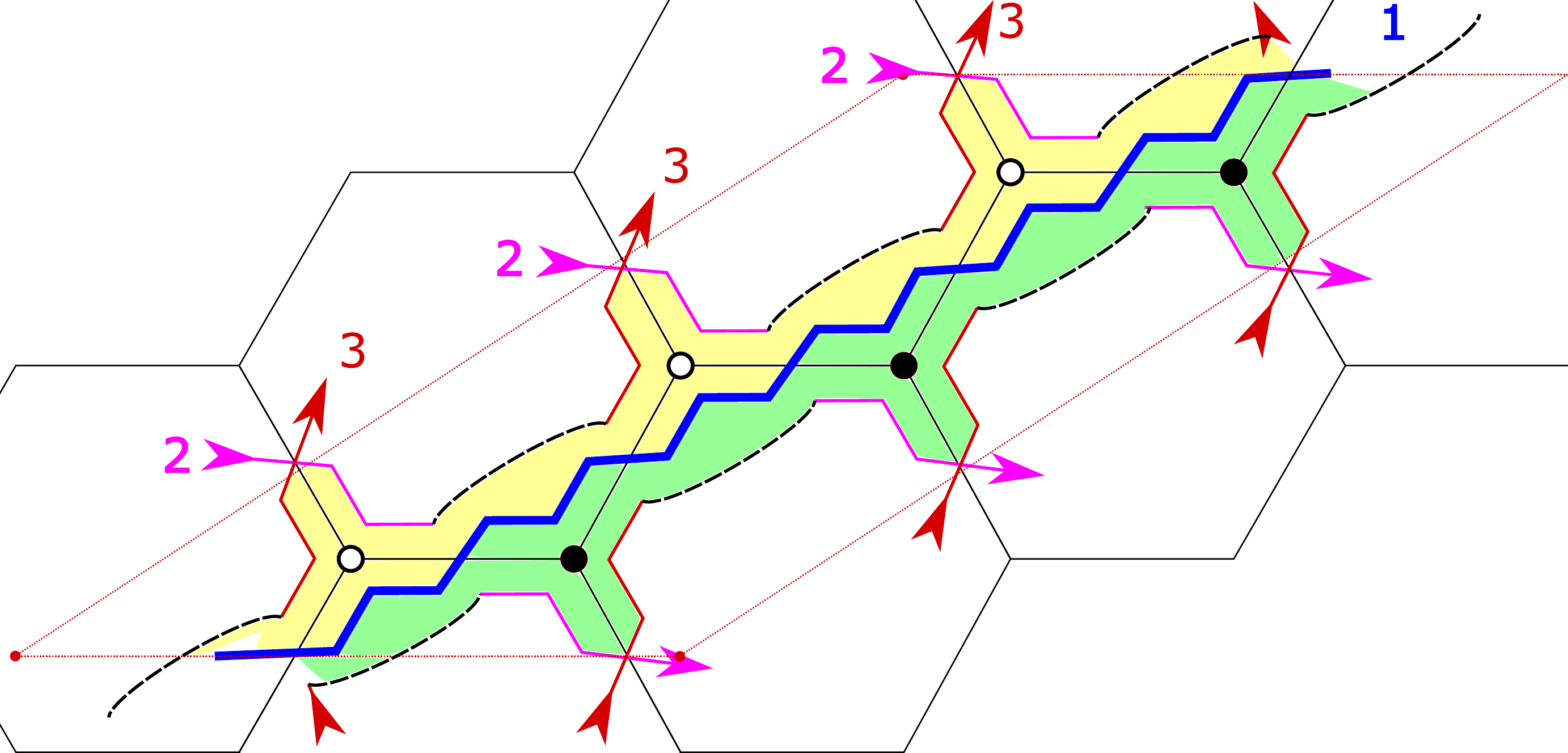}
		\caption{}
		\label{Fig:RiemanndP0ZigZag3BackreactionInst}
		\end{center}
    \end{subfigure}
    \caption{\small dP0 backreaction. a) is the quiver diagram with a D-brane instanton in node 1. Blue arrows denote fermion zero modes. b) shows $\Sigma$ before the backreaction, where the D-brane instanton is the blue zig-zag path and there are two WS instantons shaded in light colors. In c) $\Sigma$ after the backreaction is shown. The D-instanton has dissapeared from $\Sigma$ and the branes have recombined. Two WS instantons remain.  }\label{Fig:dP0Backreaction}
\end{figure}

Consider putting one D6-brane in each of the dimer diagram faces 2 and 3, and a D-brane instanton in 1. The quiver diagram is shown in Fig. \ref{Fig:QuiverDiagramdP0Backreaction}, while the mirror picture is depicted in Fig. \ref{Fig:RiemanndP0ZigZag3Inst}.  There are charged fermion zero modes $\lambda^{i}_{12}, \tilde{\lambda}^{i}_{31}, \, \, i=1,2,3$, with couplings 
\begin{equation}
\sim \epsilon_{ijk}\,{\lambda}^{i}_{12}\, X^{j}_{23}\, \tilde{\lambda}^{k}_{31}  , \quad i,j,k=1,2,3 \\
\end{equation}

The resulting field theory operator has the structure ${\cal O}_{\rm charged}=(X^1_{23} )( X^2_{23} )(X^3_{23})$.

\medskip

Let us now turn to the backreaction description.
Note that we have chosen a basis for the unit cell that shows the instanton cycle and worldsheet instantons more easily. Applying steps 1 and 2 one finds that the Riemann surface tiling after the backreaction looks as shown in \ref{Fig:RiemanndP0ZigZag3BackreactionInst}. Unlike in the previous examples, the surface is not split, but merely changed by decreasing its genus by 1, in a fashion similar to Fig. \ref{fig:recombination}. This is a general feature in backreaction of instantons not associated to complex deformations, as explained in section \ref{sec:complex-def}.

Concerning step 3, as the branes recombine, the sets of three original worldsheet instantons of each kind merge to one, thus giving rise to the following operator,
\begin{equation}
\mathcal{O}_{\rm charged} \sim (X^1_{23} )( X^2_{23})(  X^3_{23}),
\end{equation}
which indeed coincides with the superpotential obtained by saturating the fermion zero modes of the D-brane instanton. Again, the D-brane instanton disappears but the geometry and the worldsheet instantons reproduce the same field theory operator.

\subsubsection{An F$_0$ example} 
\label{Sec:F0Backreaction}

\begin{figure}
    \centering
         \begin{subfigure}[t]{0.22\textwidth }
        \begin{center} 
		\includegraphics[width=\textwidth]{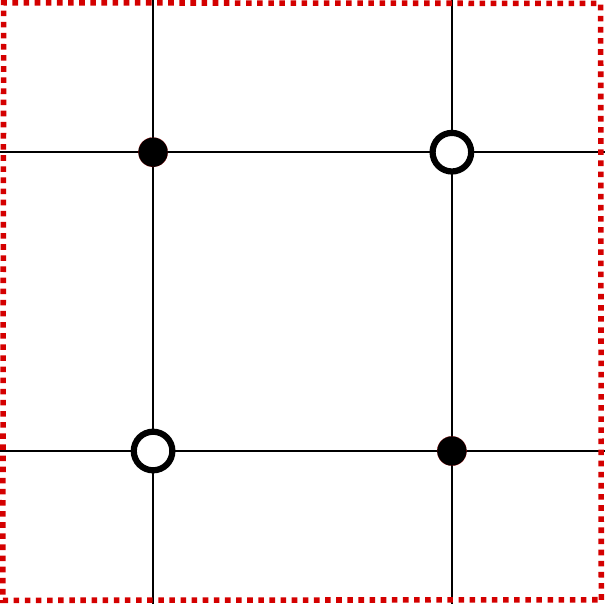}
		\caption{}
		\label{Fig:DimmerF0Unit}
		\end{center}
    \end{subfigure}  \hspace{15mm}
    \begin{subfigure}[t]{0.22\textwidth }
        \begin{center} 
		\includegraphics[width=\textwidth]{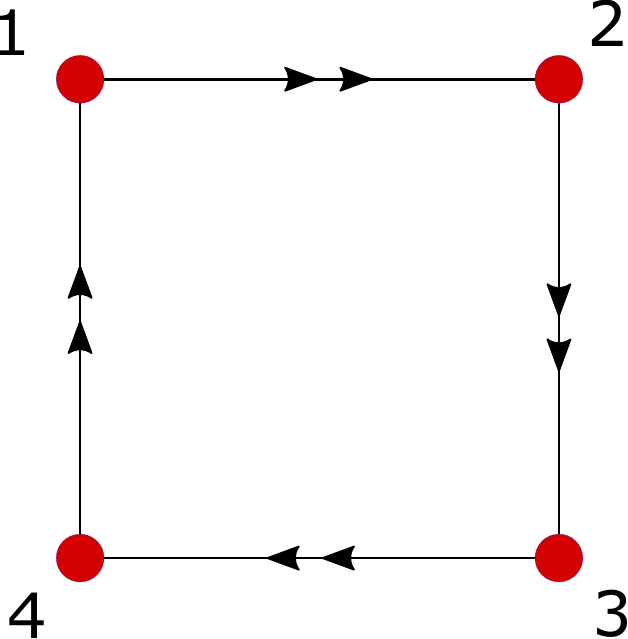}
		\caption{}
		\label{Fig:QuiverDiagramF00}
		\end{center}
    \end{subfigure}  \hspace{15mm}
    \begin{subfigure}[t]{0.22\textwidth }
        \begin{center} 
		\includegraphics[width=\textwidth]{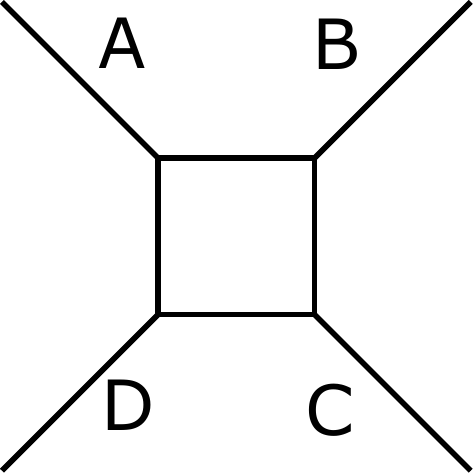}
		\caption{}
		\label{Fig:WebDiagramF0}
		\end{center}
    \end{subfigure} \hspace{10mm}
    \caption{\small Diagrams for F0 singularity, a) shows the unit cell of the dimer diagram, b) the quiver diagram and c) the web diagram. The mirror Riemann surface tiling $\Sigma$ coincides with the dimer diagram.}\label{Fig:DiagramsF0}
\end{figure}

Let us give one further simple example, based on D-branes at a complex cone over F$_0$. In Fig. \ref{Fig:DiagramsF0} we provide the dimer diagram for the F$_0$ theory, together with the quiver diagram and the web diagram. Incidentally, the tiling of the mirror Riemann surface coincides with the dimer diagram. From the web diagram one sees that this geometry admits a complex deformation by the removal of the legs A and C, which form a subweb in equilibrium. As we will see in the next section, to reproduce this complex deformation, which implies the removal of two faces in $\Sigma$, two instantons are needed. 

\begin{figure}
    \centering
         \begin{subfigure}[t]{0.16\textwidth }
        \begin{center} 
		\includegraphics[width=\textwidth]{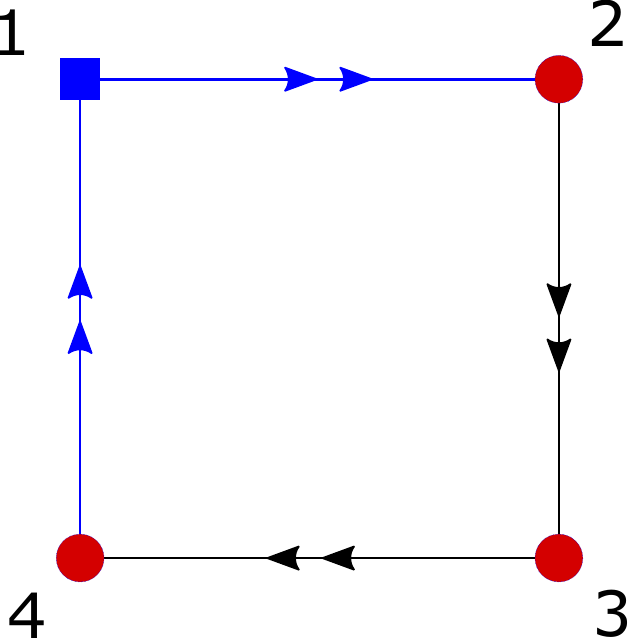}
		\caption{}
		\label{Fig:QuiverDiagramF0Backreaction}
		\end{center}
    \end{subfigure}  \hspace{10mm}
    \begin{subfigure}[t]{0.32\textwidth }
        \begin{center} 
		\includegraphics[width=\textwidth]{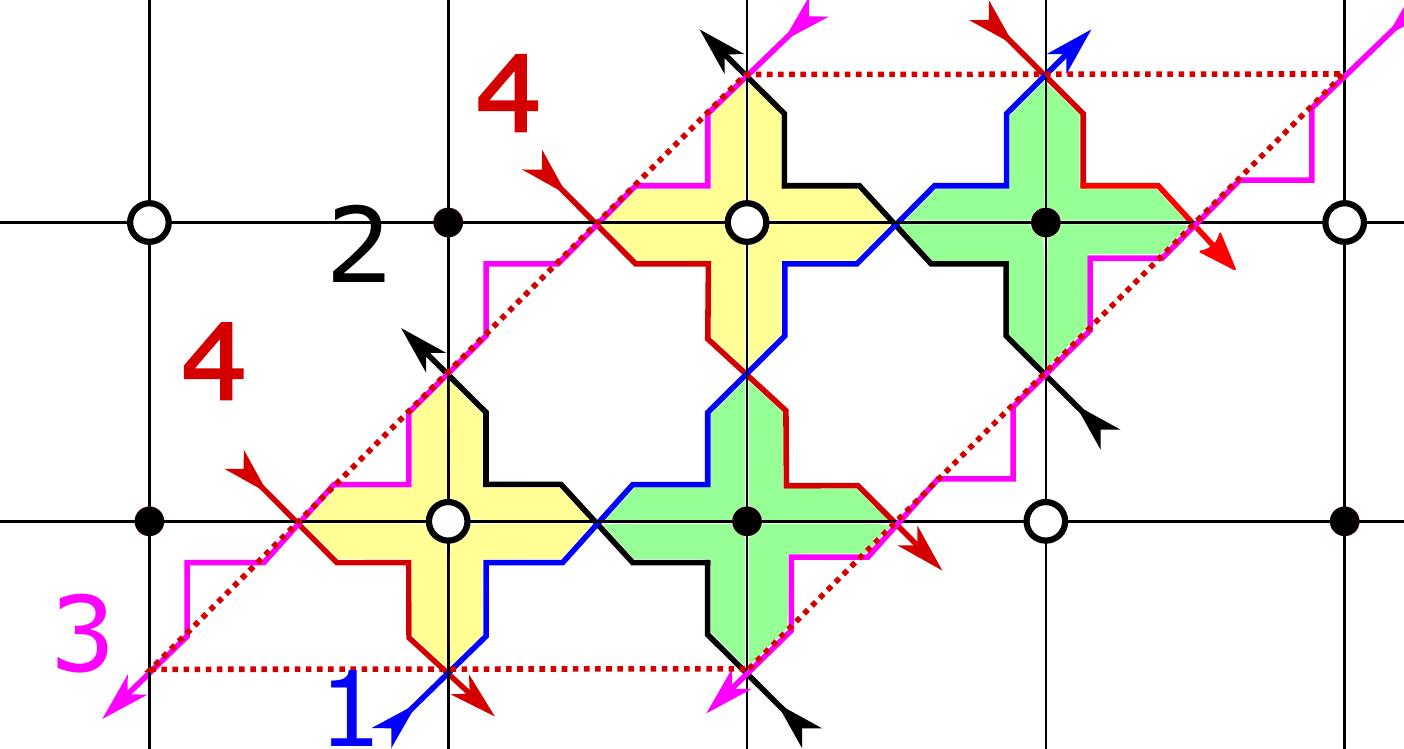}
		\caption{}
		\label{Fig:RiemannF0Inst}
		\end{center}
    \end{subfigure}  \hspace{10mm}
    \begin{subfigure}[t]{0.32\textwidth }
        \begin{center} 
		\includegraphics[width=\textwidth]{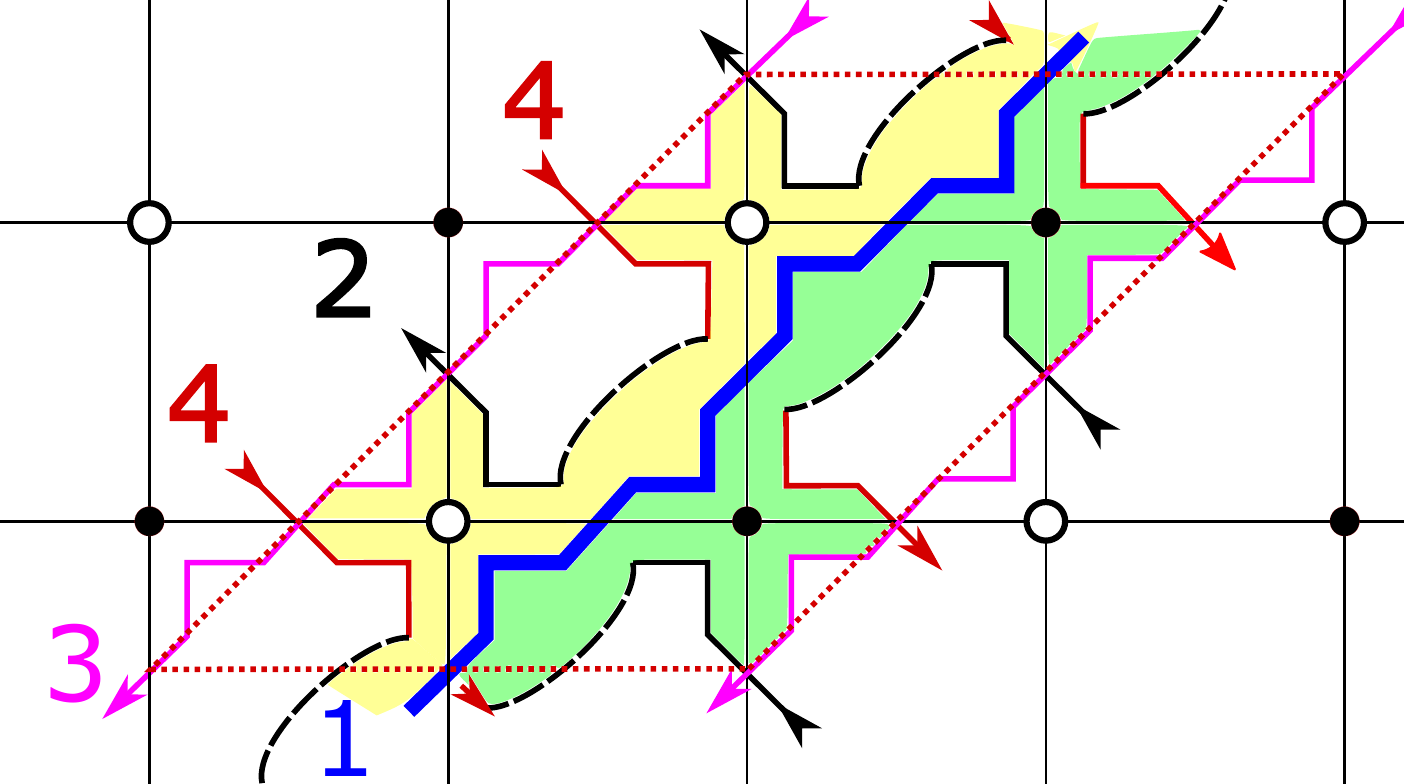}
		\caption{}
		\label{Fig:RiemannF0BackreactionInst}
		\end{center}
    \end{subfigure}
    \caption{\small F0 backreaction with a single instanton. a) is the quiver diagram with a D-brane instanton in node 1. Blue arrows denote fermion zero modes. b) shows $\Sigma$ before the backreaction, where the D-brane instanton is the blue zig-zag path and there are two WS instantons shaded in light colors. In c) $\Sigma$ after the backreaction is shown. The D-instanton has dissapeared from $\Sigma$ and the branes have recombined. Two WS instantons remain.  }\label{Fig:F0Backreaction}
\end{figure}

Let us instead study the system with a single D-brane instanton in node 1, and one gauge D-brane in each of the nodes 2, 3 and 4. The quiver diagram for this configuration is shown in Fig. \ref{Fig:QuiverDiagramF0Backreaction}. There are four chiral matter multiplets and 4 fermion zero modes $\lambda^{i}_{12}, \tilde{\lambda}^{i}_{41}, \, \, i=1,2$ that are saturated using, as usual, the couplings from the worldsheet instantons, see Fig  \ref{Fig:RiemannF0Inst},
\begin{equation}
\sim \epsilon_{ij}\epsilon_{kl} {\lambda}^{i}_{12} X^{k}_{23} X^{j}_{34} \tilde{\lambda}^{l}_{41}.
\end{equation}
The resulting field theory operator has the structure 
\beqa
\mathcal{O}_{\rm charged}\, \sim\,  (X^{1}_{23} X^{1}_{34}) (X^{2}_{23} X^{2}_{34})\, -\,  (X^{1}_{23} X^{2}_{34}) (X^{2}_{23} X^{1}_{34})\, 
\label{supo-np-f0}
\eeqa
with, as usual, parenthesis denoting determinants when promoting to the non-abelian case.

Let us now turn to the description from the backreaction perspective.
Following steps 1 and 2, the Riemann surface $\Sigma$ is cut, and the gauge D-brane 1-cycle recombine as  shown in Fig. \ref{Fig:RiemannF0BackreactionInst}. Finally, step 3 tells us that the field theory operator can be read off from the remaining worldsheet instantons, which give a contribution in agreement with (\ref{supo-np-f0}). Notice that the complete amplitude contains the sum over two worldsheet instantons, located on either side of the instanton line, which differ in orientation (hence the relative sign) and in the pairing of fields by recombination bridges (hence the change in parenthesis structures). Fields living in intersections connected by a brane, rather than a recombined brane (black dashed line), yield a determinant in the non-abelian case (arising from a sum over recombination possibilities, as discussed in section \ref{sec:multiple-inters}.

Note that, like in the dP$_0$ case, the backreaction has not split $\Sigma$ in two, but rather reduced its genus,  in agreement with our earlier spoiler about  instantons not describing complex deformations, see \ref{sec:complex-def} for additional details.

\subsection{Non-compact instantons}
\label{sec:non-compact}

\begin{figure}
    \centering
         \begin{subfigure}[t]{0.2\textwidth }
        \begin{center} 
		\includegraphics[width=\textwidth]{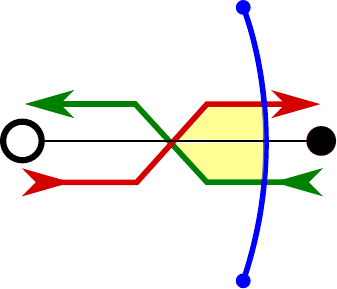}
		\caption{}
		\label{Fig:NonCompact}
		\end{center}
    \end{subfigure}  \hspace{10mm}
    \begin{subfigure}[t]{0.21\textwidth }
        \begin{center} 
		\includegraphics[width=\textwidth]{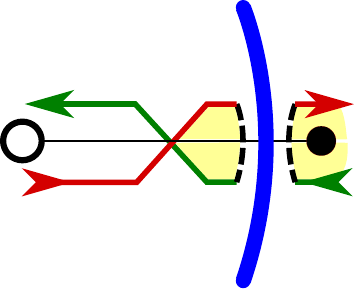}
		\caption{}
		\label{Fig:NonCompactBack}
		\end{center}
    \end{subfigure}  
    \caption{Backreaction of a non-compact instanton. In a) two zig-zag paths corresonding to branes are shown in red and green, with a non-compact D2-brane instanton in blue. In b) the instanton has backreacted and the branes recombined; worldsheet instantons are shown shaded in yellow.   }\label{Fig:NonCompactI}
\end{figure}

Let us makes one small aside to point out that the description provided can be exploited also for instantons wrapped on non-compact cycles, namely D3-brane instantons on non-compact 4-cycles passing through the singular point in the type IIB picture \cite{Forcella:2008au}. Following \cite{Franco:2013ana} in the type IIA mirror, we have D2-brane instantons on non-compact 3-cycles which on the mirror Riemann surface correspond to 1-cycles stretching between two punctures. As discussed in the references, there is one such non-compact cycle for each bi-fundamental chiral multiplet $X_{ab}$ in the gauge theory, such that the D2$_a$-D6 and D6-D2$_b$ open string sectors give fermion zero modes $\lambda_a$ and $\tilde\lambda_b$, with couplings
\beqa
\lambda_a X_{ab} \tilde\lambda_b
\eeqa
This arises from worldsheet instantons, see Figure \ref{Fig:NonCompact}. Thus, integration over these fermion zero modes gives 4d field theory operators with the structure $\det(X_{ab})$.

This structure is easily reproduced in terms of the backreaction picture, by the same steps as in earlier sections: The Riemann surface is cut along the (non-compact) D2-brane instanton 1-cycle, and the D6-brane 1-cycles are duly recombined, and define the necessary worldsheet instantons to produce the field theory operators. The result is shown in Figure \ref{Fig:NonCompactBack}

\subsection{Multiple instantons}
\label{sec:multiple}

All the examples given in the previous section correspond to systems with the D-brane instanton in a single node. In this section we consider the more general possibility of effects of multiple instantons occupying several nodes in the quiver diagram. 

To be precise, we consider the effects of the simultaneous presence of various instantons; this can be regarded as the effect of an instanton on the combined class, but it is better described as a multi-instanton effect, in the sense of \cite{GarciaEtxebarria:2007zv,GarciaEtxebarria:2008pi}. We will show that, although the system contains novel features in the open string description, like the presence of fermion zero modes stretching between pairs of instantons, the final result is nicely captured by the backreaction picture using the same recipe as in earlier examples.

The novel features in the open string description are the appearance of sectors of open string with both endpoints on the D2-brane instantons, which yield zero modes of the multi-instanton configuration. In these sectors, the 4d Minkowski dimensions yield NN boundary conditions, while the compactification space dimensions are similar to intersecting D6-branes (to which they are related by nominal T-duality along the 4d Minkowski directions). There is one complex bosonic zero mode, and two real fermion zero modes, all in the bi-fundamental of the worldvolume groups on the D2-brane instantons. This is the equivalent to a 4d $\NN=1$ chiral multiplet (namely, that in the T-dual  D6-brane system).
The couplings of this chiral multiplet of zero modes to the standard charged fermion zero modes and the actual 4d chiral multiplet can be read from the dimer diagram (or worldsheet instantons in the mirror Riemann surface), see later for explicit examples.

Finally, recall that the instanton amplitude is obtained upon integration over fermion zero modes (both in D2-D6 and D2-D2 sectors) and over bosonic zero modes. Integration over bosonic zero modes in the D2-D2 sectors effectively drags us into regimes in which the instantons are bound and act as a recombined D2-brane. This will be nicely recovered in the backreacted picture, to be discussed next.. 

\subsubsection{An F$_0$ example.}

\begin{figure}
    \centering
         \begin{subfigure}[t]{0.16\textwidth }
        \begin{center} 
		\includegraphics[width=\textwidth]{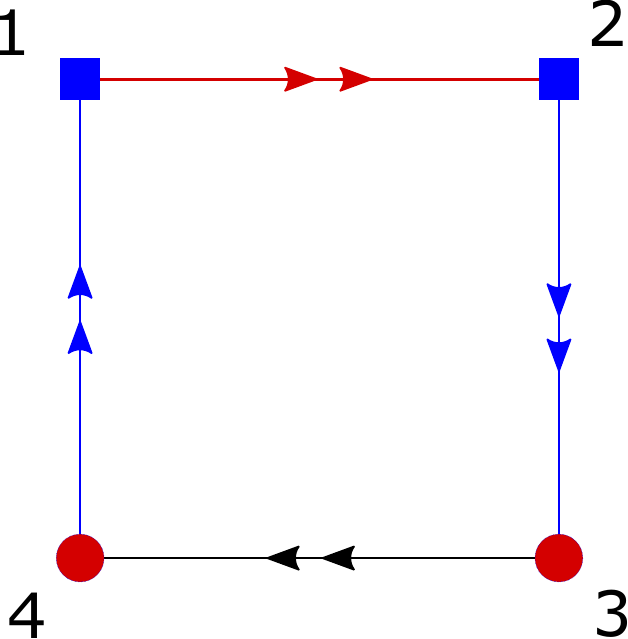}
		\caption{}
		\label{Fig:QuiverDiagramF0Backreaction3}
		\end{center}
    \end{subfigure}  \hspace{10mm}
    \begin{subfigure}[t]{0.25\textwidth }
        \begin{center} 
		\includegraphics[width=\textwidth]{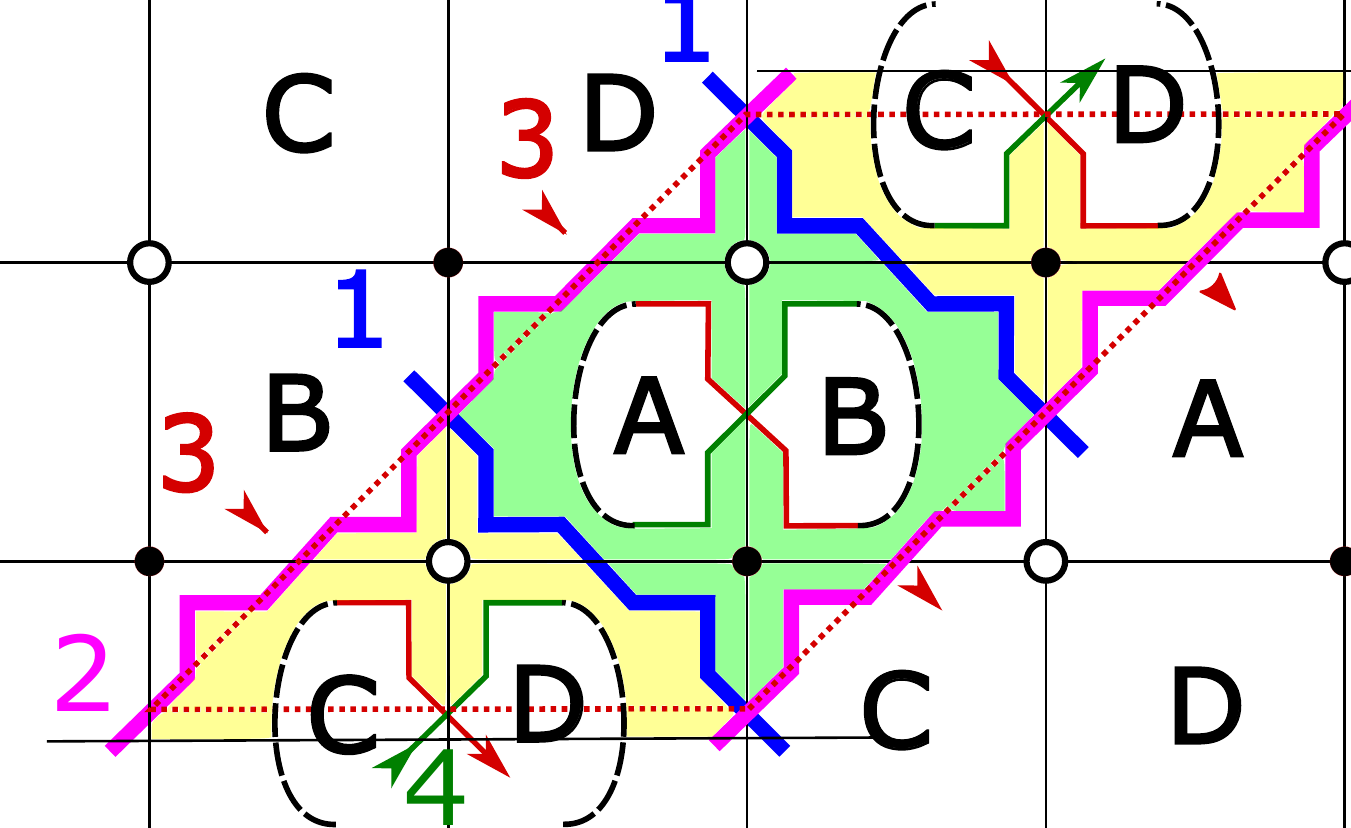}
		\caption{}
		\label{Fig:RiemannF0BakcreactionInst3}
		\end{center}
    \end{subfigure}  \hspace{10mm}
    \begin{subfigure}[t]{0.25\textwidth }
        \begin{center} 
			\includegraphics[width=\textwidth]{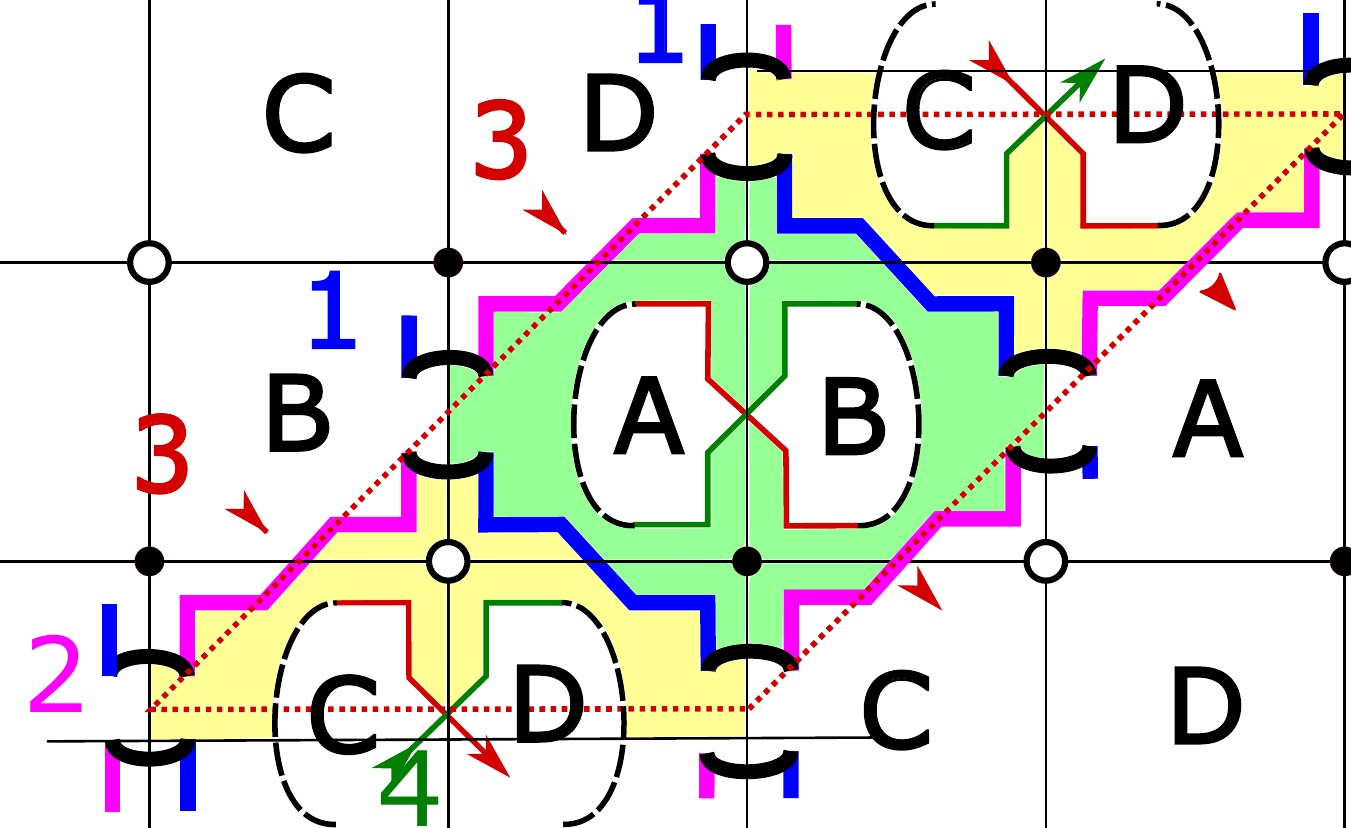}
		\caption{}
		\label{Fig:RiemannF0BakcreactionInst3Rec}
		\end{center}
    \end{subfigure}
    \caption{Backreaction due to a 2-instanton process in the F$_0$ theory. Figure a) shows the quiver, with black arrows denoting 4d chiral multiplets, blue arrows denoting  charged D2-D6 fermion zero modes, and blue arrows denoting chiral multiplets of D2-D2 zero modes. Figure b) shows the mirror Riemann surface $\Sigma$ after the backreaction and c) shows it with the instantons recombined.
      }\label{Fig:F0Backreaction3}
\end{figure}

Consider the F$_0$ theory of section \ref{Sec:F0Backreaction}, with instantons in the quiver nodes 1 and 2, one finds the quiver Fig. \ref{Fig:QuiverDiagramF0Backreaction3}, with charged fermion zero modes $\lambda^{i}_{23}, \tilde{\lambda}^{i}_{41}\, \, i=1,2$, a chiral multiplet worth of D2-D2 zero modes $\Phi^{i}_{12}$ and 4d chiral multiplets $X^{i}_{34}$. 

The couplings among these fields are 
\begin{equation}
 \sim  \epsilon_{ij} \epsilon_{kl} \lambda^{i}_{23} X^{k}_{34}  \tilde{\lambda}^{j}_{41} \Phi^{l}_{12} . 
\end{equation}

Note that in addition there are couplings e.g. between the  bosonic zero modes determining the D2-brane positions in 4d spacetime and the zero modes in the D2-D2 sector, so that the latter become `massive' upon separation of the instantons. As discussed in \cite{GarciaEtxebarria:2007zv,GarciaEtxebarria:2008pi}, the interesting multi-instanton physics arises when the two instantons are coincident  and such couplings can be safely ignored. The analysis is most simply carried out by first integrating over the bosonic D2-D2 zero modes, which effectively bind the two instantons. The combined instanton inherits the D2-D6 charged zero modes, which moreover inherit the couplings from the original ones, by simply regarding the D2-D2 chiral multiplet zero modes as a numerical factor, omitted in the following.

In total, we obtain a field theory operator with the structure
\begin{equation}
\mathcal{O}_{\rm charged} \sim (X^{1}_{34})(X^{2}_{34}) \label{Eq:OperatorF03}
\end{equation}
where brackets signal the structure of determinants in the non-abelian case, as usual.

Let us turn to the description of the multi-instanton effect in the backreaction picture. The recipe is a simple generalization of earlier ones. We take the Riemann surface and highlight the 1-cycles wrapped by the instantons, see Figure \ref{Fig:RiemannF0BakcreactionInst3}. One novelty is that in order to account for the integration over D2-D2 bosonic zero modes, the intersections between the instanton must be considered as slightly recombined, see Figure \ref{Fig:RiemannF0BakcreactionInst3}. We must cut the Riemann surface along the corresponding cycle, and recombine any D6-brane 1-cycle formerly intersecting it. As in section \ref{Sec:F0Backreaction}, the Riemann surface splits in two daughter surfaces, each of them with two punctures. This time, however, the punctures removed do not coincide with the complex deformation.

Step 3 provides the field theory operator produced from the worldsheet instantons after the backreaction. It is straightforward to see that the operator coincides with (\ref{Eq:OperatorF03}), as expected.

\subsubsection{Two-instanton effect in dP$_3$.} 

\begin{figure}
    \centering
         \begin{subfigure}[t]{0.27\textwidth }
        \begin{center} 
		\includegraphics[width=\textwidth]{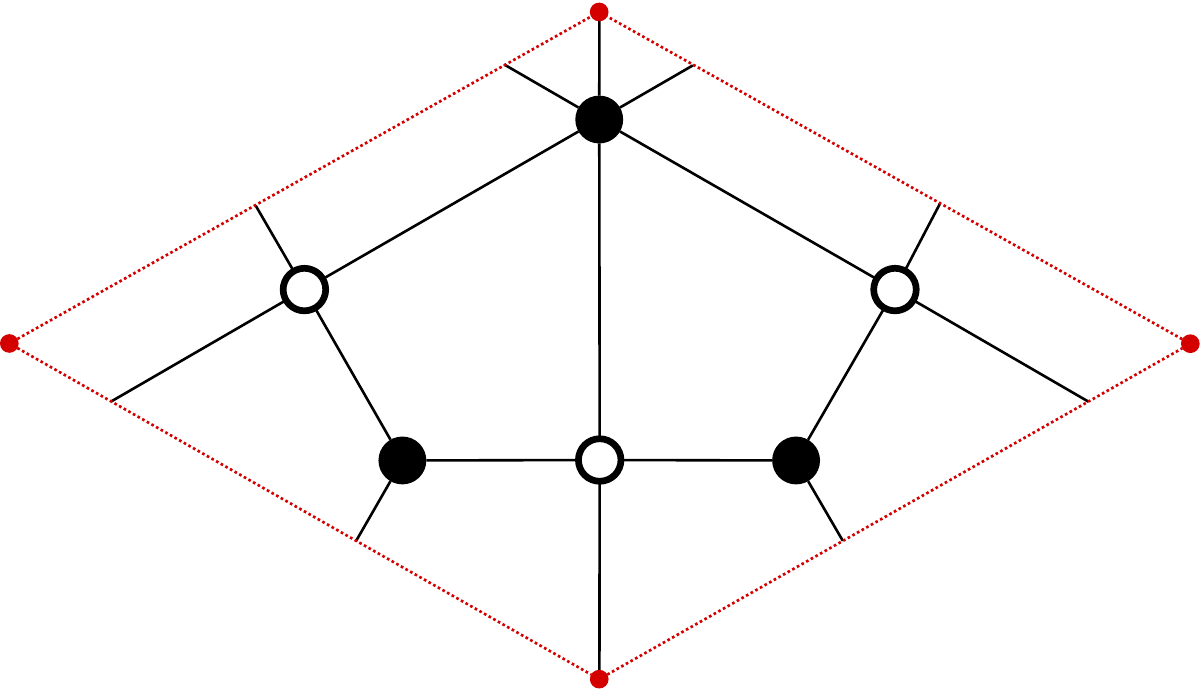}
		\caption{}
		\label{Fig:RiemanndP3Unit}
		\end{center}
    \end{subfigure}  \hspace{10mm}
    \begin{subfigure}[t]{0.20\textwidth }
        \begin{center} 
		\includegraphics[width=\textwidth]{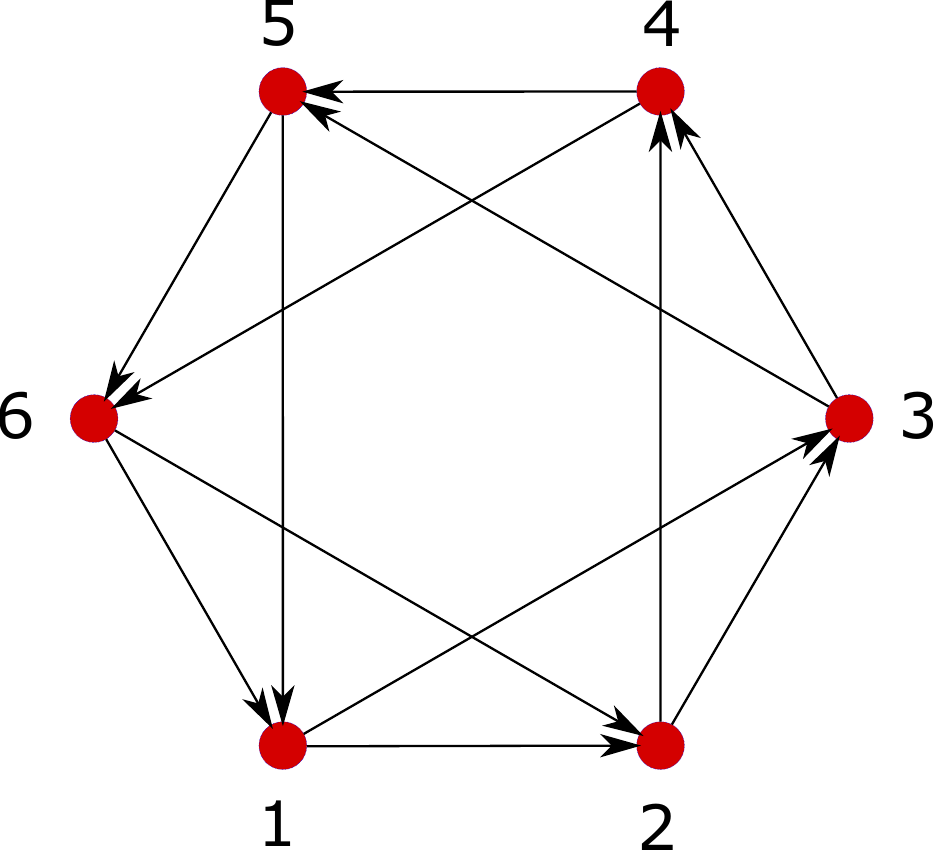}
		\caption{}
		\label{Fig:QuiverDiagramdP3}
		\end{center}
    \end{subfigure}  \hspace{10mm}
    \begin{subfigure}[t]{0.15\textwidth }
        \begin{center} 
		\includegraphics[width=\textwidth]{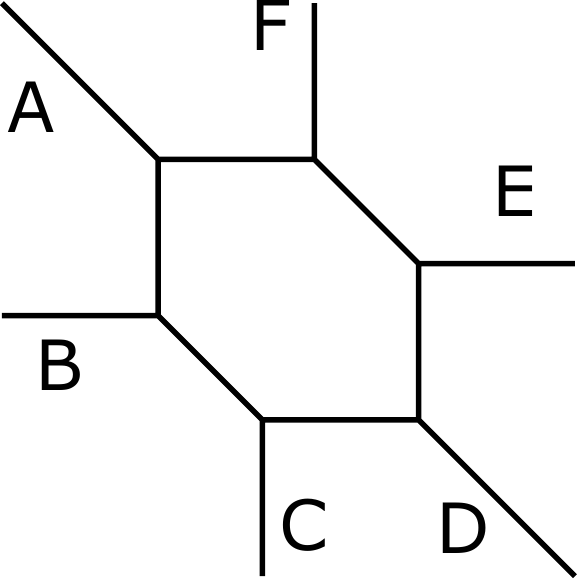}
		\caption{}
		\label{Fig:WebDiagramdP3}
		\end{center}
    \end{subfigure} \hspace{10mm}
    \begin{subfigure}[t]{0.13\textwidth }
        \begin{center} 
		\includegraphics[width=\textwidth]{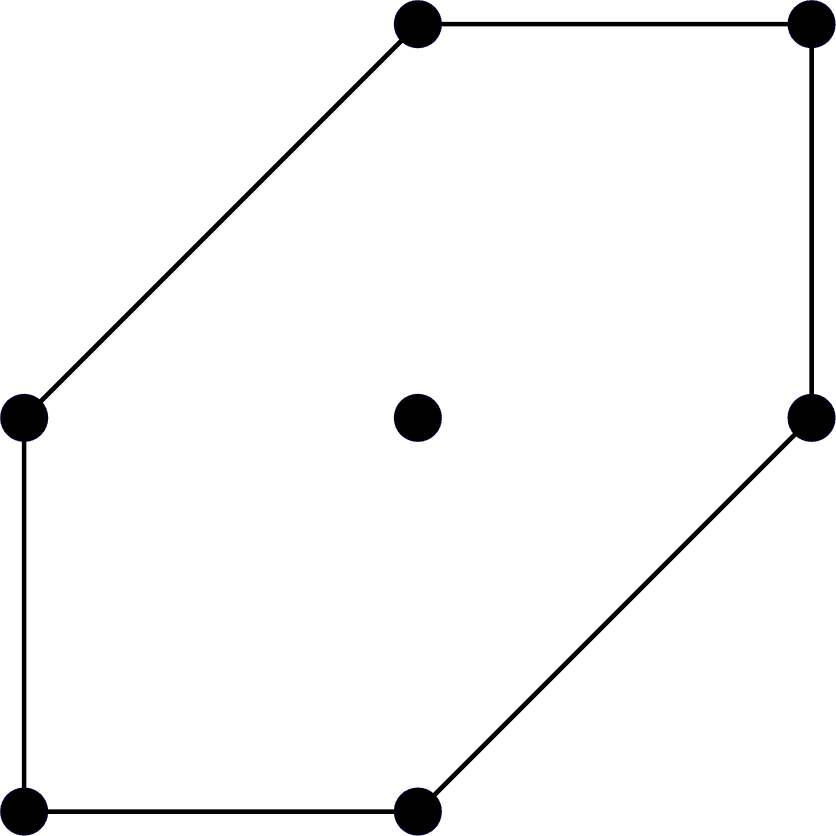}
		\caption{}
		\label{Fig:ToricDiagramdP3}
		\end{center}
    \end{subfigure}
    \caption{\small Diagrams for dP$_3$ theory, a) shows the dimer diagram, b) the quiver diagram, c) the web diagram and d) the toric diagram. The tiling of $\Sigma$ coincides with the dimer diagram.}
    \label{Fig:DiagramsdP3}
\end{figure}

Let us now consider a 2-instanton example in a configuration of branes at a complex cone over dP$_3$. The
 dimer diagram, quiver diagram, web diagram and toric diagram for the dP$_3$ theory are shown in Figs. \ref{Fig:RiemanndP3Unit}, \ref{Fig:QuiverDiagramdP3}, \ref{Fig:WebDiagramdP3}, \ref{Fig:ToricDiagramdP3}, respectively.

\begin{figure}
    \centering
         \begin{subfigure}[t]{0.20\textwidth }
        \begin{center} 
		\includegraphics[width=\textwidth]{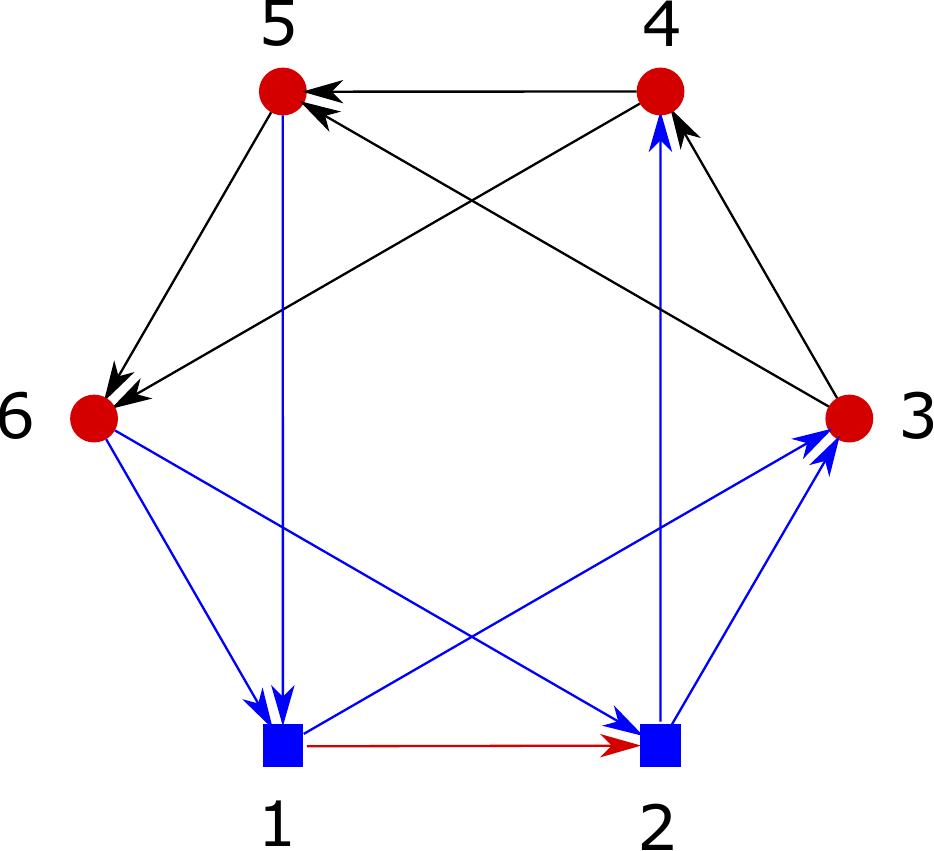}
		\caption{}
		\label{Fig:QuiverDiagramDeformation3dP3}
		\end{center}
    \end{subfigure}  \hspace{10mm}
    \begin{subfigure}[t]{0.3\textwidth }
        \begin{center} 
		\includegraphics[width=\textwidth]{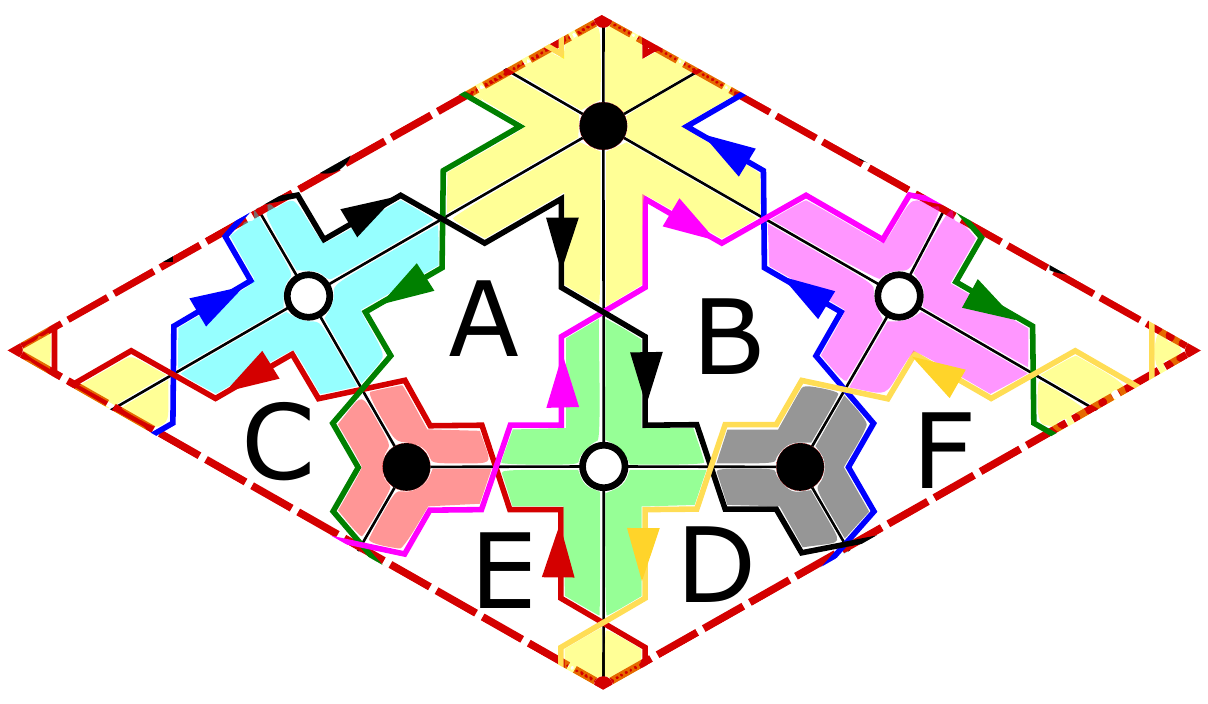}
		\caption{}
		\label{Fig:RiemannDiagramDeformation4dP3Unit}
		\end{center}
    \end{subfigure} \hspace{10mm}
    \begin{subfigure}[t]{0.3\textwidth }
        \begin{center} 
		\includegraphics[width=\textwidth]{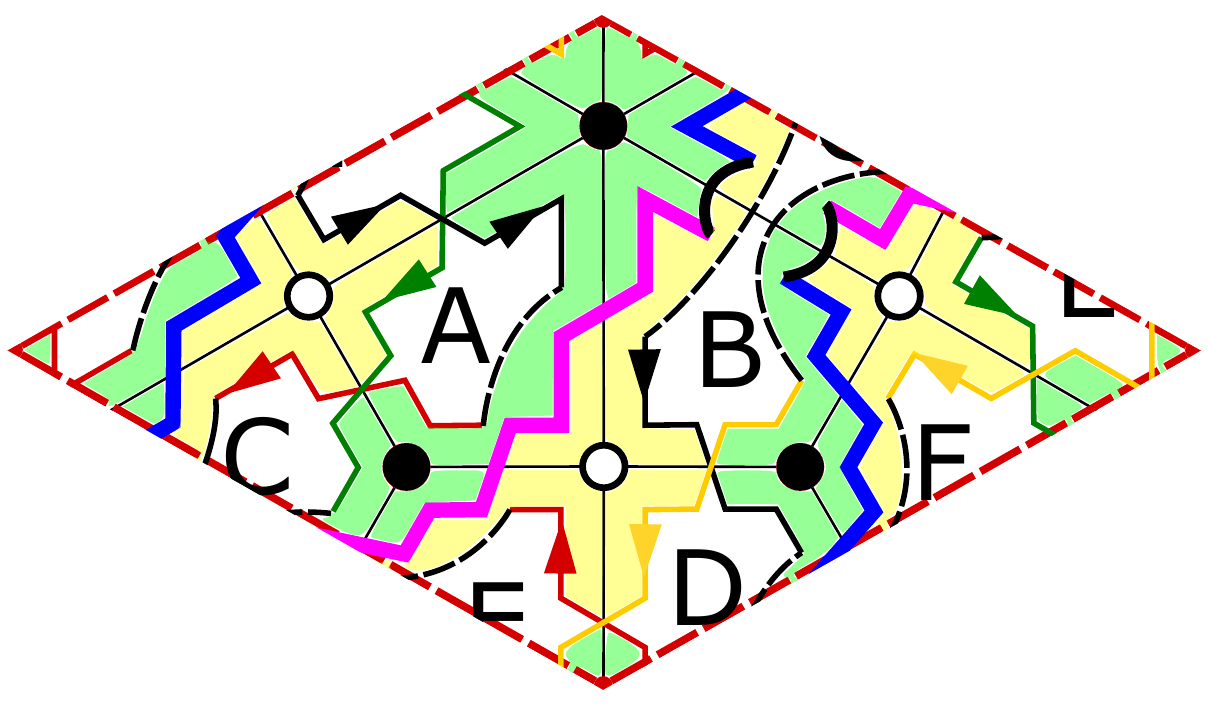}
		\caption{}
		\label{Fig:RiemannDiagramDeformation4BackdP32Unit}
		\end{center}
    \end{subfigure} \\ 
    \begin{subfigure}[t]{0.45\textwidth }
        \begin{center} 
		\includegraphics[width=\textwidth]{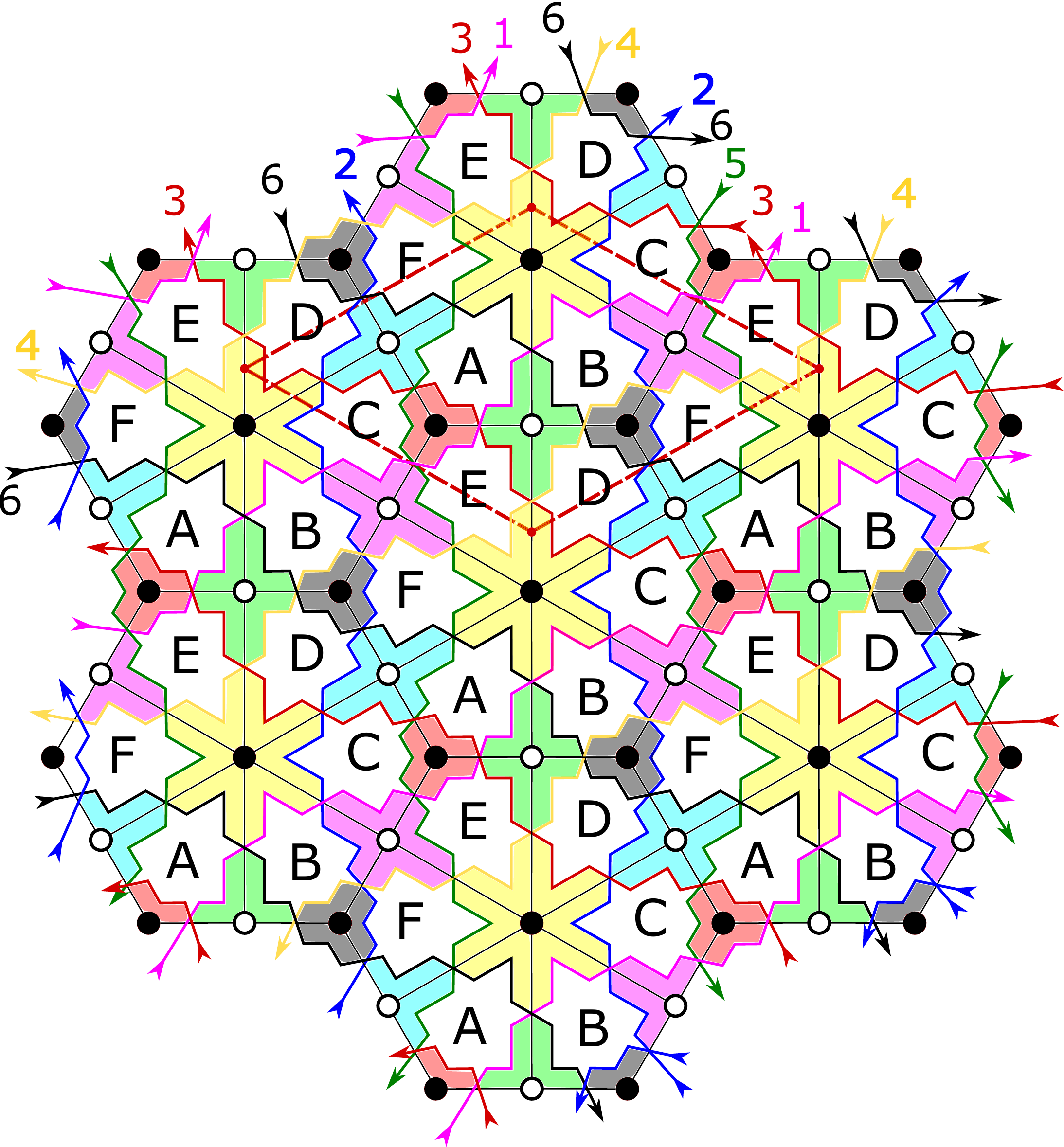}
		\caption{}
		\label{Fig:RiemannDiagramDeformation4dP3}
		\end{center}
    \end{subfigure}  \hspace{10mm}
    \begin{subfigure}[t]{0.45\textwidth }
        \begin{center} 
		\includegraphics[width=\textwidth]{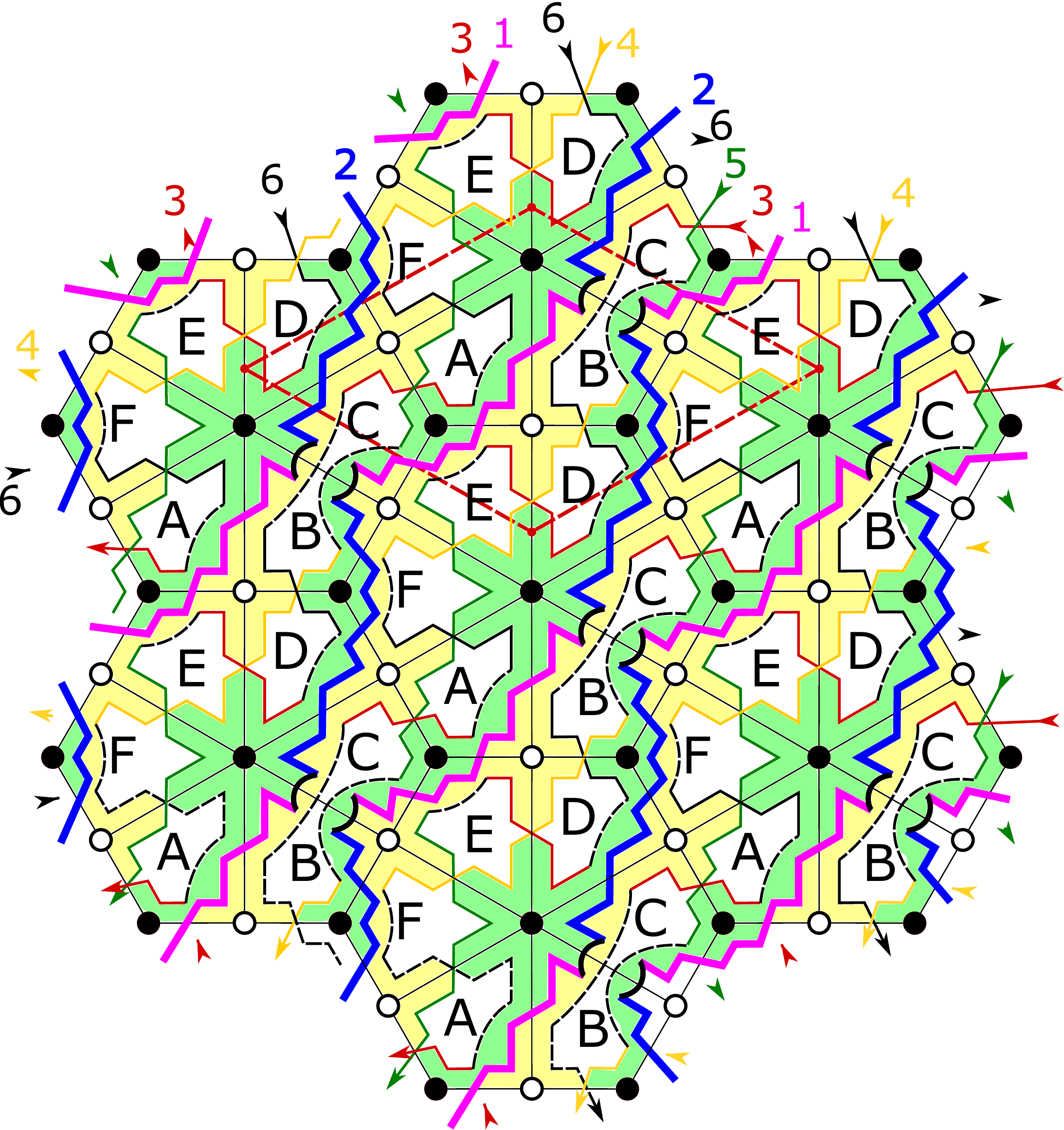}
		\caption{}
		\label{Fig:RiemannDiagramDeformation4BackdP32}
		\end{center}
    \end{subfigure} 
    \caption{\small dP3 backreaction with two instantons. a) is the quiver diagram with D2-brane instantons in nodes 1 and 2. Blue arrows denote charged D2-D6 fermion zero modes and red arrows denote a chiral multiplet worth of D2-D2  zero modes. Figure b) shows the unit cell of $\Sigma$ before the backreaction. In Figure c) we show the unit cell of $\Sigma$ after the backreaction. Note that the integration over the D2-D2 bosonic zero mode induces the recombination of instantons, shown as a thick black curved segment. Also the D6-brane 1-cycles cut by the instanton 1-cycles are duly recombined (sometimes taking advantage of the passages open due to the instanton recombination). In Figures d) and e) we provide the same pictures in the covering space description of $\Sigma$.}
    \label{Fig:dP3Backreaction2}
\end{figure}

As we will see in section \ref{sec:complex-examples}, this theory admits several complex deformations, which can be triggered by backreaction of certain multi-instantons; in this section we however focus on a multi-instanton effect whose backreaction does not correspond to a complex deformation. We consider introducing one D2-brane instanton in each of the nodes 1 and 2. The quiver diagram, see Fig. \ref{Fig:QuiverDiagramDeformation3dP3}, shows fermion zero modes $\lambda_{13},\lambda_{23}, \lambda_{24}, \tilde{\lambda}_{61}, \tilde{\lambda}_{62}, \tilde{\lambda}_{51}$, a chiral multiplet worth of zero modes $\Phi_{12}$ and matter fields $X_{34}, X_{45}, X_{56}, X_{35}, X_{46}$. The fermion zero modes have the following couplings 
\begin{eqnarray}
&&  
\Phi_{12} {\lambda}_{23} X_{34} X_{45} X_{56} \tilde\lambda_{61} \, +\,
{\lambda}_{24} X_{46} \tilde\lambda_{62} \,+ \,  {\lambda}_{13} X_{35} \tilde\lambda_{51} \,+ \nonumber \\
&&-\,  {\lambda}_{13} X_{34} X_{46} \tilde\lambda_{61} \, - \, {\lambda}_{23}X_{35} X_{56} \tilde\lambda_{62}\,-\,  \Phi_{12}  {\lambda}_{24} X_{45} \tilde\lambda_{51}  
\end{eqnarray} 	
Repeating the arguments of the previous examples about the treatment of D2-D2 zero modes, and saturating the charged D2-D6 fermion zero modes, we obtain the following 4d field theory operator structure
\begin{equation}
\mathcal{O}_{\rm charged} = (X_{34} X_{46}) (X_{35} X_{56}) (X_{45}) + (X_{46}) (X_{34} X_{45} X_{56}) (X_{35}) 
\end{equation}
where brackets indicate sets promoting to determinants in the non-abelian case, as usual.

This structure is easily recovered in the backreaction picture, see Figures \ref{Fig:RiemannDiagramDeformation4BackdP32Unit} and \ref{Fig:RiemannDiagramDeformation4BackdP32}.
Note that in order to enforce that the total homology class becomes trivial (rather than both individual classes, we have recombined the instantons at their intersections (in the only way consistent with their orientation), yielding a combined 1-cycle along which to cut $\Sigma$. This effectively reproduces the  integration over the D2-D2 bosonic zero mode.

\section{Relation to complex deformations}
\label{sec:complex-def}

\subsection{Generalities}
\label{sec:complex-gen}

Some toric geometries admit complex deformations, the canonical example being the the complex deformation of the conifold. The complex deformations a given toric geometry admits can be seen from its web diagram. They are defined by the removal of a collection of legs whose $(p,q)$ labels sums to zero, also dubbed subweb in equilibrium \cite{Franco:2005fd}. Concrete examples are the complex deformations of the conifold, the double conifold. There are richer situations, in which a given singularity admits several (incompatible) complex deformations. For instance, the complex cone over dP$_3$ admits two complex deformation, as shown in Fig. \ref{Fig:ComplexDeformationsdP3}. 

Geometric transitions corresponding to complex deformations have appeared in the context of warped throats triggered by deformation fractional branes, in the sense of \cite{Franco:2005fd, Franco:2005zu}. These correspond to anomaly-free rank assignments in the quiver diagram, which have the property that the total 1-cycle they define in the mirror Riemann surface surrounds the punctures corresponding to the sub-web in equilibrium. The geometric transition removes the corresponding cycle and seals the Riemann surface, in a combinatorial dimer prescription described in \cite{GarciaEtxebarria:2006aq}.

It is natural to wonder what happens if we consider D2-brane instantons on the same cycle that corresponds to a given deformation fractional brane, i.e. a complex deformation. In fact this already occurred in the examples in sections \ref{sec:conifold} and \ref{sec:double-con}, with the result that the backreacted geometry is the (mirror of the) complex deformation of the initial geometry. In this section we provide further examples showing that this is a general behaviour.

\subsection{Examples}
\label{sec:complex-examples}

As just explained, some of the examples already considered provide realizations that  instantons on classes corresponding to deformation fractional branes produce backreacted geometries given by the complex deformation of the original space. We now provide further examples, based on chiral quivers.

\subsubsection{Complex Deformation of cone over F$_0$}

Let us start with F$_0$, whose dimer diagram unit cell, quiver diagram and web diagram are shown in Fig. \ref{Fig:DiagramsF0}. This geometry admits a complex deformation given by the splitting of the web diagram into two subwebs in equilibrium, one with external legs A, B and the other with C and D. At the level of deformation fractional branes, this complex deformation is triggered by increasing the rank of the quiver nodes 1 and 3 simultaneously (equivalently, recalling that the sum of all nodes is topologically trivial in compact homology, of nodes 2 and 4).

\begin{figure}
    \centering
         \begin{subfigure}[t]{0.16\textwidth }
        \begin{center} 
		\includegraphics[width=\textwidth]{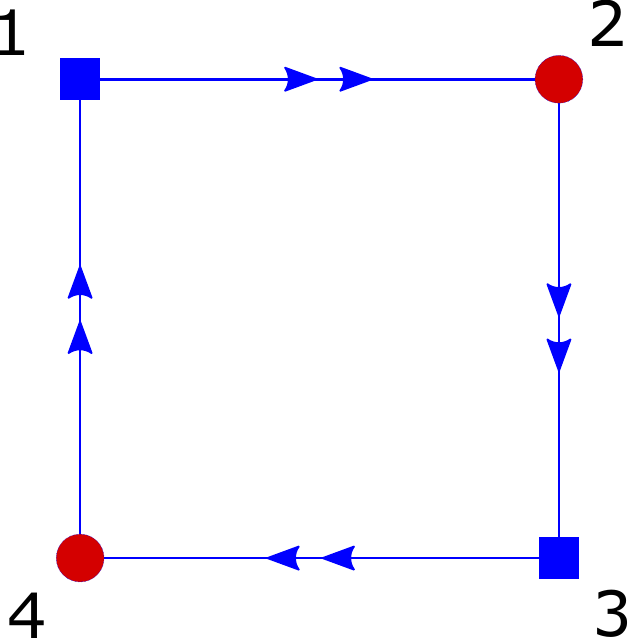}
		\caption{}
		\label{Fig:QuiverDiagramF0Backreaction2}
		\end{center}
    \end{subfigure}  \hspace{10mm}
    \begin{subfigure}[t]{0.3\textwidth }
        \begin{center} 
		\includegraphics[width=\textwidth]{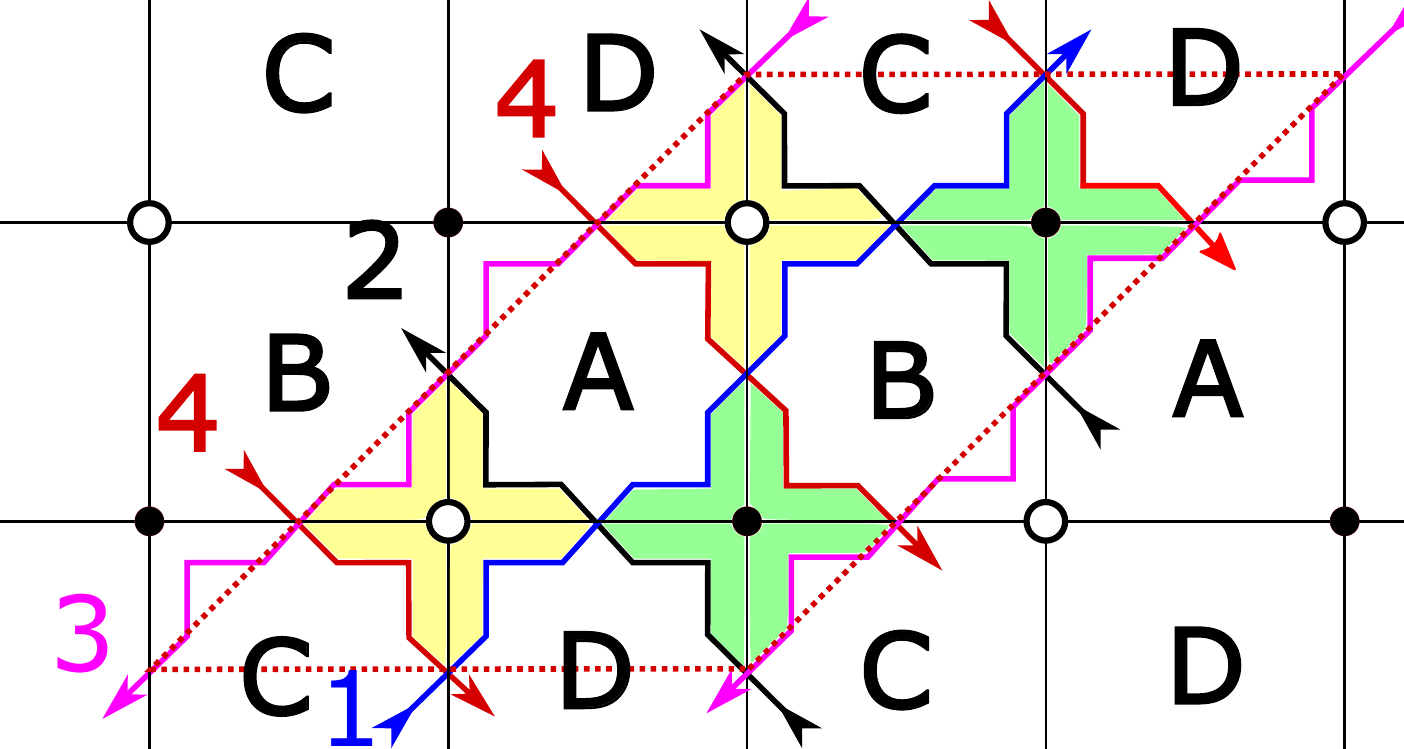}
		\caption{}
		\label{Fig:RiemannF0Inst2}
		\end{center}
    \end{subfigure}  \hspace{10mm}
    \begin{subfigure}[t]{0.3\textwidth }
        \begin{center} 
		\includegraphics[width=\textwidth]{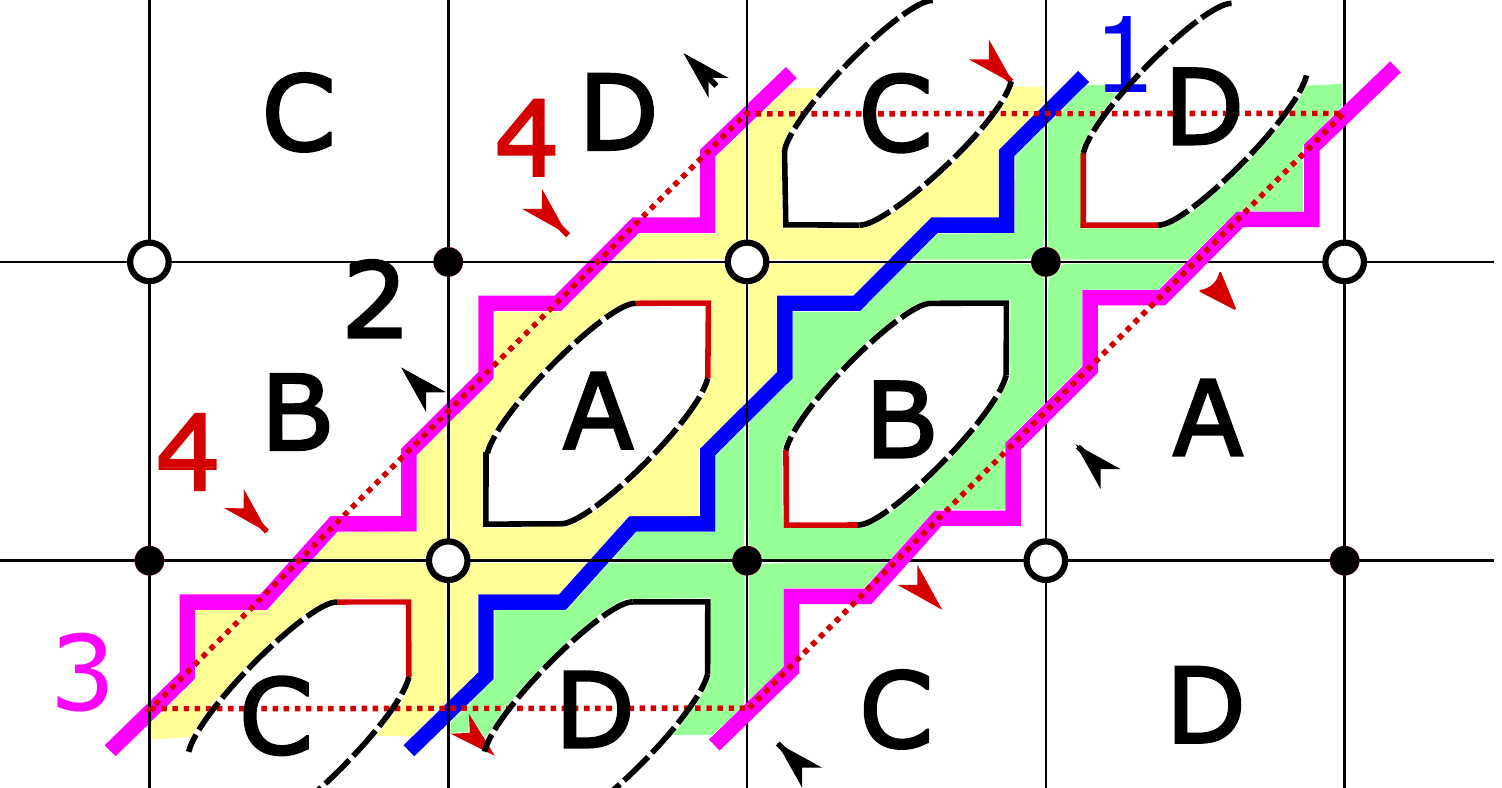}
		\caption{}
		\label{Fig:RiemannF0BakcreactionInst2}
		\end{center}
    \end{subfigure}
    \caption{\small F0 complex deformation. a) is the quiver diagram with D2-brane instantons in nodes 1 and 3. Blue arrows denote fermion zero modes. b) shows $\Sigma$ before the backreaction. In c) $\Sigma$ after the backreaction is shown. The D-instanton have dissapeared from $\Sigma$, splitting it into two daughter surfaces.}\label{Fig:F0Backreaction2}
\end{figure}

Let us consider the backreaction effect of D2-brane instantons on the corresponding class. Consider introducing one D2-brane instanton in each of the nodes 1 and 3 in the quiver, see Figure \ref{Fig:QuiverDiagramF0Backreaction2}. These correspond to the blue and pink 1-cycles in the mirror Riemann surface in Figure \ref{Fig:RiemannF0Inst2}. 

From the open string perspective, there are charged fermion zero modes $\lambda^{i}_{12}$, $\tilde{\lambda}^{i}_{41}$, $\lambda^{i}_{34}$, $\tilde{\lambda}^{i}_{23}$, for $i=1,2$, and there are no matter fields. Note also that in the present case there are no D2-D2 zero modes of the kind introduced in section \ref{sec:multiple}.
Saturation of fermion zero modes produces no charged field theory operator, like in the conifold case.

\begin{figure}
    \centering
         \begin{subfigure}[t]{0.16\textwidth }
        \begin{center} 
		\includegraphics[width=\textwidth]{WebDiagramF0.pdf}
		\caption{}
		\label{Fig:WebDiagramF02}
		\end{center}
    \end{subfigure}  \hspace{10mm}
    \begin{subfigure}[t]{0.25\textwidth }
        \begin{center} 
		\includegraphics[width=\textwidth]{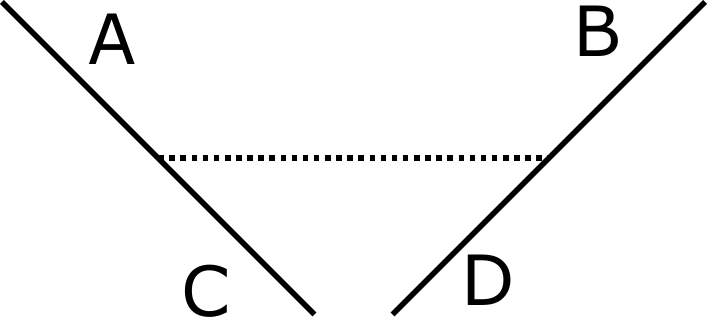}
		\caption{}
		\label{Fig:WebDiagramF0Backreaction}
		\end{center}
    \end{subfigure}  \hspace{10mm}
    \caption{\small Web diagrams of F0 a) before and b) after the complex deformation.}\label{Fig:F0BackreactionWebDiagrams}
\end{figure}

Consider now the effect of backreacting these instantons. Following step 1 in our procedure, we cut the Riemann surface along the 1-cycles defined by the D2-brane instantons, and recombine cycles following step 2. The resulting mirror Riemann surface is Figure \ref{Fig:RiemannF0BakcreactionInst2}. Unlike in section \ref{Sec:F0Backreaction}, $\Sigma$ splits in two surfaces, one with punctures A, C and the other with B, D, precisely in accordance with the complex deformation shown in Fig. \ref{Fig:WebDiagramF0Backreaction}. We thus derive the result that the backreacted geometry is the (mirror of the) complex deformation of the original singularity.

In this case, step 3 gives no operators induced in the field theory, as the worldsheet instantons after the backreaction do not touch any brane intersection with matter. This agrees with the prediction from saturating fermion zero modes. Although we obtain agreement, it is rather trivial as no charged operator is generated; we turn to a more interesting example in the following section.

\subsection{Complex deformation of complex cone over dP$_3$ into a conifold}

\begin{figure}
    \centering
    \begin{subfigure}[t]{0.20\textwidth }
        \begin{center} 
		\includegraphics[width=\textwidth]{WebDiagramdP3.pdf}
		\caption{}
		\label{Fig:WebDiagramdP32}
		\end{center}
    \end{subfigure} \hspace{10mm}
    \begin{subfigure}[t]{0.30\textwidth}
		\includegraphics[width=\textwidth]{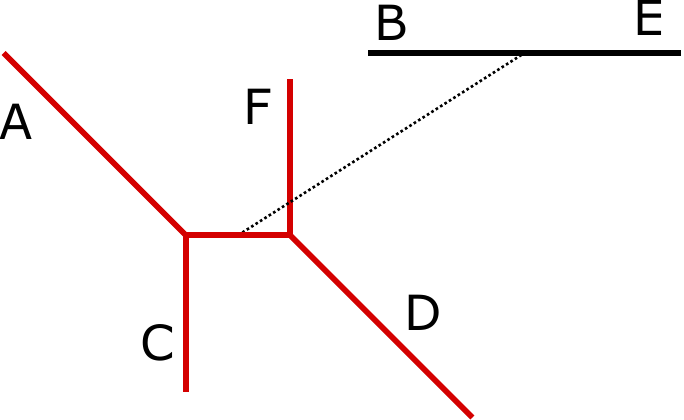}
        \caption{}
        \label{Fig:WebDiagramComplex2dP3}
    \end{subfigure} \hspace{10mm}
    \begin{subfigure}[t]{0.30\textwidth }
        \begin{center} 
		\includegraphics[width=\textwidth]{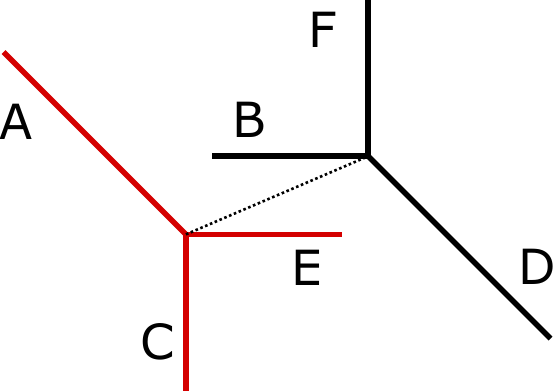}
		\caption{}
		\label{Fig:WebDiagramComplex1dP3}
		\end{center}
    \end{subfigure} 
    \caption{\small Web diagrams of the two possible complex deformations of dP3, corresponding to the removal of a subweb in equilibrium.}\label{Fig:ComplexDeformationsdP3}
\end{figure}

There are several different complex deformations for the complex cone over dP$_3$.  In this section we focus on that defined by the web diagrams splitting in Fig. \ref{Fig:WebDiagramComplex2dP3}, by removal of the external legs B and E.

At the level of deformation fractional branes, this corresponds to locating extra branes on the nodes 1 and 4 in the quiver/dimer. This is easily shown by checking that the corresponding 1-cycle in $\Sigma$ precisely surrounds the relevant punctures.

\begin{figure}
    \centering
    \begin{subfigure}[t]{0.45\textwidth }
        \begin{center} 
		\def\svgwidth{1\linewidth}
		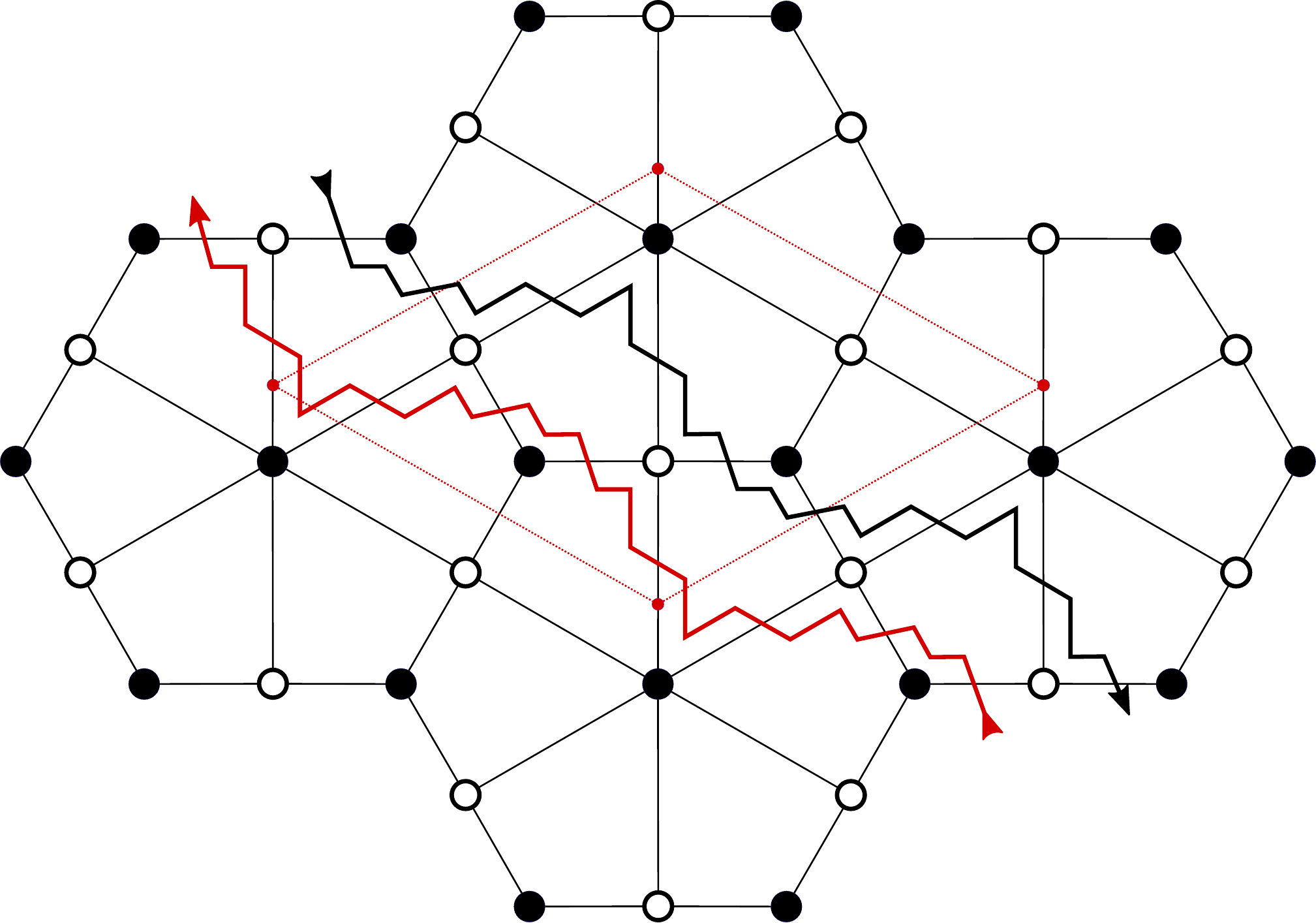
		\caption{}
		\label{Fig:DimmerdComplexDeformation2P3ZigZag}
		\end{center}
    \end{subfigure} \hspace{10mm}
    \begin{subfigure}[t]{0.3\textwidth}
		\includegraphics[width=\textwidth]{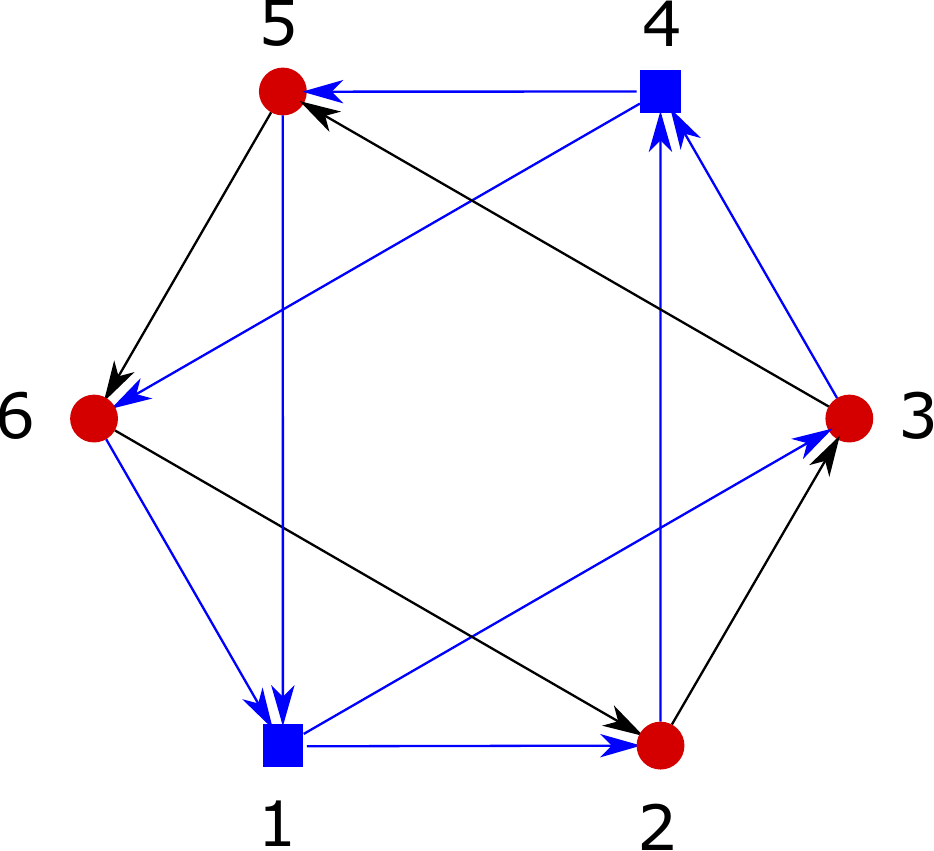}
        \caption{}
        \label{Fig:QuiverDiagramDeformation2dP3}
    \end{subfigure} 
    \caption{\small dP3 Complex Deformation with two instantons. a) shows the zig-zag paths to be removed from the dimer diagram and the corresponding faces are shaded in yellow. In b) the corresponding quiver diagram is shown.}
    \label{Fig:ComplexDeformationdP31}
\end{figure}

\begin{figure}
    \centering
    \begin{subfigure}[t]{0.45\textwidth}
		\includegraphics[width=\textwidth]{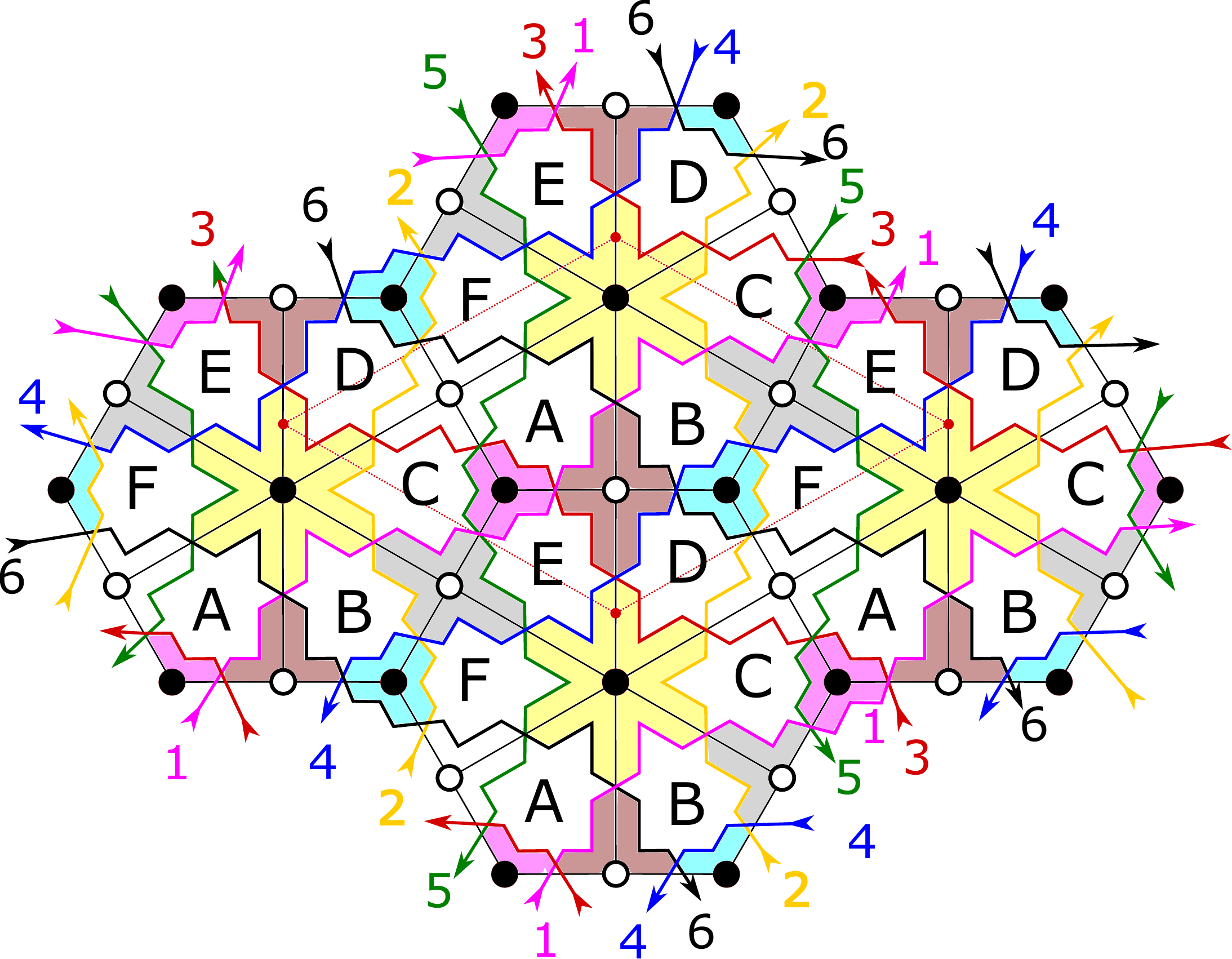}
        \caption{}
        \label{Fig:DimmerdP3ZigZagR3}
    \end{subfigure}  \hspace{10mm}
    \begin{subfigure}[t]{0.45\textwidth}
		\includegraphics[width=\textwidth]{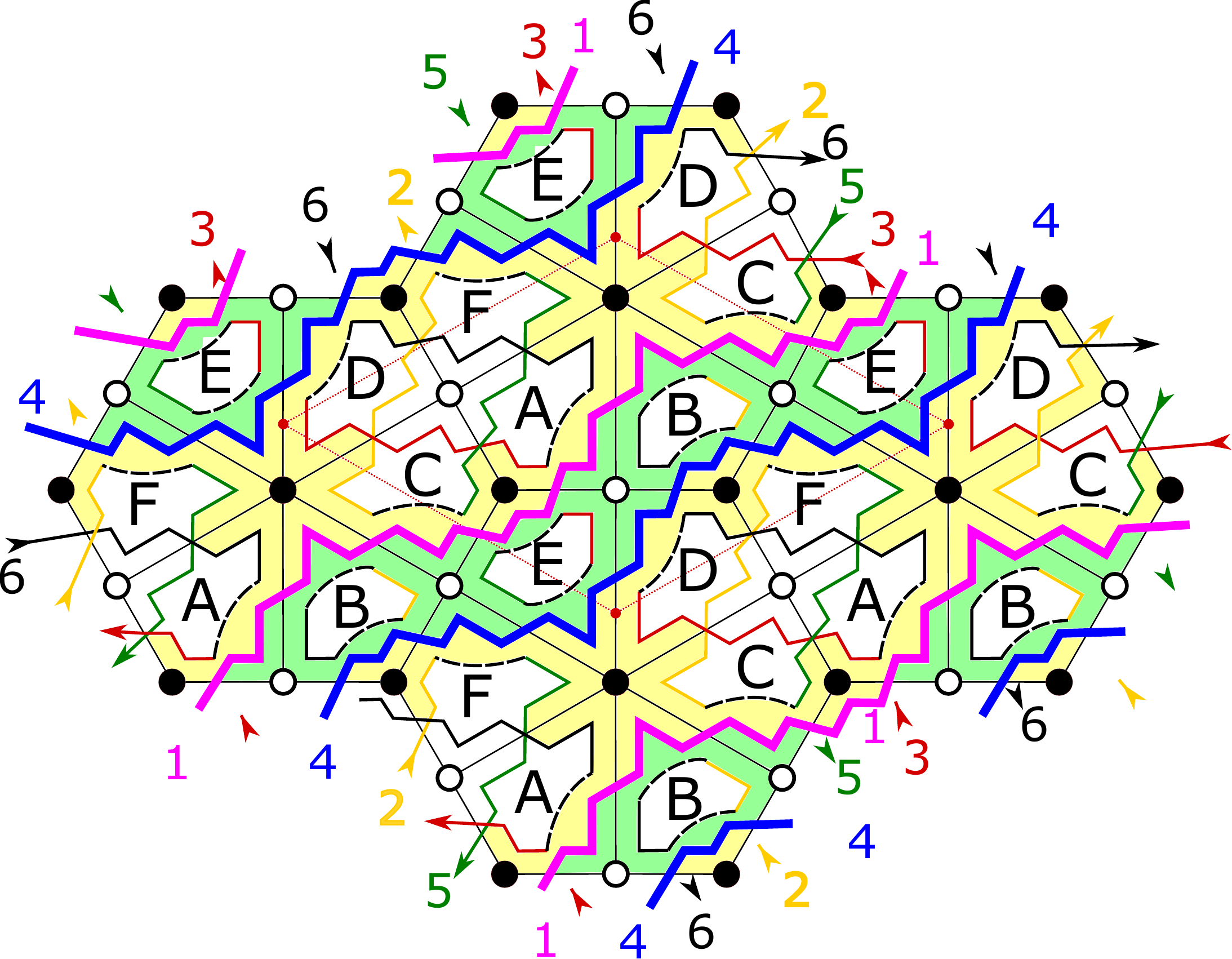}
        \caption{}
        \label{Fig:DimmerdP3ZigZagRBackreaction3}
    \end{subfigure} 
    \caption{\small dP3 Complex Deformation with two instantons. a) shows $\Sigma$ before the bacreaction, with D-brane instantons in cycles 1 and 4 and WS instantons shaded in light colors. In b) $\Sigma$ after the backreaction is shown. D-brane instantons dissapear and branes recombine. Two WS instantons remain, one in each $\Sigma^{\prime}$.}\label{Fig:ComplexDeformation1dP3Int}
\end{figure}

We would now like to consider the backreaction effect of D2-brane instantons on those nodes 1 and 4.
  The quiver diagram of this setup is shown in Fig. \ref{Fig:QuiverDiagramDeformation2dP3}. We see that there are fermion zero modes $\tilde\lambda_{61}$, $\tilde\lambda_{51}$, $\tilde\lambda_{34}$, $\tilde\lambda_{24}$, ${\lambda}_{12}$, ${\lambda}_{13}$, ${\lambda}_{45}$, ${\lambda}_{46}$, and 4d chiral multiplets $X_{56}$, $X_{62}$, $X_{23}$ and $X_{45}$. 
 The couplings are
  \beqa
  && \lambda_{12}X_{23}\tilde\lambda_{34}\lambda_{45}X_{56}\tilde\lambda_{61}\, +\, \lambda_{13}X_{35}\tilde\lambda_{51}\, +\, \lambda_{46}X_{62}\tilde\lambda_{24}\, +\nonumber \\
  && - \lambda_{12}\tilde\lambda_{24}\lambda_{45}\tilde\lambda_{51}\, -\, \lambda_{13}\tilde\lambda_{34}\lambda_{46}\tilde\lambda_{61}\, -\, X_{23}X_{35}X_{56}X_{62}
  \eeqa
  where the last term is actually a 4d coupling among chiral multiplets, which we include for completeness.
  
Saturation of fermion zero modes produces the structure
\beqa
{\cal O}_{\rm charged} \sim (X_{23}) (X_{56})(X_{35})(X_{62})\, +\, {\rm const}
\label{supo-dp3-to-con}
 \eeqa 
  
 Consider now the description in terms of the instanton backreaction see Fig. \ref{Fig:DimmerdP3ZigZagRBackreaction3}. Following steps 1 and 2 in our prescription, we cut along the pink and blue 1-cycles, and duly recombine the D6-brane 1-cycles. The mirror $\Sigma$ will split in two Riemann surfaces, one containing the punctures B and E, and the other containing the punctures A, C, D and F. Thus it reproduces the (mirror of the) complex deformation of the original singularity. Following step 3, we consider the worldsheet instantons in the resulting geometry, shown as shaded green and yellow in the Figure. The green one does not produce any charged field theory insertion, while the pink one inserts the fields $X_{56}$, $X_{62}$, $X_{23}$ and $X_{45}$; the combination of both reproduces precisely the structure in (\ref{supo-dp3-to-con}).

\subsubsection{Complex deformation of dP$_3$ into flat space}

\begin{figure}
    \centering
    \begin{subfigure}[t]{0.45\textwidth }
        \begin{center} 
		\def\svgwidth{1\linewidth}
		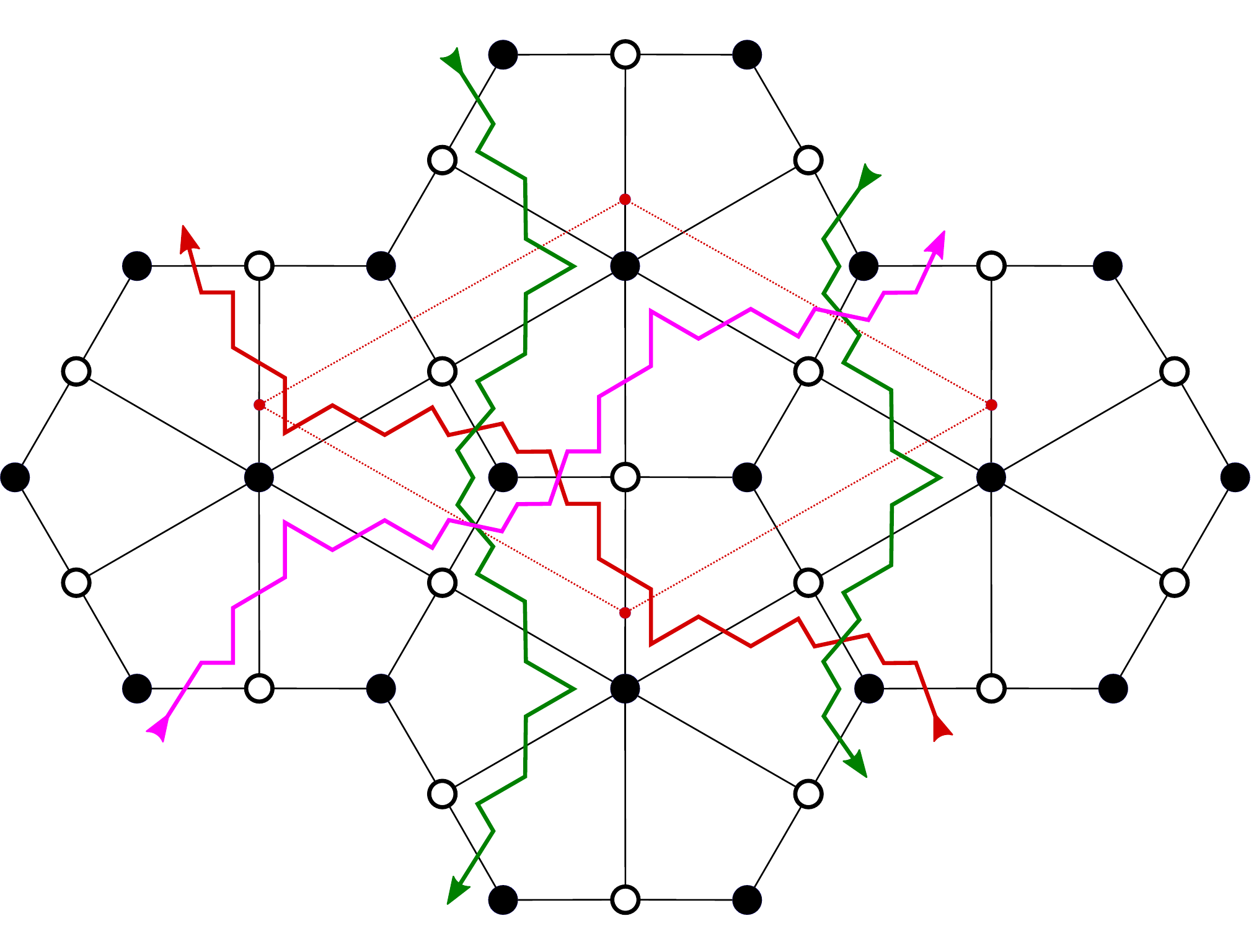
		\caption{}
		\label{Fig:DimmerdComplexDeformationP3ZigZag}
		\end{center}
    \end{subfigure} \hspace{10mm}
    \begin{subfigure}[t]{0.3\textwidth}
		\includegraphics[width=\textwidth]{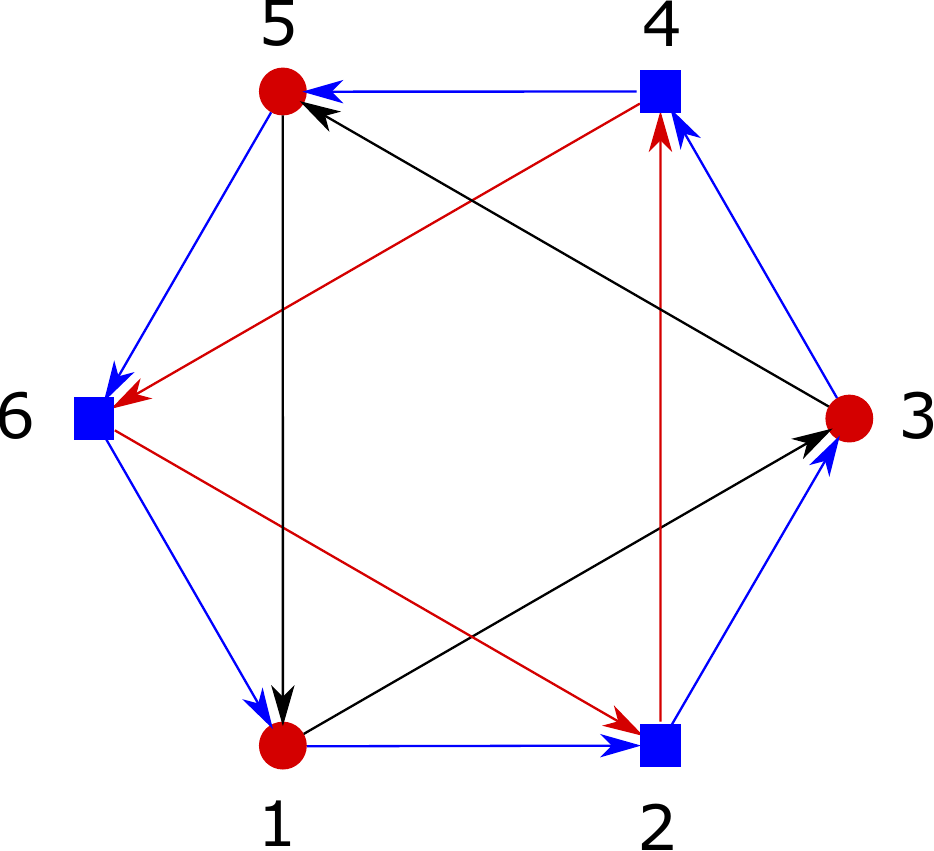}
        \caption{}
        \label{Fig:QuiverDiagramDeformationdP3}
    \end{subfigure} 
    \caption{\small Complex Deformation of the cone over dP$_3$ triggered by three instantons. Figure a) shows a set of zig-zag paths to be removed from the dimer diagram, and a set of corresponding faces  shaded in yellow. In b) the corresponding quiver diagram si shown. }\label{Fig:ComplexDeformation2dP3Int}
\end{figure}

In this section we will study a second complex deformation, defined by the removal of legs B, D and F, which form a subweb in equilibrium, as shown in Fig. \ref{Fig:WebDiagramComplex1dP3}. This corresponds to the removal of zig-zag paths with weight (-1,0) and (0,1) and (1,-1) in the dimer diagram, as shown in Fig. \ref{Fig:DimmerdComplexDeformationP3ZigZag}. In turn, this implies that faces 2, 4 and 6 must be removed from the dimer (equivalently, for 1,3 and 5). In our description this translates to putting an instanton in the class defined by these three cycles, or faces of the dimer. 

\begin{figure}
    \centering
    \begin{subfigure}[t]{0.45\textwidth}
		\includegraphics[width=\textwidth]{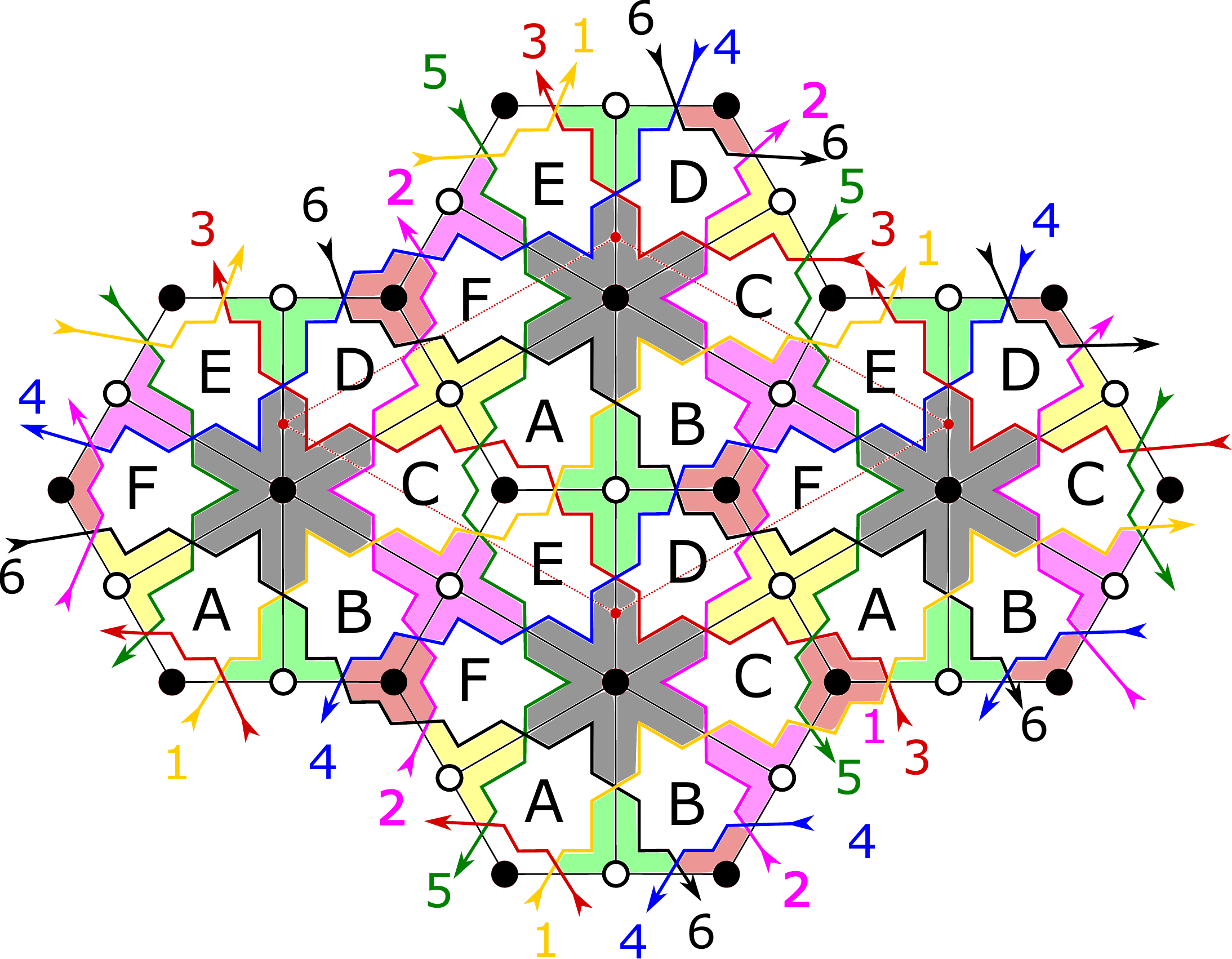}
        \caption{}
        \label{Fig:QuiverDiagramDeformation32dP3}
    \end{subfigure}  \hspace{10mm}
    \begin{subfigure}[t]{0.45\textwidth}
		\includegraphics[width=\textwidth]{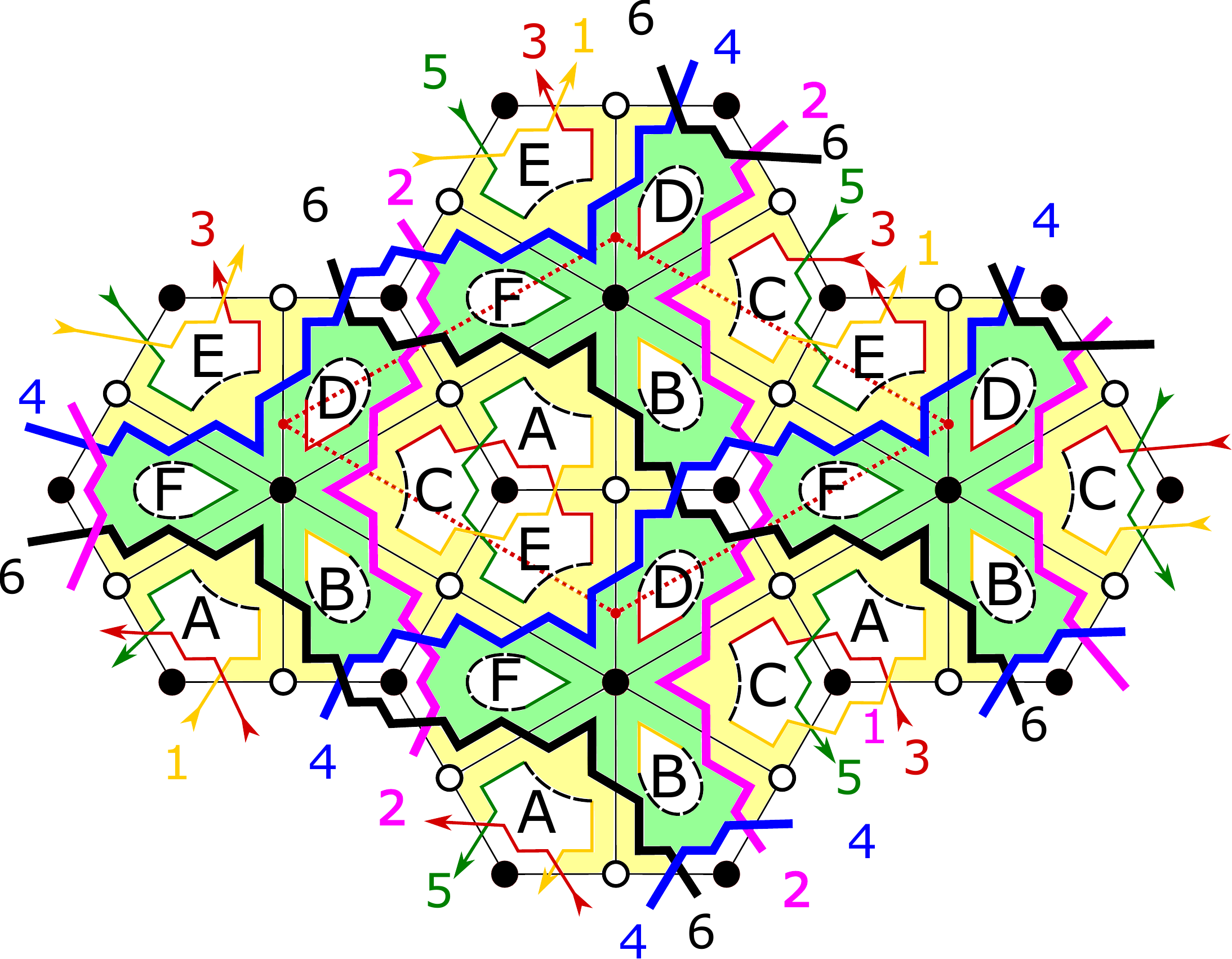}
        \caption{}
        \label{Fig:RiemannDiagramDeformationBack32dP3}
    \end{subfigure} 
    \caption{\small Complex Deformation of the cone over dP$_3$ triggered by three instantons. Figure a) shows the mirror Riemann surface $\Sigma$ before the backreaction, with D-brane instantons in cycles 2,4 and 6 and WS instantons shaded in light colors. In Figure b) we show $\Sigma$ after the backreaction. For clarity the instantons have not been recombined. The surface splits into two, in agreement with the resulting geometry being  (the mirror of) a complex deformation.}
    \label{Fig:ComplexDeformation2dP3}
\end{figure}

We thus consider the configuration with one D2-brane instanton on each of the cycles labelled 2, 4 and 6. 
The quiver diagram of this setup is shown in Fig. \ref{Fig:QuiverDiagramDeformationdP3}. We see that there are charged fermion zero modes  $\tilde\lambda_{12}$, $\lambda_{23}$, $\tilde\lambda_{34}$, $\lambda_{45}$, $\tilde\lambda_{56}$, $\lambda_{61}$, D2-D2 chiral multiplet zero modes $\Phi_{24}, \Phi_{46} \text{ and } \Phi_{62}$, and matter fields $X_{51}$, $X_{13}$, $X_{35}$. As usual, the worldsheet instantons yield the following couplings among 4d fields and the instantons zero modes 
\beqa
\sim &&  \, \tilde\lambda_{12}{\lambda}_{23} \tilde \lambda_{34}  {\lambda}_{45} \tilde \lambda_{56}  {\lambda}_{61} \, -\, \Phi_{24}\Phi_{46}\Phi_{62}\, -\, X_{13}X_{35}X_{51}\, +\nonumber \\
&& \Phi_{24}  {\lambda}_{45}X_{51} \tilde\lambda_{12}  \,+\, \Phi_{46}{\lambda}_{61} X_{13}\tilde \lambda_{34} \,  +   \, \Phi_{62} {\lambda}_{23} X_{35} \tilde \lambda_{56}  
\eeqa
The fermion zero modes can be saturated in two different ways, producing a contribution
\begin{equation}
\mathcal{O}_{\rm charged} \sim (X_{51})( X_{13} )(X_{35} )\, +\, {\rm constant}
\label{Eq:OperatordP3Deformation2}
\end{equation}
where brackets indicate determinants in the non-abelian case, as usual.

Let us consider the backreaction description, which is illustrated in Figure \ref{Fig:RiemannDiagramDeformationBack32dP3} upon application of  steps 1 and 2 of the recipe. The mirror Riemann surface has divided in two: $\Sigma^{\prime}$, with punctures 1, 3 and 5 and $\Sigma^{\prime \prime}$, with punctures 2, 4 and 6 in agreement with the complex deformation of the dP$_3$ singularity to flat space. Step 3 of the recipe allows us to find the field theory operators induced by the WS instantons in each of the new surfaces, and one finds that they indeed agree with the above operator (\ref{Eq:OperatordP3Deformation2}). Concretely, the worldsheet instanton in $\Sigma^{\prime}$ provides the non-trivial operator involving charged fields, while the worldsheet instanton in $\Sigma^{\prime \prime}$ does not produce any charge field insertion.

\bigskip

\section{Conclusions}
\label{sec:conclu}

In this paper we have studied the description of non-perturbative effects from D-brane instantons  in 4d string compactifications with gauge D-branes, from the perspective of the geometry resulting from the instanton backreaction. This extends earlier results by including the appearance of 4d charged field operators, which arise from standard perturbative couplings, like worldsheet instantons, in the non-perturbatively backreacted geometry. Along the way we have provided a novel interpretation for the appearance of worldsheet instanton amplitudes in the presence of D2-branes instantons in compactifications with D6-branes. 

We have provided a large class of examples of D-brane instanton backreaction based on (type IIA mirrors of) systems of D3-branes at toric singularities. The backreacted configuration is obtained by application of a simple set of rules on the graphs associated to the dimer diagram. This define an interesting new set of operations relating toric CY singularities with a set of generically non-CY geometries. For D2-branes on nodes associated with deformation fractional branes, the resulting geometry is still CY and corresponds to (the mirror of) complex deformations of the original singularity.

There are several interesting directions to pursue:

$\bullet$ The backreaction procedure is highly reminiscent of gauge/gravity dualities. It would be interesting to find particular setups where this analogy can be made more concrete (beyond the cases associated to complex deformations).

$\bullet$ Since our recipe is combinatoric at the level of the tiling diagrams, it may not completely capture the dynamics in configurations with arbitrary (anomaly-free) ranks in the D3-brane gauge factors. 

$\bullet$ It would be interesting to find concrete examples in global compactifications. A related issue is the description of the backreaction of instantons for instantons in the presence of non-trivial orientifold projections.

We leave these and other interesting questions for further work.

\section*{Acknowledgements}

We would like to thank L. Ib\'a\~nez,  A. Retolaza for useful discussions. E. G. and A. U. are partially supported by the grants FPA2015-65480-P from the MINECO, the ERC Advanced Grant SPLE under contract ERC-2012-ADG-20120216-320421 and the grant SEV-2012-0249 of the ``Centro de Excelencia Severo Ochoa" Programme.

\newpage

\bibliographystyle{JHEP}
\bibliography{mybib}

\providecommand{\href}[2]{#2}\begingroup\raggedright\begin{thebibliography}{10}

\bibitem{Becker:1995kb}
K.~Becker, M.~Becker, and A.~Strominger, {\it {Five-branes, membranes and
  nonperturbative string theory}},  {\em Nucl. Phys.} {\bf B456} (1995)
  130--152, [\href{http://arxiv.org/abs/hep-th/9507158}{{\tt hep-th/9507158}}].

\bibitem{Witten:1996bn}
E.~Witten, {\it {Nonperturbative superpotentials in string theory}},  {\em
  Nucl. Phys.} {\bf B474} (1996) 343--360,
  [\href{http://arxiv.org/abs/hep-th/9604030}{{\tt hep-th/9604030}}].

\bibitem{Harvey:1999as}
J.~A. Harvey and G.~W. Moore, {\it {Superpotentials and membrane instantons}},
  \href{http://arxiv.org/abs/hep-th/9907026}{{\tt hep-th/9907026}}.

\bibitem{Witten:1999eg}
E.~Witten, {\it {World sheet corrections via D instantons}},  {\em JHEP} {\bf
  02} (2000) 030, [\href{http://arxiv.org/abs/hep-th/9907041}{{\tt
  hep-th/9907041}}].

\bibitem{Kachru:2003aw}
S.~Kachru, R.~Kallosh, A.~D. Linde, and S.~P. Trivedi, {\it {De Sitter vacua in
  string theory}},  {\em Phys. Rev.} {\bf D68} (2003) 046005,
  [\href{http://arxiv.org/abs/hep-th/0301240}{{\tt hep-th/0301240}}].

\bibitem{Balasubramanian:2005zx}
V.~Balasubramanian, P.~Berglund, J.~P. Conlon, and F.~Quevedo, {\it
  {Systematics of moduli stabilisation in Calabi-Yau flux compactifications}},
  {\em JHEP} {\bf 03} (2005) 007,
  [\href{http://arxiv.org/abs/hep-th/0502058}{{\tt hep-th/0502058}}].

\bibitem{Garcia-Valdecasas:2016voz}
E.~García-Valdecasas and A.~Uranga, {\it {On the 3-form formulation of axion
  potentials from D-brane instantons}},  {\em JHEP} {\bf 02} (2017) 087,
  [\href{http://arxiv.org/abs/1605.08092}{{\tt arXiv:1605.08092}}].

\bibitem{Dvali:2005an}
G.~Dvali, {\it {Three-form gauging of axion symmetries and gravity}},
  \href{http://arxiv.org/abs/hep-th/0507215}{{\tt hep-th/0507215}}.

\bibitem{Dvali:2005ws}
G.~Dvali, R.~Jackiw, and S.-Y. Pi, {\it {Topological mass generation in four
  dimensions}},  {\em Phys. Rev. Lett.} {\bf 96} (2006) 081602,
  [\href{http://arxiv.org/abs/hep-th/0511175}{{\tt hep-th/0511175}}].

\bibitem{Kaloper:2008fb}
N.~Kaloper and L.~Sorbo, {\it {A Natural Framework for Chaotic Inflation}},
  {\em Phys. Rev. Lett.} {\bf 102} (2009) 121301,
  [\href{http://arxiv.org/abs/0811.1989}{{\tt arXiv:0811.1989}}].

\bibitem{Kaloper:2011jz}
N.~Kaloper, A.~Lawrence, and L.~Sorbo, {\it {An Ignoble Approach to Large Field
  Inflation}},  {\em JCAP} {\bf 1103} (2011) 023,
  [\href{http://arxiv.org/abs/1101.0026}{{\tt arXiv:1101.0026}}].

\bibitem{Marchesano:2014mla}
F.~Marchesano, G.~Shiu, and A.~M. Uranga, {\it {F-term Axion Monodromy
  Inflation}},  {\em JHEP} {\bf 09} (2014) 184,
  [\href{http://arxiv.org/abs/1404.3040}{{\tt arXiv:1404.3040}}].

\bibitem{McAllister:2014mpa}
L.~McAllister, E.~Silverstein, A.~Westphal, and T.~Wrase, {\it {The Powers of
  Monodromy}},  {\em JHEP} {\bf 09} (2014) 123,
  [\href{http://arxiv.org/abs/1405.3652}{{\tt arXiv:1405.3652}}].

\bibitem{Retolaza:2015sta}
A.~Retolaza, A.~M. Uranga, and A.~Westphal, {\it {Bifid Throats for Axion
  Monodromy Inflation}},  {\em JHEP} {\bf 07} (2015) 099,
  [\href{http://arxiv.org/abs/1504.02103}{{\tt arXiv:1504.02103}}].

\bibitem{Bielleman:2015ina}
S.~Bielleman, L.~E. Ibanez, and I.~Valenzuela, {\it {Minkowski 3-forms, Flux
  String Vacua, Axion Stability and Naturalness}},  {\em JHEP} {\bf 12} (2015)
  119, [\href{http://arxiv.org/abs/1507.06793}{{\tt arXiv:1507.06793}}].

\bibitem{Ibanez:2015fcv}
L.~E. Ibanez, M.~Montero, A.~Uranga, and I.~Valenzuela, {\it {Relaxion
  Monodromy and the Weak Gravity Conjecture}},  {\em JHEP} {\bf 04} (2016) 020,
  [\href{http://arxiv.org/abs/1512.00025}{{\tt arXiv:1512.00025}}].

\bibitem{Koerber:2007xk}
P.~Koerber and L.~Martucci, {\it {From ten to four and back again: How to
  generalize the geometry}},  {\em JHEP} {\bf 08} (2007) 059,
  [\href{http://arxiv.org/abs/0707.1038}{{\tt arXiv:0707.1038}}].

\bibitem{Koerber:2008sx}
P.~Koerber and L.~Martucci, {\it {Warped generalized geometry
  compactifications, effective theories and non-perturbative effects}},  {\em
  Fortsch. Phys.} {\bf 56} (2008) 862--868,
  [\href{http://arxiv.org/abs/0803.3149}{{\tt arXiv:0803.3149}}].

\bibitem{Blumenhagen:2006xt}
R.~Blumenhagen, M.~Cvetic, and T.~Weigand, {\it {Spacetime instanton
  corrections in 4D string vacua: The Seesaw mechanism for D-Brane models}},
  {\em Nucl.Phys.} {\bf B771} (2007) 113--142,
  [\href{http://arxiv.org/abs/hep-th/0609191}{{\tt hep-th/0609191}}].

\bibitem{Ibanez:2006da}
L.~Ibanez and A.~Uranga, {\it {Neutrino Majorana Masses from String Theory
  Instanton Effects}},  {\em JHEP} {\bf 0703} (2007) 052,
  [\href{http://arxiv.org/abs/hep-th/0609213}{{\tt hep-th/0609213}}].

\bibitem{Florea:2006si}
B.~Florea, S.~Kachru, J.~McGreevy, and N.~Saulina, {\it {Stringy Instantons and
  Quiver Gauge Theories}},  {\em JHEP} {\bf 0705} (2007) 024,
  [\href{http://arxiv.org/abs/hep-th/0610003}{{\tt hep-th/0610003}}].

\bibitem{Blumenhagen:2009qh}
R.~Blumenhagen, M.~Cvetic, S.~Kachru, and T.~Weigand, {\it {D-Brane Instantons
  in Type II Orientifolds}},  {\em Ann.Rev.Nucl.Part.Sci.} {\bf 59} (2009)
  269--296, [\href{http://arxiv.org/abs/0902.3251}{{\tt arXiv:0902.3251}}].

\bibitem{Ibanez:2012zz}
L.~E. Ibanez and A.~M. Uranga, {\it {String theory and particle physics: An
  introduction to string phenomenology}}, .

\bibitem{Ibanez:2007rs}
L.~Ibanez, A.~Schellekens, and A.~Uranga, {\it {Instanton Induced Neutrino
  Majorana Masses in CFT Orientifolds with MSSM-like spectra}},  {\em JHEP}
  {\bf 0706} (2007) 011, [\href{http://arxiv.org/abs/0704.1079}{{\tt
  arXiv:0704.1079}}].

\bibitem{Blumenhagen:2007zk}
R.~Blumenhagen, M.~Cvetic, D.~Lust, R.~Richter, and T.~Weigand, {\it
  {Non-perturbative Yukawa Couplings from String Instantons}},  {\em
  Phys.Rev.Lett.} {\bf 100} (2008) 061602,
  [\href{http://arxiv.org/abs/0707.1871}{{\tt arXiv:0707.1871}}].

\bibitem{Cvetic:2010dz}
M.~Cvetic, J.~Halverson, and P.~Langacker, {\it {Singlet Extensions of the MSSM
  in the Quiver Landscape}},  {\em JHEP} {\bf 1009} (2010) 076,
  [\href{http://arxiv.org/abs/1006.3341}{{\tt arXiv:1006.3341}}].

\bibitem{Cvetic:2010mm}
M.~Cvetic, J.~Halverson, P.~Langacker, and R.~Richter, {\it {The Weinberg
  Operator and a Lower String Scale in Orientifold Compactifications}},  {\em
  JHEP} {\bf 1010} (2010) 094, [\href{http://arxiv.org/abs/1001.3148}{{\tt
  arXiv:1001.3148}}].

\bibitem{Argurio:2006ny}
R.~Argurio, M.~Bertolini, S.~Franco, and S.~Kachru, {\it {Gauge/gravity duality
  and meta-stable dynamical supersymmetry breaking}},  {\em JHEP} {\bf 0701}
  (2007) 083, [\href{http://arxiv.org/abs/hep-th/0610212}{{\tt
  hep-th/0610212}}].

\bibitem{Argurio:2007qk}
R.~Argurio, M.~Bertolini, S.~Franco, and S.~Kachru, {\it {Meta-stable vacua and
  D-branes at the conifold}},  {\em JHEP} {\bf 0706} (2007) 017,
  [\href{http://arxiv.org/abs/hep-th/0703236}{{\tt hep-th/0703236}}].

\bibitem{Cvetic:2008mh}
M.~Cvetic and T.~Weigand, {\it {A String theoretic model of gauge mediated
  supersymmetry beaking}},  \href{http://arxiv.org/abs/0807.3953}{{\tt
  arXiv:0807.3953}}.

\bibitem{Buican:2008qe}
M.~Buican and S.~Franco, {\it {SUSY breaking mediation by D-brane instantons}},
   {\em JHEP} {\bf 0812} (2008) 030,
  [\href{http://arxiv.org/abs/0806.1964}{{\tt arXiv:0806.1964}}].

\bibitem{Blumenhagen:2010dt}
R.~Blumenhagen, A.~Deser, and D.~Lust, {\it {FCNC Processes from D-brane
  Instantons}},  {\em JHEP} {\bf 1102} (2011) 079,
  [\href{http://arxiv.org/abs/1007.4770}{{\tt arXiv:1007.4770}}].

\bibitem{Addazi:2014ila}
A.~Addazi and M.~Bianchi, {\it {Neutron Majorana mass from exotic instantons}},
   {\em JHEP} {\bf 1412} (2014) 089,
  [\href{http://arxiv.org/abs/1407.2897}{{\tt arXiv:1407.2897}}].

\bibitem{Addazi:2015rwa}
A.~Addazi and M.~Bianchi, {\it {Un-oriented Quiver Theories for Majorana
  Neutrons}},  \href{http://arxiv.org/abs/1502.01531}{{\tt arXiv:1502.01531}}.

\bibitem{Addazi:2015hka}
A.~Addazi and M.~Bianchi, {\it {Neutron Majorana mass from Exotic Instantons in
  a Pati-Salam model}},  \href{http://arxiv.org/abs/1502.08041}{{\tt
  arXiv:1502.08041}}.

\bibitem{Argurio:2007vqa}
R.~Argurio, M.~Bertolini, G.~Ferretti, A.~Lerda, and C.~Petersson, {\it
  {Stringy instantons at orbifold singularities}},  {\em JHEP} {\bf 0706}
  (2007) 067, [\href{http://arxiv.org/abs/0704.0262}{{\tt arXiv:0704.0262}}].

\bibitem{Bianchi:2007wy}
M.~Bianchi, F.~Fucito, and J.~F. Morales, {\it {D-brane instantons on the T**6
  / Z(3) orientifold}},  {\em JHEP} {\bf 0707} (2007) 038,
  [\href{http://arxiv.org/abs/0704.0784}{{\tt arXiv:0704.0784}}].

\bibitem{Aldazabal:2000cn}
G.~Aldazabal, S.~Franco, L.~E. Ibanez, R.~Rabadan, and A.~M. Uranga, {\it
  {Intersecting brane worlds}},  {\em JHEP} {\bf 02} (2001) 047,
  [\href{http://arxiv.org/abs/hep-ph/0011132}{{\tt hep-ph/0011132}}].

\bibitem{Aldazabal:2000dg}
G.~Aldazabal, S.~Franco, L.~E. Ibanez, R.~Rabadan, and A.~M. Uranga, {\it {D =
  4 chiral string compactifications from intersecting branes}},  {\em J. Math.
  Phys.} {\bf 42} (2001) 3103--3126,
  [\href{http://arxiv.org/abs/hep-th/0011073}{{\tt hep-th/0011073}}].

\bibitem{Beasley:2004ys}
C.~Beasley and E.~Witten, {\it {New instanton effects in supersymmetric QCD}},
  {\em JHEP} {\bf 01} (2005) 056,
  [\href{http://arxiv.org/abs/hep-th/0409149}{{\tt hep-th/0409149}}].

\bibitem{GarciaEtxebarria:2008pi}
I.~Garcia-Etxebarria, F.~Marchesano, and A.~M. Uranga, {\it {Non-perturbative
  F-terms across lines of BPS stability}},  {\em JHEP} {\bf 0807} (2008) 028,
  [\href{http://arxiv.org/abs/0805.0713}{{\tt arXiv:0805.0713}}].

\bibitem{Blumenhagen:2007bn}
R.~Blumenhagen, M.~Cvetic, R.~Richter, and T.~Weigand, {\it {Lifting
  D-Instanton Zero Modes by Recombination and Background Fluxes}},  {\em JHEP}
  {\bf 10} (2007) 098, [\href{http://arxiv.org/abs/0708.0403}{{\tt
  arXiv:0708.0403}}].

\bibitem{Tripathy:2005hv}
P.~K. Tripathy and S.~P. Trivedi, {\it {D3 brane action and fermion zero modes
  in presence of background flux}},  {\em JHEP} {\bf 06} (2005) 066,
  [\href{http://arxiv.org/abs/hep-th/0503072}{{\tt hep-th/0503072}}].

\bibitem{Bergshoeff:2005yp}
E.~Bergshoeff, R.~Kallosh, A.-K. Kashani-Poor, D.~Sorokin, and A.~Tomasiello,
  {\it {An Index for the Dirac operator on D3 branes with background fluxes}},
  {\em JHEP} {\bf 10} (2005) 102,
  [\href{http://arxiv.org/abs/hep-th/0507069}{{\tt hep-th/0507069}}].

\bibitem{Petersson:2007sc}
C.~Petersson, {\it {Superpotentials From Stringy Instantons Without
  Orientifolds}},  {\em JHEP} {\bf 05} (2008) 078,
  [\href{http://arxiv.org/abs/0711.1837}{{\tt arXiv:0711.1837}}].

\bibitem{Freed:1999vc}
D.~S. Freed and E.~Witten, {\it {Anomalies in string theory with D-branes}},
  {\em Asian J. Math.} {\bf 3} (1999) 819,
  [\href{http://arxiv.org/abs/hep-th/9907189}{{\tt hep-th/9907189}}].

\bibitem{Maldacena:2001xj}
J.~M. Maldacena, G.~W. Moore, and N.~Seiberg, {\it {D-brane instantons and K
  theory charges}},  {\em JHEP} {\bf 11} (2001) 062,
  [\href{http://arxiv.org/abs/hep-th/0108100}{{\tt hep-th/0108100}}].

\bibitem{Witten:1998xy}
E.~Witten, {\it {Baryons and branes in anti-de Sitter space}},  {\em JHEP} {\bf
  07} (1998) 006, [\href{http://arxiv.org/abs/hep-th/9805112}{{\tt
  hep-th/9805112}}].

\bibitem{Hanany:2005ve}
A.~Hanany and K.~D. Kennaway, {\it {Dimer models and toric diagrams}},
  \href{http://arxiv.org/abs/hep-th/0503149}{{\tt hep-th/0503149}}.

\bibitem{Franco:2005rj}
S.~Franco, A.~Hanany, K.~D. Kennaway, D.~Vegh, and B.~Wecht, {\it {Brane dimers
  and quiver gauge theories}},  {\em JHEP} {\bf 0601} (2006) 096,
  [\href{http://arxiv.org/abs/hep-th/0504110}{{\tt hep-th/0504110}}].

\bibitem{Kennaway:2007tq}
K.~D. Kennaway, {\it {Brane Tilings}},  {\em Int. J. Mod. Phys.} {\bf A22}
  (2007) 2977--3038, [\href{http://arxiv.org/abs/0706.1660}{{\tt
  arXiv:0706.1660}}].

\bibitem{Feng:2005gw}
B.~Feng, Y.-H. He, K.~D. Kennaway, and C.~Vafa, {\it {Dimer models from mirror
  symmetry and quivering amoebae}},  {\em Adv. Theor. Math. Phys.} {\bf 12}
  (2008), no.~3 489--545, [\href{http://arxiv.org/abs/hep-th/0511287}{{\tt
  hep-th/0511287}}].

\bibitem{Aharony:1997ju}
O.~Aharony and A.~Hanany, {\it {Branes, superpotentials and superconformal
  fixed points}},  {\em Nucl. Phys.} {\bf B504} (1997) 239--271,
  [\href{http://arxiv.org/abs/hep-th/9704170}{{\tt hep-th/9704170}}].

\bibitem{Aharony:1997bh}
O.~Aharony, A.~Hanany, and B.~Kol, {\it {Webs of (p,q) five-branes,
  five-dimensional field theories and grid diagrams}},  {\em JHEP} {\bf 01}
  (1998) 002, [\href{http://arxiv.org/abs/hep-th/9710116}{{\tt
  hep-th/9710116}}].

\bibitem{Leung:1997tw}
N.~C. Leung and C.~Vafa, {\it {Branes and toric geometry}},  {\em Adv. Theor.
  Math. Phys.} {\bf 2} (1998) 91--118,
  [\href{http://arxiv.org/abs/hep-th/9711013}{{\tt hep-th/9711013}}].

\bibitem{Hanany:2005ss}
A.~Hanany and D.~Vegh, {\it {Quivers, tilings, branes and rhombi}},  {\em JHEP}
  {\bf 10} (2007) 029, [\href{http://arxiv.org/abs/hep-th/0511063}{{\tt
  hep-th/0511063}}].

\bibitem{Franco:2005fd}
S.~Franco, A.~Hanany, and A.~M. Uranga, {\it {Multi-flux warped throats and
  cascading gauge theories}},  {\em JHEP} {\bf 09} (2005) 028,
  [\href{http://arxiv.org/abs/hep-th/0502113}{{\tt hep-th/0502113}}].

\bibitem{GarciaEtxebarria:2006aq}
I.~Garcia-Etxebarria, F.~Saad, and A.~M. Uranga, {\it {Quiver gauge theories at
  resolved and deformed singularities using dimers}},  {\em JHEP} {\bf 0606}
  (2006) 055, [\href{http://arxiv.org/abs/hep-th/0603108}{{\tt
  hep-th/0603108}}].

\bibitem{Franco:2007ii}
S.~Franco, A.~Hanany, D.~Krefl, J.~Park, A.~M. Uranga, and D.~Vegh, {\it
  {Dimers and orientifolds}},  {\em JHEP} {\bf 0709} (2007) 075,
  [\href{http://arxiv.org/abs/0707.0298}{{\tt arXiv:0707.0298}}].

\bibitem{Aharony:2007pr}
O.~Aharony and S.~Kachru, {\it {Stringy Instantons and Cascading Quivers}},
  {\em JHEP} {\bf 0709} (2007) 060, [\href{http://arxiv.org/abs/0707.3126}{{\tt
  arXiv:0707.3126}}].

\bibitem{Amariti:2008xu}
A.~Amariti, L.~Girardello, and A.~Mariotti, {\it {Stringy Instantons as Strong
  Dynamics}},  {\em JHEP} {\bf 0811} (2008) 041,
  [\href{http://arxiv.org/abs/0809.3432}{{\tt arXiv:0809.3432}}].

\bibitem{Argurio:2012iw}
R.~Argurio, D.~Forcella, A.~Mariotti, D.~Musso, and C.~Petersson, {\it {Field
  Theory Interpretation of N=2 Stringy Instantons}},  {\em JHEP} {\bf 1302}
  (2013) 002, [\href{http://arxiv.org/abs/1211.1884}{{\tt arXiv:1211.1884}}].

\bibitem{Franco:2015kfa}
S.~Franco, A.~Retolaza, and A.~Uranga, {\it {D-brane Instantons as Gauge
  Instantons in Orientifolds of Chiral Quiver Theories}},  {\em JHEP} {\bf 11}
  (2015) 165, [\href{http://arxiv.org/abs/1507.05330}{{\tt arXiv:1507.05330}}].

\bibitem{Uranga:2002pg}
A.~M. Uranga, {\it {Local models for intersecting brane worlds}},  {\em JHEP}
  {\bf 12} (2002) 058, [\href{http://arxiv.org/abs/hep-th/0208014}{{\tt
  hep-th/0208014}}].

\bibitem{Franco:2005zu}
S.~Franco, A.~Hanany, F.~Saad, and A.~M. Uranga, {\it {Fractional branes and
  dynamical supersymmetry breaking}},  {\em JHEP} {\bf 01} (2006) 011,
  [\href{http://arxiv.org/abs/hep-th/0505040}{{\tt hep-th/0505040}}].

\bibitem{Klebanov:2000hb}
I.~R. Klebanov and M.~J. Strassler, {\it {Supergravity and a confining gauge
  theory: Duality cascades and chi SB resolution of naked singularities}},
  {\em JHEP} {\bf 0008} (2000) 052,
  [\href{http://arxiv.org/abs/hep-th/0007191}{{\tt hep-th/0007191}}].

\bibitem{Forcella:2008au}
D.~Forcella, I.~Garcia-Etxebarria, and A.~Uranga, {\it {E3-brane instantons and
  baryonic operators for D3-branes on toric singularities}},  {\em JHEP} {\bf
  03} (2009) 041, [\href{http://arxiv.org/abs/0806.2291}{{\tt
  arXiv:0806.2291}}].

\bibitem{Franco:2013ana}
S.~Franco and A.~Uranga, {\it {Bipartite Field Theories from D-Branes}},  {\em
  JHEP} {\bf 04} (2014) 161, [\href{http://arxiv.org/abs/1306.6331}{{\tt
  arXiv:1306.6331}}].

\bibitem{GarciaEtxebarria:2007zv}
I.~Garcia-Etxebarria and A.~M. Uranga, {\it {Non-perturbative superpotentials
  across lines of marginal stability}},  {\em JHEP} {\bf 0801} (2008) 033,
  [\href{http://arxiv.org/abs/0711.1430}{{\tt arXiv:0711.1430}}].

\end{thebibliography}\endgroup

\end{document}